%% file: main.tex
\documentclass[twocolumn]{aastex631}
\usepackage{lineno,xspace}
\usepackage{graphicx}

\newcommand{\factornewclustersmax}{8.5$\times$\xspace}
\newcommand{\factornewclustersmin}{1.8$\times$\xspace}

\shorttitle{Compact 3.3~$\mu$\lowercase{m} PAH emitters in the PHANGS-JWST galaxies}

\shortauthors{Rodríguez, Lee, Indebetouw et al.}
\begin{document}

\title{Tracing the earliest stages of star and cluster formation in 19 nearby galaxies with PHANGS-JWST and HST: compact 3.3~$\mu$m PAH emitters and their relation to the optical census of star clusters}

\input{authors}

\begin{abstract}
The earliest stages of star and cluster formation are hidden within dense cocoons of gas and dust, limiting their detection at optical wavelengths. With the unprecedented infrared capabilities of JWST, we can now observe dust-enshrouded star formation 
with $\sim$10 pc resolution out to $\sim$20~Mpc. Early findings from PHANGS-JWST suggest that 3.3~$\mu$m polycyclic aromatic hydrocarbon (PAH) emission can identify star clusters in their dust-embedded phases. Here, we extend this analysis to 19 galaxies from the PHANGS-JWST Cycle 1 Treasury Survey, providing the first characterization of compact sources exhibiting 3.3~$\mu$m PAH emission across a diverse sample of nearby star-forming galaxies.  We establish selection criteria,
a median color threshold of $\rm F300M–F335M = 0.67$ at F335M $= 20$, and identify of 1816 sources. These sources are predominantly located in dust lanes, spiral arms, rings, and galaxy centers, with $\sim$87\% showing concentration indices similar to optically detected star clusters. Comparison with the PHANGS-HST catalogs suggests that PAH emission fades within $\sim$3 Myr.
The H$\alpha$ equivalent width of PAH emitters is 1–2.8 times higher than that of young PHANGS-HST clusters, providing evidence that PAH emitters are on average younger. 
Analysis of the bright portions of luminosity functions (which should not suffer from incompleteness) shows that young dusty clusters may increase the number of optically visible $\leq$~3~Myr-old clusters in PHANGS-HST by a factor between 
$\sim$\factornewclustersmin -\factornewclustersmax.
\end{abstract}



\section{Introduction} 
\label{sec:intro}

The formation of star clusters, the timescales over which they clear their natal interstellar environments, and the dependencies of these processes on stellar mass, the properties of the natal molecular clouds, and galactic environment are key 
for understanding galaxy evolution and the observed properties of galaxies. Clusters are sites of intense massive star formation, and as they age, they inject energy, momentum, and heavy elements into their surroundings. The feedback from these clusters can drive changes in the structure and composition of the interstellar medium, influencing subsequent star formation and reshaping the gas in the galaxy
\citep[e.g.,][and references therein]{hopkins14,2016SAAS...43...85K, Chevance20, grudic21, evaadam24}. 

Previous studies, notably those based on the morphology of HII regions based on H$\alpha$ imaging with HST which is required to resolve clusters in most nearby galaxies, have estimated a timescale $\sim$3 Myr for clusters to clear the dust and gas from their surroundings \citep[e.g.,][]{hollyhead15, hannon19, hannon22}.
These clusters, which are detected in the optical with HST, have cleared enough of their natal dust to allow the photospheres of the massive stars to be observed, 
and the age and mass of the cluster to be estimated based on our understanding of stellar evolution in the context of simple stellar populations \citep[e.g.,][]{adamo17, turner21}.  
Earlier evolutionary phases 
can be sufficiently dust embedded that they are not detected in the UV, and the very youngest are not detected in the optical at all.  Progress requires a large sample of young embedded clusters, selected at infrared or longer wavelengths to overcome the limitations of optical and UV observations  and to provide statistically robust constraints on the timescales for clearing natal dust and gas across a range of environments.

The sensitivity and resolution of JWST NIRCam enable 
embedded clusters and their environments to be studied with unprecedented detail in
nearby ($\lesssim$20~Mpc) galaxies, with a resolution of $\lesssim$10~pc. 
Thus, very recent studies on this topic have used JWST infrared observations to study dusty young clusters in select nearby galaxies (NGC~7496, NGC~1365, NGC~3351, NGC~628, NGC~3256) using Pa$\alpha$, Br$\alpha$, and/or 3.3 $\mu$m PAH (Polycyclic Aromatic Hydrocarbon) emission \citep[][]{2023ApJ...944L..26R,2023ApJ...944L..14W, sun24, pedrini24, linden24}.  Interestingly, short dust clearing timescales (few Myr) have also been reported from these studies, supporting the findings based on HST observations.

In our own work early in the JWST mission \citep{2023ApJ...944L..26R}, we demonstrated that 3.3~$\mu$m PAH emission is a particularly powerful tracer of the youngest star clusters that are still enveloped by dust and gas, and are not detected in the optical with HST.  The analysis was based on a sample of compact 3.3~$\mu$m PAH emitters in the spiral galaxy NGC~7496, the first galaxy to be observed for the PHANGS-JWST Cycle 1 Treasury program \citep{phangs-jwst, phangs-jwst-pipeline} in July 2022. 


PAHs are molecules made up of hydrogen and carbon atoms arranged in multiple aromatic rings. They are commonly found in various environments including molecular clouds, disks around young stars, the diffuse interstellar medium, star-forming regions, HII regions, and planetary and reflection nebulae. PAHs play a pivotal role in galaxies, contributing substantially to their total integrated infrared luminosity, with PAHs contributing up to 20\% of the total infrared luminosity in some galaxies \citep{Smith2007}. These molecules exhibit prominent infrared (IR) emission features at wavelengths of 3.3, 6.2, 7.7, 8.6, 11.3, and 12.7~$\mu$m \citep[e.g.][]{1989A&A...216..148L, 2008ARA&A..46..289T, li20}. This distinctive emission is produced by vibrational modes excited by the absorption of 5-15eV ultraviolet (UV) photons \citep[][and refs therein]{dh21_dielectric}. 
As most of this UV is emitted by intermediate and massive stars, PAHs have been investigated as star formation indicators, which works well near massive star formation regions and needs calibration for larger regions including diffuse gas \citep[e.g., ][and references therein]{kenicutt12, zhang23, belfiore23}.  

In particular, the 3.3~$\mu$m PAH feature arises from a C–H stretching vibrational mode of small PAHs which have limited heat capacities \citep{1993ApJ...415..397S}, which makes them easily excitable by single UV photons or prone to destruction. In addition, this feature generally requires more intense or harder radiation fields than the longer-wavelength PAH features. Consequently, the 3.3~$\mu$m PAH emission is particularly sensitive to the radiation environment, offering a distinctive signature compared to other PAH emission features at longer wavelengths \citep{2013PASJ...65..103Y}. 

The 3.3~$\mu$m PAH feature has not been as extensively studied other PAH features \citep[][]{li20}. Spitzer's InfraRed Spectrograph (IRS) covered a spectral range between 5.2 and 38 $\mu$m.   Although it was captured by Spitzer InfraRed Array Camera (IRAC),  the wide bandwidth of IRAC1 ($\sim$20\%; $\delta\lambda/\lambda = 0.75/3.55$) was not useful for isolating the 3.3~$\mu$m emission \footnote{\url{https://irsa.ipac.caltech.edu/data/SPITZER/docs/irac/iracinstrumenthandbook/6/}}.  Observations of this feature had previously been carried out by AKARI (68.5~cm primary mirror) and the Infrared Space Observatory (60~cm primary mirror).  The resulting studies provide most of our pre-JWST knowledge of the nature of this emission.  Given the small apertures of these earlier IR observatories, studies of the 3.3~$\mu$m PAH feature generally were largely limited to studies of the Milky Way and Magellanic Clouds, or to spatially unresolved studies of the brightest galaxies and AGN \citep[][]{spoon2000,sturm2000,kim12,li20}.
As far as the authors know, the first attempts to use 3.3 $\mu$m PAH emission as a star formation indicator were based on AKARI spectra of a small number of star-forming galaxies \citep[e.g.][]{Imanishi_2010,kim12,2013PASJ...65..103Y}. 
Later \cite{Mori2014} found this emission in Galactic HII regions.


 With JWST NIRCam, 3.3~$\mu$m PAH emission can be imaged with the F335M filter at resolutions of 2.6--10.6~pc (PSF FWHM 0\farcs11) at 5--20 Mpc, the range of distances of the PHANGS galaxies \citep{phangs-jwst, phangs-jwst-pipeline}.    Given the diffraction limit, JWST observations of the 3.3~$\mu$m feature also have 2-3 times better angular resolution than PAH features at longer wavelengths. 
 
 Here, we expand our initial study of compact sources with 3.3~$\mu$m PAH emission \citep{2023ApJ...944L..26R} to the full sample of 19 galaxies from the PHANGS-JWST Cycle 1 Treasury Program, and provide the largest study to-date of embedded star clusters and their dust clearing timescales.  
We aim to check the timescale of $\sim$2~Myr based on our previous analysis of NGC~7496, using an order-of-magnitude larger sample drawn from a diversity of galactic environments.  We identify compact 3.3~$\mu$m PAH sources and separate them from older infrared emitting sources (e.g. AGB stars, PNe).  With this expanded sample, we seek to characterize the observed properties of the youngest star clusters as traced by 3.3~$\mu$m PAH emission, and to gain insight into their evolutionary status through comparison with published catalogs of star clusters from PHANGS-HST, and gas+dust maps from PHANGS-JWST and PHANGS-ALMA. We compute the fraction of the compact 3.3~$\mu$m population associated with young visible clusters identified by the PHANGS-HST survey \citep{phangs-hst, Maschmann2024}, and the fraction that can be considered to be dust embedded.   


The outline of the paper is as follows.  Section~\ref{sec:Data} provides an overview of the data employed in our analysis. The source detection and photometry procedure are detailed in Sect.~\ref{sec:det&phot}, while Sect.~\ref{sec:Identification} outlines the methods employed for identifying the PAH emitters. The properties of these objects are explored in Sect.~\ref{sec:Properties}. The comparison of these objects with the clusters from HST cluster catalogs is presented in Sect.~\ref{sec:HST-clusters-comparison} and a discussion of our findings is presented in Sect.~\ref{sec:Discussion}. Finally, our key conclusions are summarized in Sect.~\ref{sec:Conclusions}.


\section{Data} \label{sec:Data}


This paper is based on data from three PHANGS HST and JWST Treasury programs which provide UV-IR imaging from 0.27~$\micron$ to 21~$\micron$ for 19 galaxies \citep[][]{phangs-hst, phangs-jwst, chandar24}\footnote{HST15654, PI Lee; JWST2107, PI Lee; HST 17126, PI Chandar}.  The galaxy sample is representative of spiral galaxies along the star-forming main sequence, which have solar gas-phase metallicities.    The sample consists of the  
19 galaxies observed by PHANGS in the first year of JWST science operations, i.e., those for which observations are available from all of the PHANGS principal surveys with HST, ALMA, and MUSE \citep{phangs-hst, phangs-alma, phangs-muse}.  The list of the galaxies and their properties can be found in \citet[Table 1]{phangs-jwst}.  The galaxies and their distances are also provided in Table~\ref{tab:source_detection} in this paper for the reader's convenience.

Infrared observations with NIRCam and MIRI obtained through the PHANGS-JWST Cycle 1 Treasury program included imaging in 8 filters, spanning 2.0~$\micron$ to 21~$\micron$ (NIRCam: F200W, F300M, F335M, and F360M; MIRI: F770W, F1000W, F1130W, and F2100W). Detailed information regarding the JWST observing strategy and data reduction procedures can be found in \citet{phangs-jwst} and \citet{phangs-jwst-pipeline}.  We use the v1p1 version of the PHANGS-JWST reduced images, which were released in 2024 January and available at \url{https://archive.stsci.edu/hlsp/phangs}.   

A companion HST Cycle 30 Treasury program obtained H$\alpha$ narrow-band imaging for the sample \citep{chandar24}.  As of the writing of this paper, H$\alpha$ observations have been completed for all galaxies except for NGC4535 and the northern pointing for NGC~2835, which is scheduled to be re-observed in January 2025 due to a guide star acquisition failure.  A release of the available H$\alpha$ reduced and continuum subtracted images is expected in the later half of 2024 and will be available at \url{https://archive.stsci.edu/hlsp/phangs}.

Basic characteristics of the PHANGS-JWST Cycle 1 imaging, including total exposure times and PSF size, are provided in Table~\ref{tab:JWST-observations}.  For the PHANGS-HST survey, which provide the broadband UV-optical imaging, we provide a quick summary of the corresponding quantities here.  The Cycle 26 PHANGS-HST Treasury survey obtained NUV-U-B-V-I observations of 38 galaxies from 2019-2021.  The exposure times were $\sim$2200 s (NUV), $\sim$1100 s (U), $\sim$1100 s (B), $\sim$670 s (V), and $\sim$830 s (I). Exposure times varied depending on whether appropriate imaging was already available in the archive, and exact values for each galaxy are provided in \citet[][Table 1]{Maschmann2024}.  The HST images were drizzled with a pixel size of 0\farcs04, and the WFC3 UVIS PSF size in the V-band is in principle the same as the NIRCam PSF size (0\farcs067)\footnote{\url{https://hst-docs.stsci.edu/wfc3ihb/chapter-6-uvis-imaging-with-wfc3/6-6-uvis-optical-performance}}, though detector under-sampling degrades the resolution.

PHANGS-HST has recently completed the largest census to-date of $\sim$100,000 optically-selected star clusters and compact associations across 38 spiral galaxies \citep[][]{Maschmann2024, thilker24} \footnote{\url{https://archive.stsci.edu/hlsp/phangs}}.  Catalogs of the observed properties were released in 2024 January, 
and are available at PHANGS high level science product (HLSP) website at MAST\footnote{\url{https://archive.stsci.edu/hlsp/phangs}}. 
The catalogs are the result of pipeline efforts as summarized in \cite{phangs-hst} to establish improved techniques for cluster candidate detection and selection \citep{whitmore21, 2022Thilker}, photometry \citep{deger22}, and automated morphological classification using machine learning techniques \citep{wei20, whitmore21, hannon23}.  

As in \citet{2023ApJ...944L..26R}, we use the PHANGS-HST cluster catalog to place our study of the early stages of star, cluster, and ISM evolution in the context of the greater population which is no longer enshrouded by dust.  The PHANGS-HST cluster catalog provides ages, stellar mass, and dust reddening from spectral-energy distribution (SED) fitting of 5-band NUV, U, B, V, I photometry \citep[][the latter paper will support the catalog release of SED-inferred physical properties expected by late 2024]{turner21, thilker24}.  As effectively single-age populations, the clusters can be used as clocks to constrain the timescale of compact 3.3~$\mu$m PAH emission, and ultimately of the embedded phase.

The PHANGS-HST optically-selected cluster census includes populations with ages spanning from $\sim$1~Myr to the age of the universe, and over four decades of stellar mass up to $\sim10^{6}M_\odot$.  Separate catalogs are available for clusters that have been morphologically classified by human and convolutional neural networks \citep{wei20, whitmore21, hannon23}.  Here, we consider clusters in both ``human classified'' and ``machine learning'' catalogs which are classified as Class 1 (symmetric compact cluster), and Class 2 (asymmetric compact cluster).  The number of clusters identified in each galaxy is provided in \citet[][Table 3]{Maschmann2024}.  The number of human (machine) classified clusters in each of the 19 PHANGS-JWST galaxies studied here ranges from 182 to 774 (182 to 2828), with a median of 480 (777) and a total sample size of 8608 (21353).  We note that these numbers are slightly different from those listed in \cite{Maschmann2024} because of small differences in the HST and JWST footprints.




\begin{figure*}
    \centering
    \includegraphics[width=2\columnwidth]{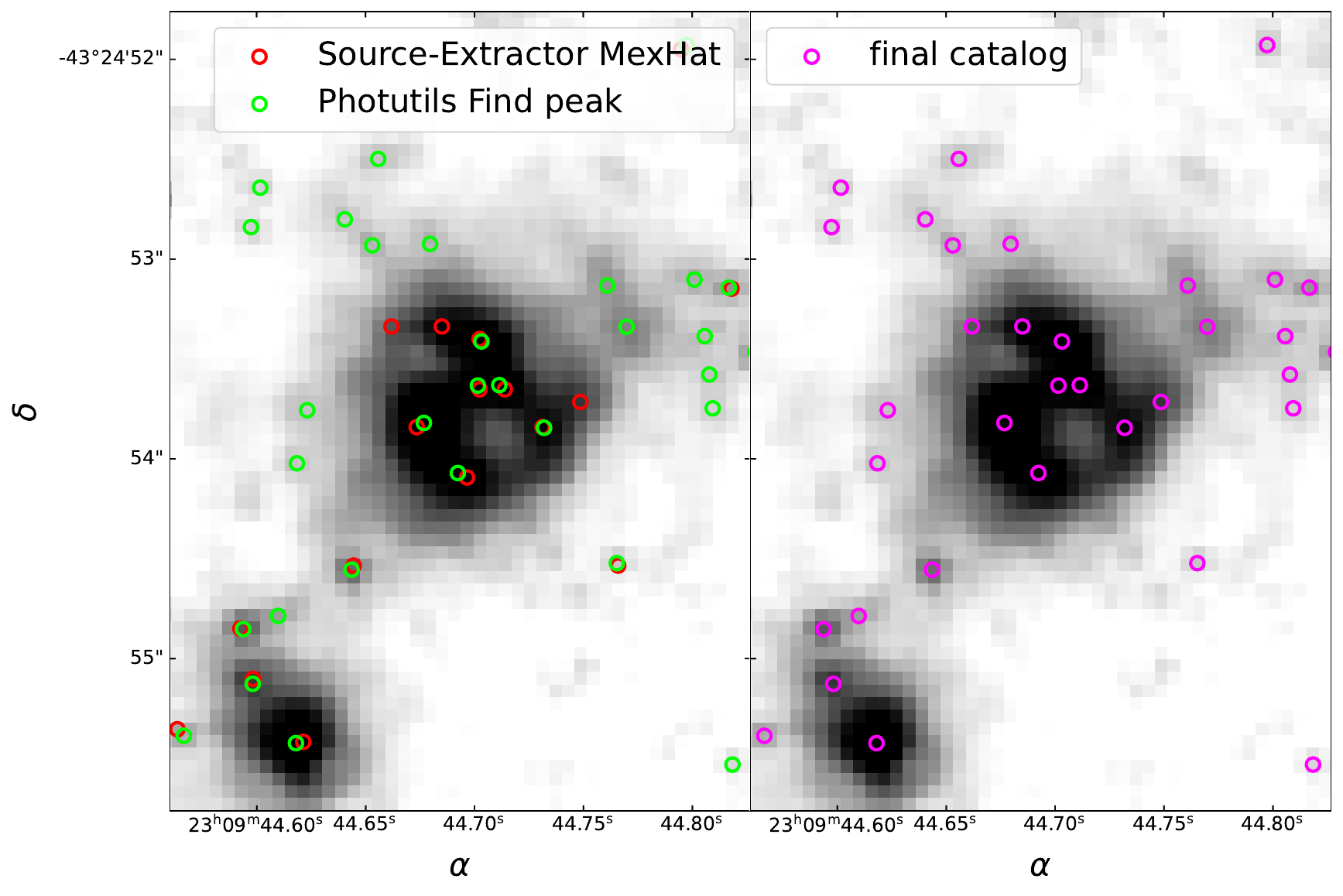}
    \caption{F335M image of a crowded region in the galaxy NGC~7496. The left panel displays the sources detected using both the \textit{find\_peaks} algorithm of Photutils and the \textsc{SExtractor} Mexican\-hat filter. It is noticeable that the \textsc{SExtractor} detects three sources (red circles) missed by \textit{find\_peaks} (green circles) which can be attributed to its better performance in deblending sources. In the right panel, we present the final catalog of sources for the same region, produced as a combination of the two aforementioned methods (see Sect~\ref{sec:detection}) for details.}
    \label{fig:source-selection}
\end{figure*}

\section{Source identification and photometry} 
\label{sec:det&phot}

\subsection{NIRCam F335M Source Detection}
\label{sec:detection}

Our procedures build on the methods used to study dust-embedded star clusters in NGC~7496 \citep{2023ApJ...944L..26R}, the first galaxy observed for the PHANGS-JWST Treasury program in the first month of JWST science operations.  We perform source detection on images taken with the NIRCam F335M filter, which captures the 3.3~$\mu$m PAH feature as well as the stellar continuum.


We employ the {\it find\_peaks} algorithm of the {\sc Photutils} \citep{larry_bradley_2022_6825092} Astropy package. 
This algorithm scans the image for local maxima above a specified intensity threshold value, separated by a specified minimum number of pixels. We adopted a value of 3 pixels, equivalent to 1.7 times the full width at half maximum (FWHM) of the point-spread function (PSF) in F335M (Table~\ref{tab:JWST-observations}), which corresponds to  $\sim$ 5 to 18~pc for the range of distance of the studied galaxies (Table~\ref{tab:source_detection}). 
This choice proved effective, particularly given that the objects of interest are predominantly situated within crowded regions.

To determine appropriate local intensity threshold values in the images, we produced 2D background images for each galaxy. We estimated the background in boxes of 85~pc$^2$, using the {\it SExtractorBackground} function of {\sc Photutils}. As described in the {\sc Photutils} documentation (\url{https://photutils.readthedocs.io/en/stable/index.html}), the background in each box is calculated as $(2.5 \times median) - (1.5 \times mean)$, except in cases where $(mean - median) / std > 0.3$ in which the median is used instead.
To choose the background box size, we considered several factors. The region should be larger than the typical size of compact sources, small enough to capture the local variations of the background, but also large enough to include enough pixels to able statistics to be robustly calculated. Additionally, to be consistent across our galaxy sample, we choose a common physical box size that meets these conditions. To determine this box size, we tested different values in the galaxies NGC~628 (9.84~Mpc) and NGC~7496 (18.72~Mpc).
After visually inspecting the results in these two galaxies, we found that using a box size in pixels corresponding to 85~pc (Table~\ref{tab:source_detection}, column 4)  successfully identified all or most of the visually recognizable objects in both galaxies. In contrast, larger or smaller box sizes resulted in numerous spurious detections at the noise level or the loss of faint objects in one or both galaxies.
Finally, the background image was processed with a median filter
to suppress local overestimations or underestimations, such as those caused by bright regions within specific boxes.
The detection threshold value was set at the background level plus 3 times the sigma-clipped standard deviation.

Based on visual inspection (e.g.,  Fig.~\ref{fig:source-selection} left panel), the {\it find\_peaks} algorithm provides good results in the detection of faint and extended sources as well as point-like bright sources but appears to miss sources in crowded regions. 
In those regions {\sc SExtractor} \citep[Source-Extractor][]{1996A&AS..117..393B} using a Mexican hat filter produces improved results, since the algorithm includes the option for source deblending.   Thus, we also run {\sc SExtractor} using a Mexican hat filter with 64 deblending sub-thresholds and the same background parameters used with the {\it find\_peaks} method. 
In the left panel of Fig.~\ref{fig:source-selection} we show the sources detected using {\it find\_peaks} (green) and {\sc SExtractor} (red) over a crowded region in the northern area of the galaxy NGC~7496.  We can see here that {\sc SExtractor} detects three additional sources in the ring, but does not capture many of the faint sources beyond the ring. For this reason, our final source catalog is a combination of the sources detected using both algorithms. We perform a cross-matching of the sources positions in both catalogs using a search radius of 0\farcs126 (2 pixels in the F335M image). We keep all sources detected using {\it find\_peaks} and add the sources that were only detected with {\sc SExtractor} for the final source catalog. 
In the right panel of Fig.~\ref{fig:source-selection}, we show the sources in the final catalog after the cross-correlation of both methods. 

In Table~\ref{tab:source_detection} we list the number of sources detected with each method, and the number of sources detected only with {\sc SExtractor}. Again, the number of sources in the final catalog is generally the number of sources identified with {\it find\_peaks} plus those that were only found with {\sc SExtractor}.  The fraction of sources added with {\sc SExtractor} ranges from less than 1\% in IC5332 to 36\% and in NGC3627 with a median of 8\%.  


\begin{table*}[]
\centering
\begin{tabular}{l|cccccccc}
\hline
Galaxy&d$[Mpc]$& (1px=X~pc) &box[px]&N$_{FPeak}$ & N$_{SE}$ & N$_{SE-add}$ & N$_{total}$ & $\rm F300M-F335M>5\sigma$\\
\hline 
NGC~5068 & 5.2 & 1.59 & 54  & 136547 & 30691 & 327 & 136874 & 96\\
IC~5332 & 9.01 & 2.75 & 31  & 43845 & 11992 & 50 & 43895 & 30\\
NGC~0628 & 9.84 & 3.01 & 28  & 119008 & 56592 & 877 & 119885 & 83\\
NGC~3351 & 9.96 & 3.04 & 28 & 80583 & 64513 & 5267 & 85850 & 45\\
NGC~3627 & 11.32 & 3.46 & 25  & 79559 & 112383 & 40699 & 120258 & 105\\
NGC~2835 & 12.22 & 3.73 & 23 & 34345 & 16454 & 136 & 34481 & 41 \\
NGC~4254 & 13.1 & 4.0 & 21 & 106078 & 108340 & 22009 & 128087 & 188\\
NGC~4321 & 15.21 & 4.65 & 18 & 59883 & 19128 & 1023 & 60906 & 154\\
NGC~4535 & 15.77 & 4.82 & 18  & 39321 & 5073 & 189 & 39510 & 49\\
NGC~1087 & 15.85 & 4.84 & 18  & 23155 & 15989 & 1765 & 24920 & 112 \\
NGC~4303 & 16.99 & 5.19 & 16  & 65738 & 72403 & 17938 & 83676 & 135\\
NGC~1385 & 17.22 & 5.26 & 16  & 34268 & 17179 & 1978 & 36246 & 239\\
NGC~1566 & 17.69 & 5.4 & 16  & 27318 & 21898 & 3606 & 30924 & 134\\
NGC~1433 & 18.63 & 5.69 & 15  & 71194 & 6078 & 487 & 71681 & 12 \\
NGC~7496 & 18.72 & 5.72 & 15 & 8248 & 5327 & 559 & 8807 & 34\\
NGC~1512 & 18.83 & 5.75 & 15 & 41319 & 20884 & 1596 & 42915 & 19\\
NGC~1300 & 18.99 & 5.8 & 15  & 39210 & 8742 & 555 & 39765  & 27\\
NGC~1672 & 19.4 & 5.93 & 14  & 72109 & 34974 & 2411 & 74520 & 265\\
NGC~1365 & 19.57 & 5.98 & 14  & 22142 & 15542 & 2709 & 24851 & 294\\
\end{tabular}
\caption{Parameters used in the sources detection and number of sources detected. The columns refer to galaxy's name, distance \citep[][both from PHANGS-HST TRGB measurements and compiled from the literature]{phangs-jwst, anand20}, physical scale corresponding to 1 pixel in the F335M image (NIRCam long wavelength channel, 0.031 $^{\prime\prime}$ pix$^{-1}$)\footnote{\url{https://jwst-docs.stsci.edu/jwst-near-infrared-camera/nircam-observing-modes/nircam-imaging\#gsc.tab=0}}, box size used for the background estimation (in pixels corresponding to 85~pc), number of sources detected using \textit{find\_peaks}, number of sources detected using \textsc{SExtractor}, number of sources only detected with \textsc{SExtractor} and number of sources in the combined catalog. The last column lists the number of objects detected with $\rm F300M - F335M > 5\sigma$ and F300M and F335M above the $\rm 5\sigma$ detection limits (see Section~\ref{sec:CMD}, and green points in Figure~\ref{fig:selection}).}
\label{tab:source_detection}
\end{table*}

\begin{table*}[]
\centering
\begin{tabular}{l|ccccccc}
\hline
Filter & detector& t$_{exp}$ & PSF FWHM & PSF FWHM & Aperture radius&Aperture radius & Aperture Correction\\
 & & [s] & [arcsec] & [px] & [arcsec]& [px]& [mag]\\
\hline 
F200W  & NIRCam & 1202.5 & 0.066 & 2.129 & 0.124 & 4 & -0.63 \\
F300M & NIRCam & 386.5 & 0.100 & 1.587 & 0.124 & 2 & -0.68 \\
F335M & NIRCam & 386.52 & 0.111	& 1.762 & 0.124 & 2 & -0.66 \\
F360M & NIRCam & 429.5 & 0.120 & 1.905 & 0.124 & 2 & -0.67 \\
F770W & MIRI & 88.8 & 0.269 & 2.445 & 0.168 & 1.5 & -0.75\\
F1000W & MIRI & 122.1 & 0.328 & 2.982 & 0.209 & 1.9 & -0.75\\
F1130W & MIRI & 310.8 & 0.375 & 3.409 & 0.236 & 2.14 & -0.75\\
F2100W & MIRI & 321.9 & 0.674 & 6.127 & 0.420 & 3.8 & -0.75\\
\end{tabular}
\caption{JWST observations.
PSF FWHM NIRCam: https://jwst-docs.stsci.edu/jwst-near-infrared-camera/nircam-performance/nircam-point-spread-functions
PSF MIRI: https://jwst-docs.stsci.edu/jwst-mid-infrared-instrument/miri-performance/miri-point-spread-functions}
\label{tab:JWST-observations}
\end{table*}

\subsection{Photometry \& Aperture Corrections}
\label{Sect:Potometry}


We perform photometry using {\sc Photutils} \citep{larry_bradley_2022_6825092} using circular apertures centered on the position of the sources detected on the F335M image as described in the previous section. Photometry is measured in up to 6 HST and 8 JWST filters.  For HST these include F275W, F336W, F438W or F435W, F555W and F814W; F658N or F657N.  For JWST the bandpasses are F200W, F300M, F335M, F360M, F770W, F1000W, F1130W, F2100W. 

For the HST images we use apertures with radii of 0\farcs158 (equivalent to 4 WFC3 pixels) and select annuli to compute the background between 0\farcs277 and 0\farcs356 (7-9 WFC3 pixels), consistent with the aperture sizes used for the PHANGS-HST clusters catalog, which roughly correspond to the half-light radius of clusters \citep{deger22}. Similarly, for the NIRCam bands we use a circular aperture with radius of 0\farcs124, corresponding to four pixels for the NIRCam short wavelength channel (F200W) and 2 pixels for the long wavelength channel (F300M, F335M, F360M).  For the MIRI bands, the PSF is significantly larger (up to a factor of ten larger for F2100W relative to F200W), and we simply adopt an aperture corresponding to the 50\% of the encircled energy\footnote{https://jwst-docs.stsci.edu/jwst-mid-infrared-instrument/miri-performance/miri-point-spread-functions}. The width of the background annulus for both NIRCam and MIRI bands is set to be equal to the radius of the aperture as listed in Table~\ref{tab:JWST-observations}. To exclude sources from the measurement of the background we performed sigma clipping with a maximum of 5 iterations to remove pixels in the annulus above the 3$\sigma$ level.
Detailed analysis of spectral energy distributions with such a significant change in angular resolution will be carried out in the future (e.g. by convolving models to the appropriate resolutions);  in this work we only use the longer-wavelength photometry in F2100W and F1000W to impose a conservative color cut and remove a few percent of our sample as suspected red evolved stars ($\S$\ref{sec:filtering}), and to qualitatively examine the shape of the SEDs of compact 3.3 $\mu$m emitters ($\S$\ref{sec:Pah_emitters} and $\S$\ref{sec:categories}).

In optical images, photometric uncertainty is usually dominated by Poisson noise, and by how well one can estimate a locally flat noise background measured in an annulus.  In the mid-infrared, photometric uncertainty is usually dominated by the estimate of the background, which is highly structured and varies significantly even on the scales of the background annulus.  
We have tested several methods to determine this contribution and derive appropriate photometric uncertainties.  For H$\alpha$ and JWST images F300M and longer wavelengths, we perform sigma clipping of the distribution of pixel values in the annulus, and then take the difference between the 0.1 and 0.9 quantiles. Sigma clipping and using the standard deviation instead of quantiles have only modest effect on the result -- these measures do not decrease as the square root of the annulus size, as does the estimation for a locally flat background (as used in the optical).


For the  HST imaging, we applied aperture corrections derived in \cite{deger22} using a carefully selected sample of bright isolated PHANGS-HST star clusters, including both young and old clusters. For NIRCam bands we derived aperture corrections following the same procedure as in \cite{deger22}, but for photometry obtained using a radius of 4 NIRCAM pixels (i.e., 0\farcs124) using isolated old globular clusters in NGC~628. The values (listed in Table~\ref{tab:JWST-observations}) show good consistency ranging from 0.63~mag in F200W to 0.68~mag in F300M. This is similar to the average aperture correction from \cite{deger22} (i.e., 0.67~mag), which is reasonable since the spatial resolution in the NIRCAM short wavelength channel is similar to HST. Only the brightest old globular clusters have sufficient IR flux in NIRCAM and MIRI to be used for this measurement. 

For MIRI images, we applied a factor of 2.0 to the flux 
measured from a circular aperture which captures 50\% of the encircled energy of a point source. This simplification, compared to NIRCAM, is motivated by the much poorer spatial resolution of MIRI, which results in spatial profiles for clusters that are essentially the same as for stars.


\section{Selection of compact 3.3$\mu$\lowercase{m} PAH emitters}
\label{sec:Identification}

\begin{figure*}
    \centering    \includegraphics[width=2\columnwidth,height=0.90\textheight]{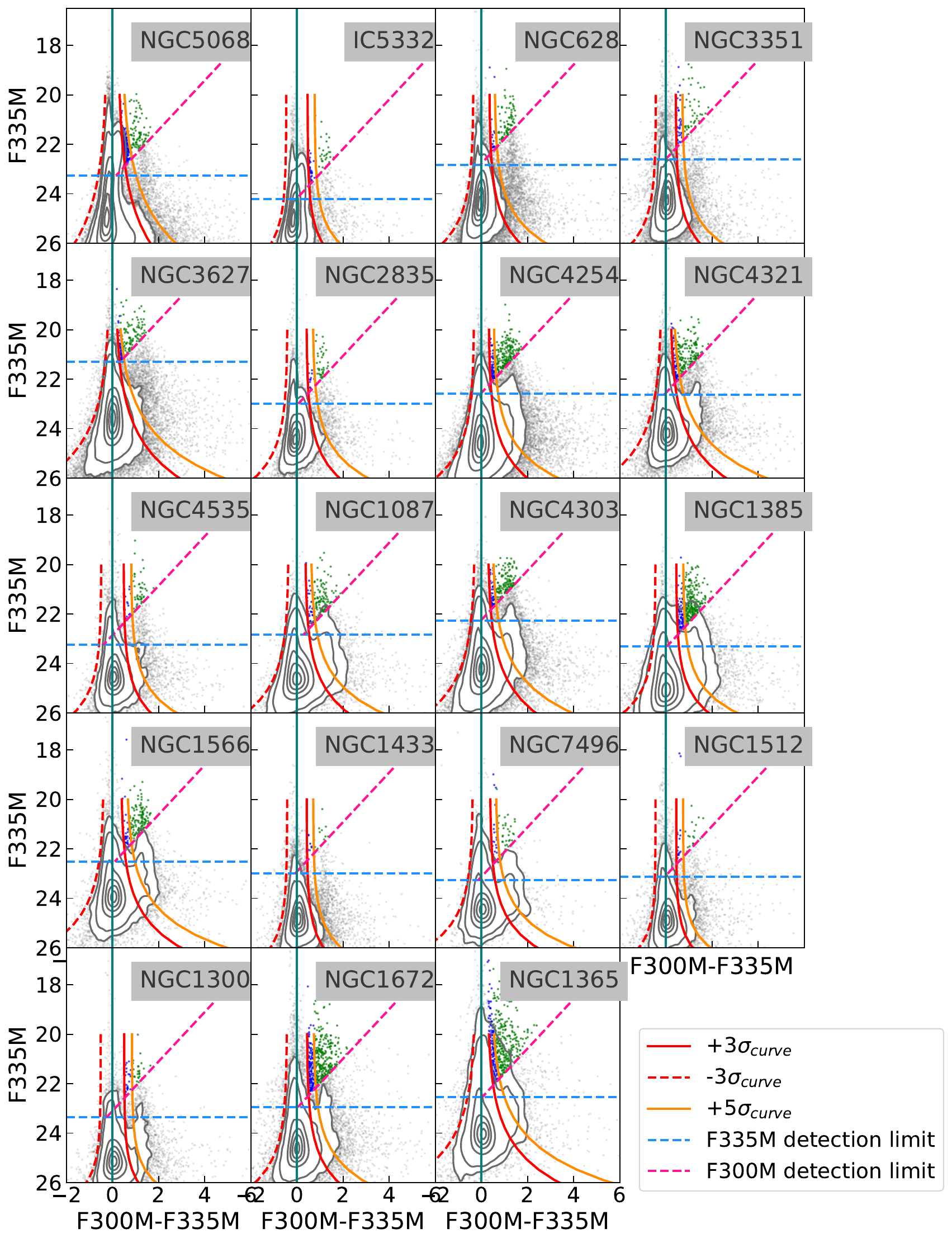}
    \caption{Color-magnitude diagrams (CMDs) F335M vs. $\mathrm{F300M-F335M}$ for the 19 galaxies included in the PHANGS--JWST cycle 1 Treasury Program. The plots are organized by increasing galaxy distance from the upper left corner to the bottom right. The contours represent the 5th, 10th, 30th, 50th, 70th, and 90th density percentiles. The small grey dots indicate regions where the density is lower than the 5th percentile. The red and orange curves indicate $\pm 3\sigma$ and +5$\sigma$ in color dispersion respectively. The dashed horizontal light blue line indicates the adopted F335M detection limit for each galaxy, while the F300M limit is shown with a dashed pink line. These values were derived from 5$\sigma$ measurements relative to the mean obtained via random apertures. The vertical line indicates $\rm F300M-F335M=0$. The blue dots show sources with color excess between 3-5$\sigma$, while the green dots are sources with $\rm F300M-F335M > 5\sigma$ and comprise our primary compact PAH emitter sample.}
    \label{fig:selection}
\end{figure*}

\subsection{Color-Magnitude Diagram and Sample Selection}
\label{sec:CMD}

Using the aperture-corrected photometry described in the previous section, we construct F335M vs. $\mathrm{F300M-F335M}$ color-magnitude diagrams (CMDs) for all 19 PHANGS-JWST Cycle 1 galaxies (Figure~\ref{fig:selection}).  F335M covers the 3.3$\mu$m PAH feature, while F300M primarily probes stellar and dust continuum.  In our previous work, we used this CMD to identify objects with significant PAH emission through detection of a $\mathrm{F300M-F335M}$ color excess \citep[Figure 2 in][]{2023ApJ...944L..26R}. These CMDs are analogous to those used to select emission-line galaxies in narrowband imaging surveys \citep[e.g.,][]{ly11,jlee12}, and similar strategies can be adopted to identify and characterize stellar populations associated with PAH emission.


To select PAH emitters, we employ a combination of two thresholds: (1) the significance of the color excess (2) the F335M and F300M detection limits.  
The locus of points around $\rm F300M-F335M\sim0$ represents continuum sources with little-to-no PAH emission.  The distribution of continuum sources broadens at fainter F335M magnitudes, due to increasing noise in the $\rm F300M-F335M$ measurement \citep[compare with Fig.~3 in][, and see $\S$\ref{Sect:Potometry}]{jlee12}. To select sources with significant color excess relative to the combined intrinsic color spread plus noise, we assume symmetry in the color distribution of continuum sources and measure the dispersion of negative $\rm F300M-F335M$ values, in bins of F335M. We then determine the 1$\sigma$ curve by fitting an exponential function of the form 
$\mathrm{\sigma_{(F300M-F335M)} = a e^{b(F335M)}} + c $. 
The curves in Figure~\ref{fig:selection} represent values 3 and 5 times the dispersion computed in this way.  


We select a primary sample of compact PAH emitters above 5 $\sigma$ (green points in Fig.~\ref{fig:selection}). The number of objects in this sample is listed in the last column of Table~\ref{tab:source_detection}. Sources with color excess between 3-5 $\sigma$ are also highlighted in Fig.~\ref{fig:selection} (blue points).  Later in the paper we examine whether there differences in the properties of these weaker PAH emitters.

The 5$\sigma$ F335M and F300M detection limit for each galaxy was derived by measuring photometry using 500 randomly positioned apertures on the F335M and F300M images respectively, and computing the dispersion of the resulting distribution of measurements. 
The F335M limit is represented with a light blue dashed horizontal line in Fig.~\ref{fig:selection}, while the F300M limit is represented by the dashed pink line. 

There is considerable variation in the F335M detection limit from galaxy to galaxy - the median 5$\sigma$ limit is 23.6 AB mag with a range of 22.0-24.9.  Computing the detection limit for stars and clusters in PHANGS imaging is complicated.  These small structures must be detected above the background light of the galaxy, which can be highly non-uniform.  This threshold will be higher and more variable from galaxy to galaxy (and indeed among different regions within the galaxies themselves) than a limit based on the instrumental and Poisson noise from the source and sky background, which should be relatively consistent since the exposure time and other observational parameters are fixed for all targets \citep[$\sim$26 AB; $t_{exp}\sim$400 s; Table 4,][]{phangs-jwst}. In F335M, the galaxy background is dominated by diffuse PAH emission and the unresolved light from stars.  As expected the F335M detection limit correlates with the star formation rate of the galaxy \cite[Table 1][]{phangs-hst}.  In Section~\ref{sec:spatial} the F335M images of the galaxy sample makes clear the substantial range in galaxy surface brightness, corresponding to lowest F335M detection limit for IC~5332, and the highest in NGC~3627.

With the aim of providing a general recipe for selecting compact 3.3$\mu$m PAH emitters that readers can apply to other galaxies, we report the median of the 5$\sigma$ curve (Fig.~\ref{fig:selection}) at a magnitude of F335M$ = $20. This corresponds to $\rm F300M–F335M = 0.67$, with a range of 0.35–0.84, using apertures of 0\farcs{124}. It  should be noted that this limit is highly sensitive to the background of the host galaxy as discussed above.


\begin{figure*}
    \centering
    \includegraphics[width=2\columnwidth]{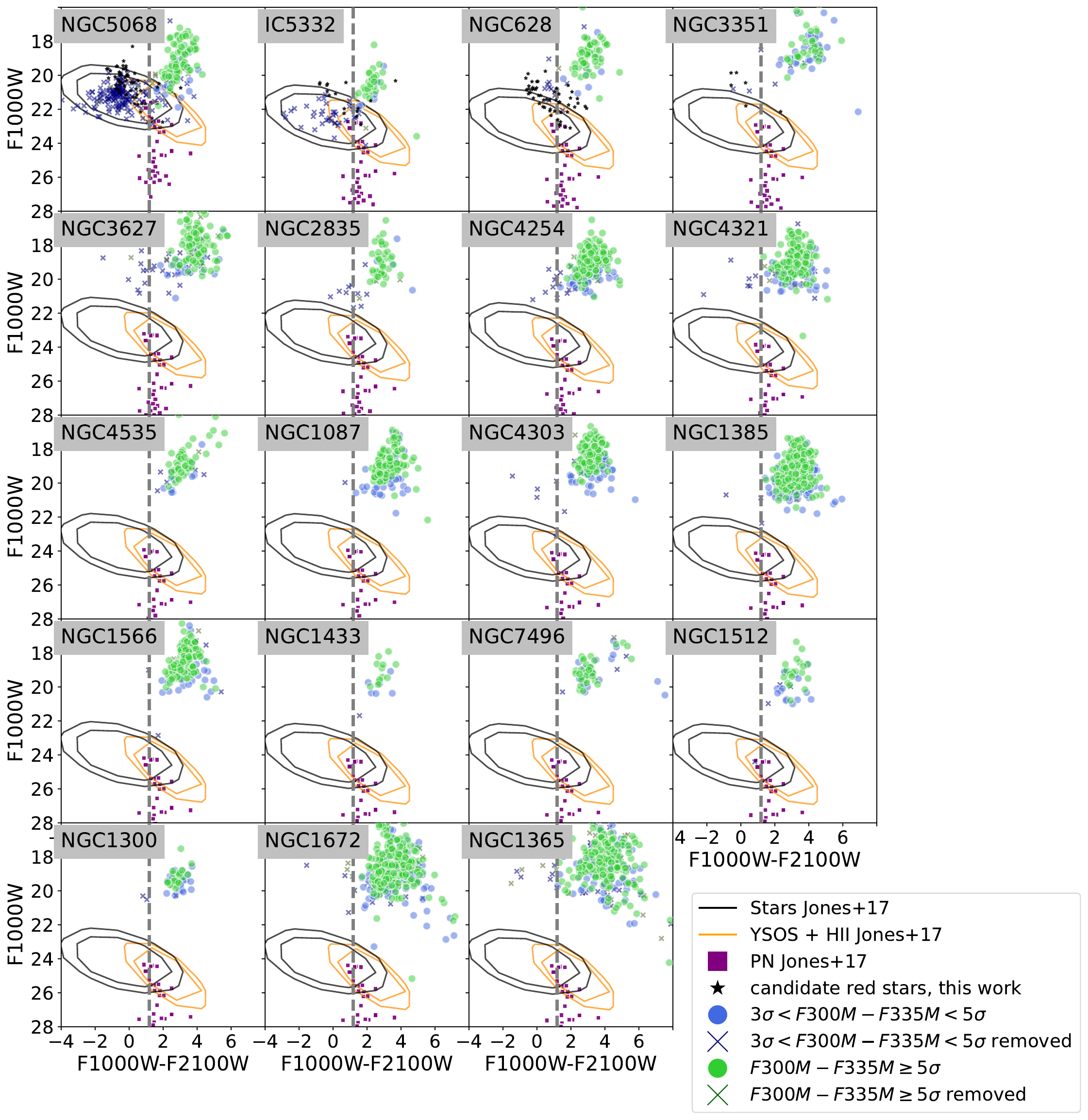}
    \caption{F1000W vs. $\mathrm{F1000W-F2100W}$ CMDs showing the compact 3.3~$\mu$m PAH emitters (green and blue points) from this analysis and a comparison sample from \cite{Jones2017} which includes point-like sources in the LMC: PN (purple points), YSOS and HII regions (orange contours) and different kind of stars (black contours representing 98th and 99.9th percentiles). Objects that were removed from our 3.3~$\mu$m PAH emitters sample are shown with crosses.  Our sample of red evolved stars (see Appendix A) are shown with black points for galaxies at distances $<$ 10~Mpc. The vertical line at $\mathrm{F1000W-F2100W}$ = 1.2~mag ($F_{\rm 21\mu m}/F_{\rm 10\mu m}$ = 3) corresponds to the adopted limit (see text for details). As in all the plots, galaxies are sorted by distance. The fact that the angular resolution scales with wavelength for both the JWST and Spitzer \citep{Jones2017} data points should be kept in mind when interpreting this plot.}
    \label{fig:CMD_21_10_jones_strong}
\end{figure*}

\subsection{Exclusion of Old Objects} 
\label{sec:filtering}

Our sample of 3.3~$\mu$m PAH emitters may encompass a variety of objects, among which are evolved stars that have undergone significant mass loss \citep[e.g.,][]{galliano08}, resulting in high optical obscuration and pronounced brightness in the near and mid-infrared due to the circumstellar envelope. These objects include carbon-rich AGB stars \citep[e.g.,][]{2022Groenewegen} and planetary nebulae (PNe) \citep[e.g.][]{2013Ohsawa}. As the primary focus of this work is on young dusty objects, we wish to remove these older objects from our compact 3.3~$\mu$m PAH sample.  Star forming regions will contain a larger amount of cooler dust, while the high optical depth dust surrounding AGB stars will be warmer. Thus, we expect that dusty young objects should be distinguished by a rising SED in the mid-infrared, and the ratio $F_{\nu}(F2100W)/F_{\nu}(F1000W)$ should be larger than 1.  A conservative choice of removing sources with $F_{\nu}(F2100W)/F_{\nu}(F1000W)<$3 is justified in Appendix~\ref{append:AGBs}. The $\sim$2 times
broader PSF 
at F2100W compared to F1000W increases $F_{\nu}(F2100W)/F_{\nu}(F1000W)$, so we may retain some evolved stars in confused regions, but we won't accidentally exclude any {\em bona fide} young clusters by applying this color cut.

We also apply criteria to remove objects from the sample whose $\mathrm{F300M-F335M}$ color may reflect a rising near-infrared SED rather than PAH emission,  
and limit the sample to sources with $F_{\nu}(F360M)/F_{\nu}(F335M) < 1$.

We compare the 10$\mu$m and 21$\mu$m properties of our final cleaned selection of 3.3~$\mu$m PAH emitters with a sample from \cite{Jones2017}, which includes more than 1000 point sources in the Large Magellanic Clouds (LMC) observed with the Spitzer IRS and reduced by the SAGE-Spec Spitzer legacy program \citep{Kemper2010}. \cite{Jones2017} derived MIRI synthetic photometry for this sample, which encompasses various types of objects, including young stellar objects (YSO), HII regions, main-sequence stars, AGB stars, post-AGB, and PNe. 

Fig.~\ref{fig:CMD_21_10_jones_strong} shows the F1000W vs. F1000W-F2100W CMD \citep[similar to Fig.~3c of ][]{Jones2017}, where in each panel we place the SAGE LMC sample at the distance of parent galaxy, to compare the distributions of the various populations with our final selection of PAH emitters  (Fig.~\ref{fig:CMD_21_10_jones_strong}). 
We show the PNe, all the different stellar objects from \cite{Jones2017} under the `star' category, and combine all the types of YSOs$+$HII regions into the `YSO and HII regions' category. We also plot a PHANGS-HST candidate sample of red evolved stars  (Appendix~\ref{append:AGBs}) for galaxies at distances less than 10 Mpc (first row of panels only), where individual red old stars may still dominate the light subtended by the 10$\micron$ and 21$\micron$ PSF (0\farcs328 and 0\farcs674 FWHM, respectively).  

The vertical line in these plots corresponds to the adopted limit of $F_\nu(\mathrm{F2100W})/F_\nu(\mathrm{F1000W})=$ 3 (or F1000W-F2100W=1.2 AB mag).
We observe from these plots that this adopted value helps to separate the \cite{Jones2017} LMC sample of stars from YSOs$+$HII regions (both populations shown as contours in Fig.~\ref{fig:CMD_21_10_jones_strong}). While PAH emitters tend to exhibit similar $\mathrm{F1000W-F2100W}$ colors but brighter F1000W magnitudes compared to YSOs$+$HII regions, there is overlap between these populations in the closest galaxies. This suggests that the physical resolution and depth in NGC~5068 and IC~5332, enables the detection of compact clusters dominated by a single massive YSO as well as individual HII regions which we discuss further in the next section (\ref{sec:myso_perhaps}).

We also observe that PNe are very faint in F1000W, much lower than the point source detection limit in the emptiest areas of the images\citep[$\sim$23 AB;][Table 4]{phangs-jwst}. Therefore, it is unlikely that our sample would be contaminated by these objects. 
We also conducted a visual inspection of the images to identify potential background galaxies within our sample. However, no clear contaminants of this type were observed.

\subsection{The 3.3~$\mu$m PAH Young Cluster Candidate Sample}
\label{sec:Pah_emitters}
\begin{figure*}
    \centering
    \includegraphics[width=\columnwidth]{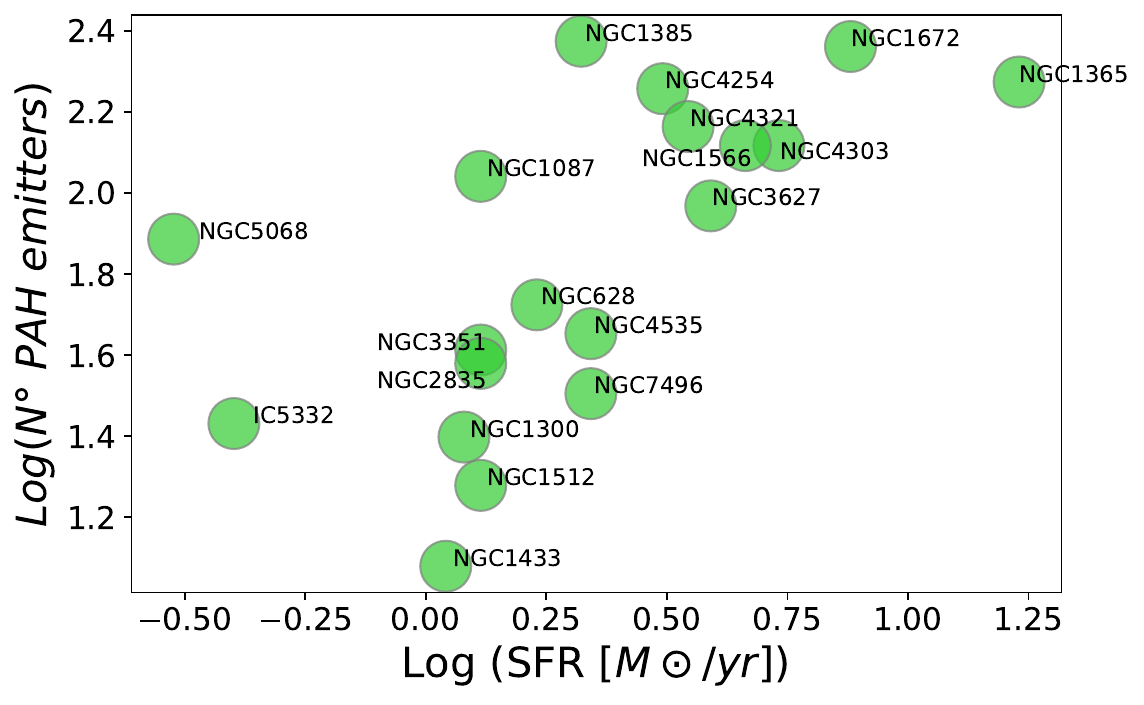}
    \includegraphics[width=\columnwidth]{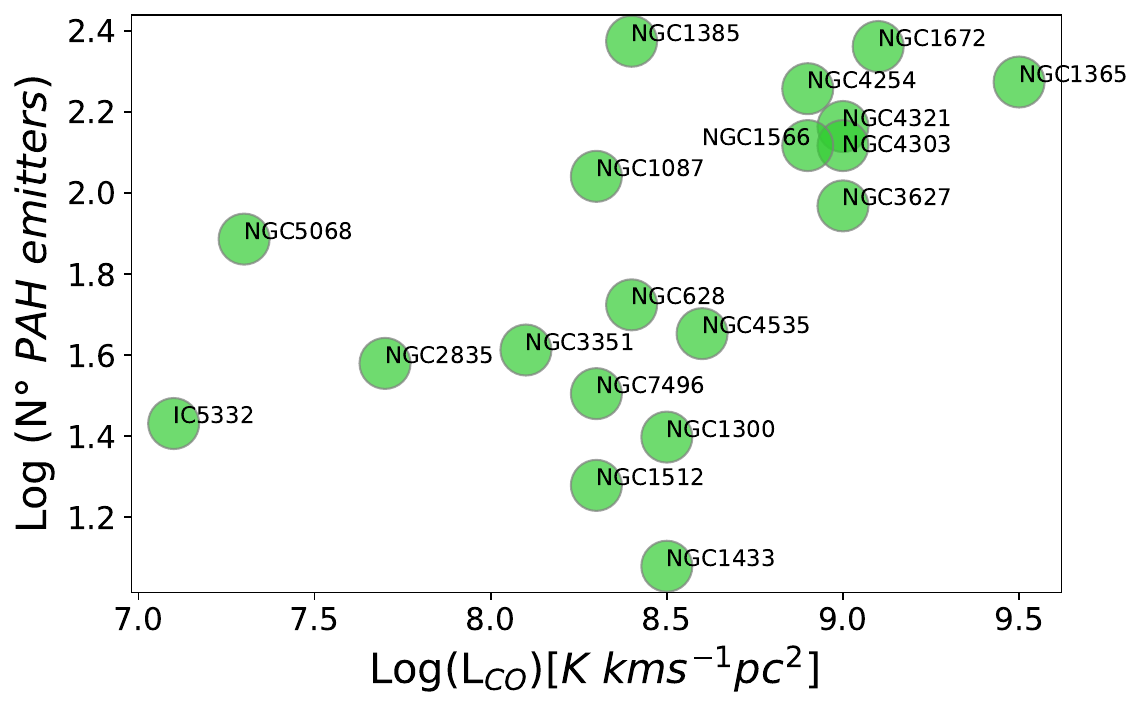}
    \caption{Number of PAH emitters detected above the 5$\sigma$ color dispersion (see Sect~\ref{sec:CMD} and \ref{sec:filtering}) against the total galaxy SFR, ({\it left}) and the integrated CO (2-1) luminosity  (L$_{CO}$, {\it right}). Values for SFR and L$_{CO}$ are from \citet[][Table~2]{phangs-jwst}. The number of compact PAH  correlates with the SFR and L$_{CO}$ as would be expected.}
    \label{N_pah_galaxies_properties}
\end{figure*}

\begin{figure*}
    \centering
    \includegraphics[width=2\columnwidth]{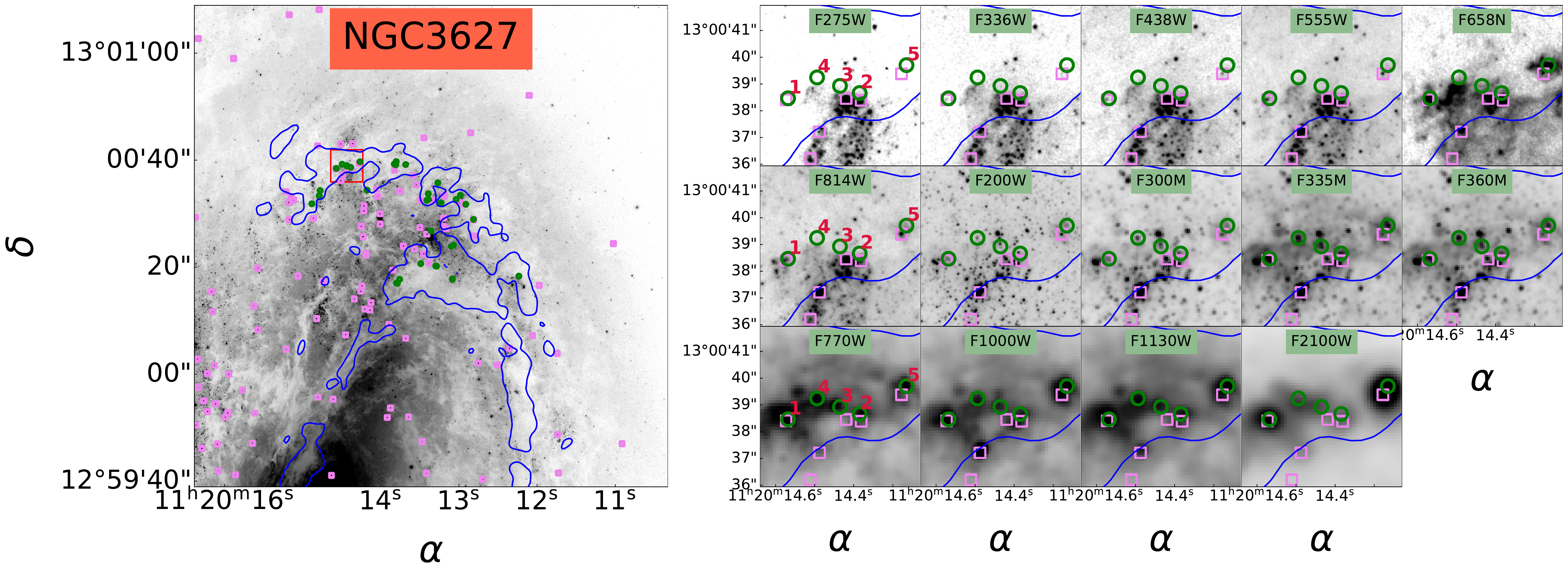}
    \caption{
    Example of compact 3.3$\mu$m PAH emitters (green) detected in a northern region of the galaxy NGC3627, located at a distance of 11.32Mpc. The left panel shows an HST F555W image of the northern part of the galaxy, with the red box indicating the region displayed in the zoomed-in view on the right, across the 14 observed bands. The sample includes objects which are completely undetected in the HST broad bands (objects 2, 3 and 4), others which are undetected or very faint in the HST broad bands but appear very bright in the H$\alpha$ narrow band (5), as well as objects which are visible in all or almost all of the HST broad bands (1). PHANGS-HST clusters from \cite{Maschmann2024} younger than 10~Myr are shown with pink squares and ALMA CO(2-1) intensity contours are shown in blue. We will discuss the spatial distribution of PAH emitters in relation to both the CO contours and HST clusters in $\S$~\ref{sec:spatial_dist_comparison}. }
    \label{fig:example_region}
\end{figure*}

\begin{figure}
    \centering
    \includegraphics[width=\columnwidth]{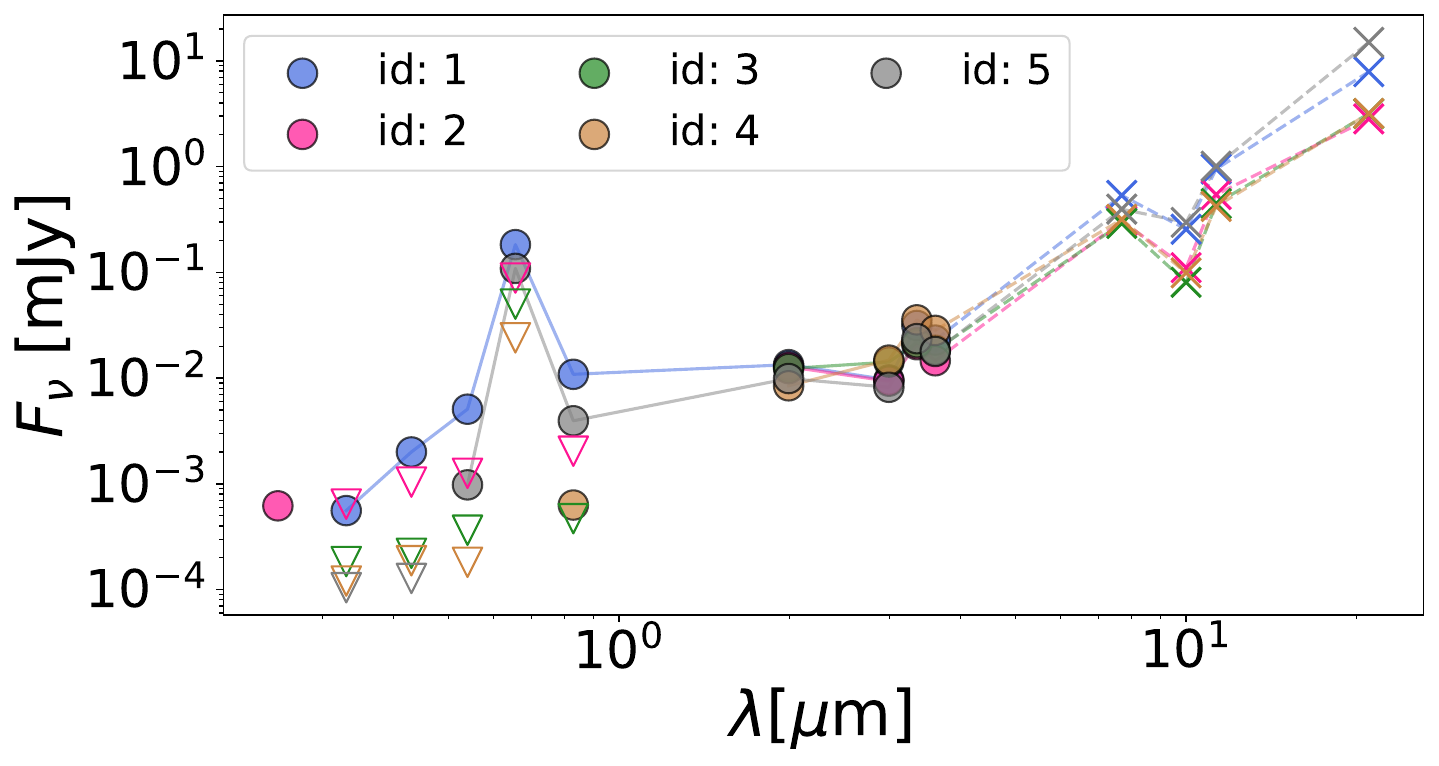}
    \caption{Observed SED for the five objects shown in Fig~\ref{fig:example_region}. Upper limits indicated by open downward-pointing triangles are presented for objects that are not detected in the HST bands.  For MIRI bands detections, upper limits are shown by Xs. In this case there is a detection, but due to the coarser resolution of MIRI, the resulting fluxes correspond to a larger physical scale than those shown for the HST and NIRCam bands.
    }
    \label{fig:sed_example_region}
\end{figure}

After applying the $F_\nu(\mathrm{F360M})/F_\nu{\mathrm{F335M}} < 1$ and $F_\nu(\mathrm{F2100W})/F_\nu(\mathrm{1000W}) = 3$ conditions to remove contaminants, we keep a total of 1816 objects detected above the 5$\sigma$ color dispersion 
across the 19 galaxies; this is 88\% of the original color magnitude selected sample. 
Tables~\ref{tab:PAH-HST-fractions} 
provides the number of compact 3.3~$\mu$m PAH emitters which are likely associated with young dusty stellar populations for each of the galaxies. 

In Fig.~\ref{N_pah_galaxies_properties}, we show the number of PAH emitters against the total galaxy star formation rate (SFR) and the galaxy integrated CO~(2-1) luminosity L$_{CO}$ from \citet{phangs-jwst}. The number of PAH emitters generally increases with both the SFR and L${CO}$ as may be expected, highlighting that the number of compact PAH emitters should scale with the star formation activity and gas supply of the galaxy.  The sample size spans from 12-237 (NGC~1433-NGC~1385), with a median of 77.  These sample sizes will be discussed later in the paper in relation to the number of optically detected young clusters as reported in the PHANGS-HST star cluster catalog \citep{Maschmann2024}.
  
Figure~\ref{fig:example_region} presents examples of the detected compact PAH emitters across the 14 observed bands from HST-UV to JWST 21$\mu$m. These objects are located in a northern region of the galaxy NGC~3627 (see right panel).
Some of them are undetected or very faint in the HST bands (objects 2, 3 and 4), while others, despite not being detected or being very faint in the HST broad bands, appear clearly in the HST H$\alpha$ narrow band (5). Other objects in our sample such as object 1 in this example, are detected in the HST broad bands. All these objects show increasing brightness observed in the near-IR and mid-IR bands. We will discuss these different categories in $\S$~\ref{sec:categories}.
Figure~\ref{fig:sed_example_region} shows the observed SED for these five objects. Objects not detected in HST broad bands are shown with upper limits.  The flux densities corresponding to the MIRI bands are also shown as "x"s, since we adopted apertures covering larger physical scales than those used for the HST and NIRCam bands (due to the lower MIRI resolution; Table~\ref{tab:source_detection}).
We can distinguish the emission feature at H${\alpha}$ and 3.3~$\mu$m as well as an apparent dip at 10~$\mu$m, compared to the neighboring 7.7 and 11.3$\mu$m filters. 
This spectral shape is likely due to the PAH emission at 7.7~$\mu$m and 11.3~$\mu$m that does not contribute significantly to F1000W. It is possible that some objects have true silicate absorption at 10$\mu$m. However, the 9.7$\mu$m silicate absorption optical depth is $\sim$1/10 $A_V$ \citep{1994A&A...291..943O,RiekeLebofsky,dh21_dielectric}. Absorption is sometimes seen in spectra of individual YSOs at high resolution \citep[e.g.][]{sagespec}, and towards AGN \citep{spoon2000}, so unlikely to be common in these star formation regions with larger apertures.  Followup mid-IR spectroscopy will be required to establish whether the F1000W point in these objects represents actually silicate absorption or is simply the dust continuum in between the PAH features.



\subsection{Possible Detection of Individual Massive YSOs \& OB Stars in the Nearest Galaxies}
\label{sec:myso_perhaps}

Careful examination of the CMD for NGC~5068 and IC~5332 (first two panels in Figure~\ref{fig:selection}) reveals an additional population which manifests as a knobby overdensity, generally located between the 3$\sigma$ and 5$\sigma$ curves; i.e., in the region of 0.5 $\lesssim$ F335M-F300M $\lesssim$ 1, and a limited range of magnitudes: 21 $\lesssim$ F335M $\lesssim$ 23 in NGC~5068, and 22.5 $\lesssim$ F335M $\lesssim$ 24.5 in IC~5332.  One can even make out signs of such a  population in NGC~0628, although it is fainter and blends into the noise.  

To investigate further, we examined the Milky Way ISO HII region sample described in \citet{peeters-iso}, using the high level data products retrieved from the ISO archive.  The sources are a few to 10~pc in size, and have bolometric luminosities of a few $\times$ 10$^4$ to a few times 10$^6$~L$\odot$ and ionizing photon luminosities measured from centimeter continuum of 10$^{47}$ to a few times 10$^{50}$ photons$\;s^{-1}$ -- exactly what is expected for one or a few OB stars still associated with their nascent gas \citep{martins05}.  
We convolved the spectra with JWST filter profiles to synthesize F335M and F300M fluxes, and find that these MW ISO HII regions have F335M-F300M color of 0.5-1 magnitudes, and at the distances of NGC~5068 and IC~5332, F335M magnitudes consistent with the location of the knobby overdensity of 3-5$\sigma$ PAH sources.  The nature of this population in the most nearby galaxies will be further investigated in future work, but the properties are consistent with the detection of individual OB stars (or clusters with a handful of them) beginning to clear their natal material.

\begin{figure*}
    \centering
    \includegraphics[width=2\columnwidth]{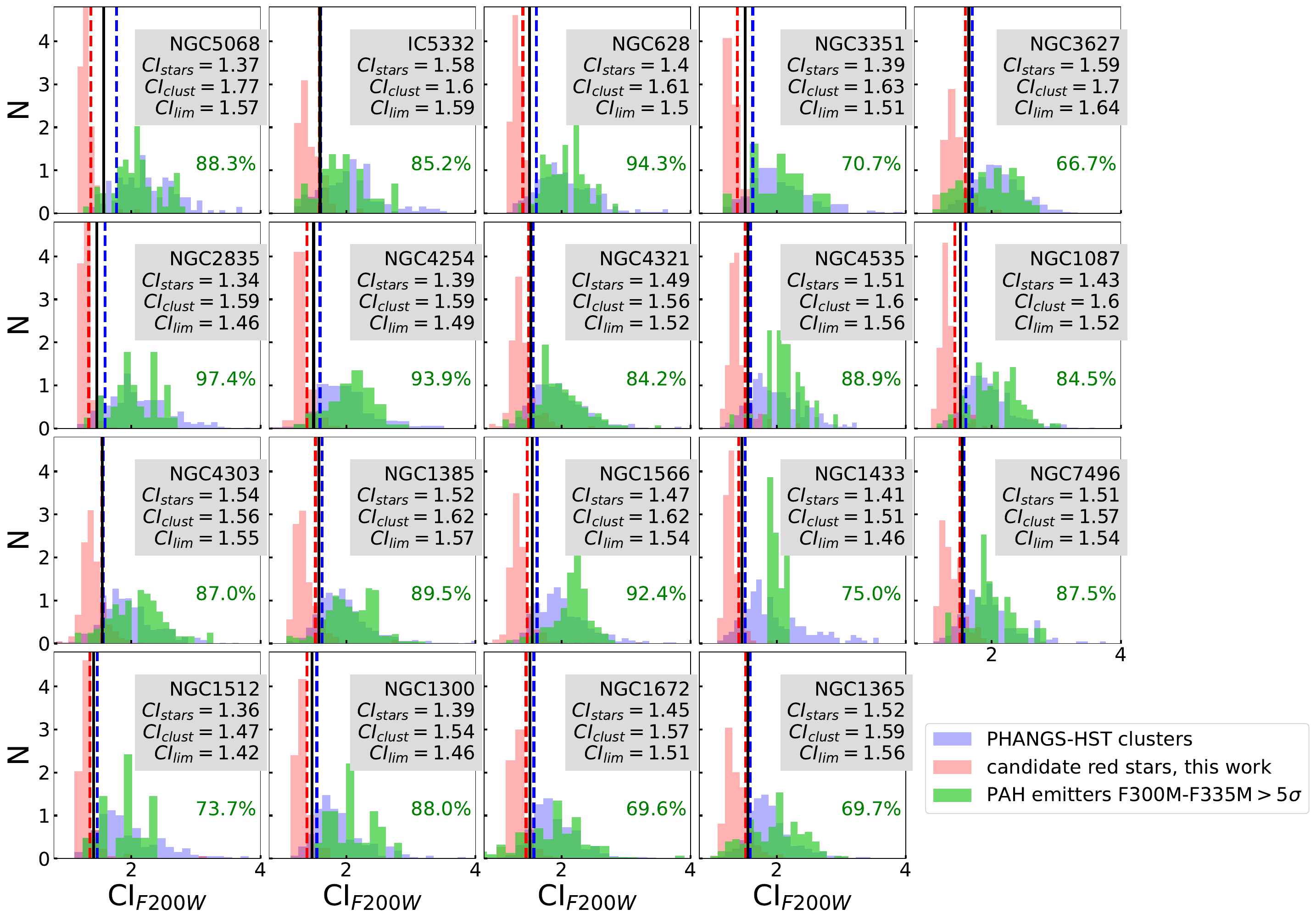}
    \caption{F200W concentration index (CI) -- difference between F200W photometry measured in circular apertures with radii of 1 and 4 pixels-- distribution for PHANGS-HST Class 1 and 2 star clusters (blue), candidate red evolved stars (red) and $>5\sigma$ PAH emitters (green).  The red dashed line indicates the 84\% quantile of the candidate red stars, denoted as CI$_{\rm stars}$ in the top right corner of each plot. The blue dashed line corresponds to the 16\% quantile of the cluster population, and its value is indicated as CI$_{\rm clust}$. The black solid line indicates the mean value between these two quantiles, indicated as CI$_{\rm lim}$, which can be used to distinguish cluster-dominated and point source/star-dominated samples. Each panel includes in green the percentage of PAH emitters with CI values consistent with star clusters, which varies across galaxies with a median of $\sim$70\%. The plots are organized by increasing galaxy distance from the upper left corner to the bottom right. Histograms are normalized by the total number of objects and the width of the bin.}
    \label{fig:hist_ci}
\end{figure*}



\section{Properties of 3.3~$\mu$\MakeLowercase{m} PAH emitters}\label{sec:Properties}

\subsection{Concentration Index Analysis}
\label{sec:concentration_index}


To investigate the nature of the 3.3~$\mu$m PAH emitters, and assess the likelihood that they are star clusters, we conducted a Concentration Index (CI) analysis. The CI serves as a metric to quantify the degree of central concentration in the light distribution of an object, providing an indication of its compactness. This method has been widely employed in various HST studies of star clusters in nearby galaxies to differentiate them from individual stars \cite[e.g.,][]{chandar10b,whitmore14,cook19,deger22}. In \cite{2023ApJ...944L..14W}, published as part of the ApJL PHANGS-JWST First Results Issue, we found that a CI computed from JWST F200W imaging for NGC~1365 improves the distinction between stars and clusters compared with CI computed from HST F814W imaging. The CI for star clusters in nearby galaxies is computed as the difference in photometry measured using circular apertures. \cite{2023ApJ...944L..14W} adopts radii of 1 and 4 pixels for the JWST F200W imaging, and 1 and 3 pixels for the HST F814W imaging.  The outer radii both subtend 0\farcs12, given the smaller PSF of the NIRCam short wavelength detectors compared with HST WFC3 UVIS (pixel scales are 0\farcs03 vs. 0\farcs0.4, respectively).  Based on this definition, we found that a value of CI = 1.4, based on JWST F200W imaging, was determined to be a good threshold between point source/star-dominated and cluster-dominated populations in that one galaxy.

Now expanding the analysis to all galaxies, we computed CI$_\mathrm{F200W}$ for all clusters in the PHANGS-HST clusters catalogs \citep[][]{Maschmann2024} and for red evolved star candidates (selected as explained in Appendix~\ref{append:AGBs}). 
Aperture photometry for these small apertures was conducted using a fractional pixels method to determine the overlap of the aperture on the pixel grid.
In Fig.~\ref{fig:hist_ci} we present the distribution of CI values for the red evolved star candidates (red), HST star clusters (blue) and the $>5\sigma$ PAH emitters (green) for each galaxy. The red dashed line indicates the 84\% quantile of the red evolved star candidates, the blue dashed line indicates the 16\% quantile of the HST cluster population, and the black solid line indicates the mean between these two values. We adopt the mean value to distinguish between stars and clusters. These three CI values are also indicated in the top corner of each panel in Fig.~\ref{fig:hist_ci}, and are slightly larger ($\sim$10\%), but generally consistent with the value of 1.4 reported in \citet{2023ApJ...944L..14W}.  As expected, it becomes more difficult to distinguish clusters from point sources as the distance of the parent galaxy increases from 5 Mpc to 20 Mpc.  At 10 Mpc, the F200W PSF (FWHM 0\farcs066) subtends 3.2 pc, and the 1 pixel and 4 pixel radii used to compute CI subtend 1.5 pc and 6 pc.
Star clusters typically have half-light radii of a few parsecs \citep{pz10,ryon17,krumholz19,brown2021}.

To the left of  the vertical black line, we find the most compact objects with CI values typically associated with stars, while to the right, slightly more extended objects which exhibit CI values consistent with star clusters. The figure provides the percentage of PAH emitters located on the right side of each panel, indicating those within the cluster regime.  

The percentage of objects in the star cluster CI regime varies from galaxy to galaxy, and ranges from 66-97\% with a median of 87\%.  The percentage does not show a correlation with galaxy distance, but recall that many of the sources which are potentially individual evolved red stars shown in the histograms in Figure~\ref{fig:hist_ci}, are removed from the PAH emitter sample in the closer galaxies by the F1000W-F2100W color cut imposed in Sec.~\ref{sec:filtering}.  

PAH emitters with smaller CI than the values indicated by this limit (black line)
do not necessarily imply their exclusion as clusters.  As depicted in Fig.\ref{fig:hist_ci}, there remains a tail in the cluster population with small CI values, suggesting the possibility of very compact star clusters, or clusters dominated by a single massive young stellar object, which are also interesting for our analysis. Moreover, younger clusters are also expected to be more compact, as clusters should expand as they age, due to mass loss and two-body relaxation \citep[][, and references therein]{brown2021}.  For these reasons, we opt to keep the objects with small CI (again which already have been filtered for possible old red stars) in the sample.  

A key question is how source crowding will affect the F200W CI values presented here. This is particularly important for deep, highly resolved imaging at 2$\micron$ which will be far more sensitive to the large population of older stars compared with HST optical imaging.  This issue can be investigated in future work, for example when PSF-fitting catalogs have been produced for our F200W imaging.


\begin{figure*}[p]
    \centering
    \includegraphics[width=2\columnwidth]{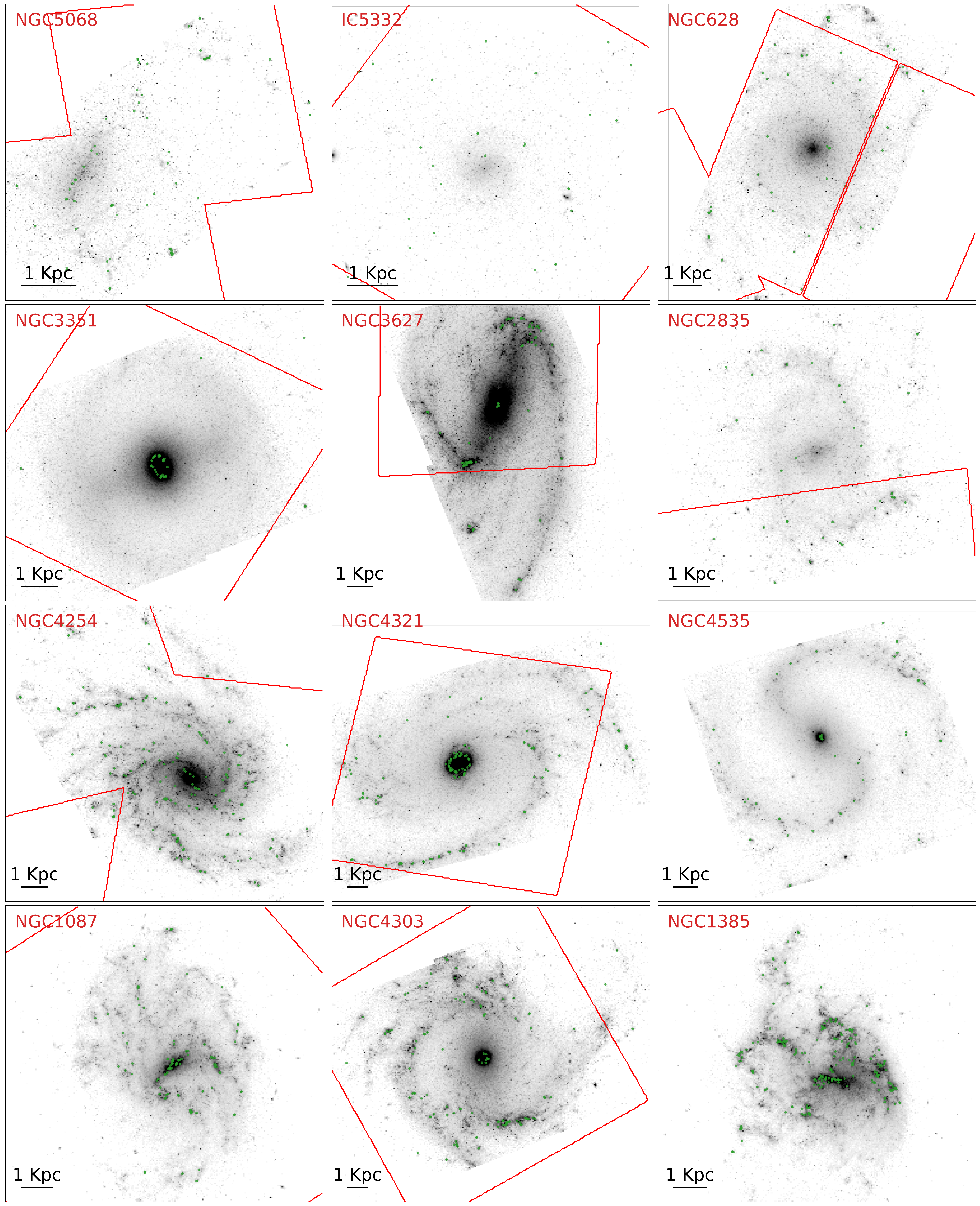}
    \caption{Spatial distribution of  the $<5\sigma$ PAH emitters overlaid in a JWST F335M image of the galaxies. The plots are organized by increasing galactic distance from the upper left corner to the bottom right. The red footprints shows the HST H-$\alpha$ observations.}
    \label{fig:spatial_dist_1}
\end{figure*}

\begin{figure*}[p]
    \centering
\includegraphics[width=2\columnwidth]{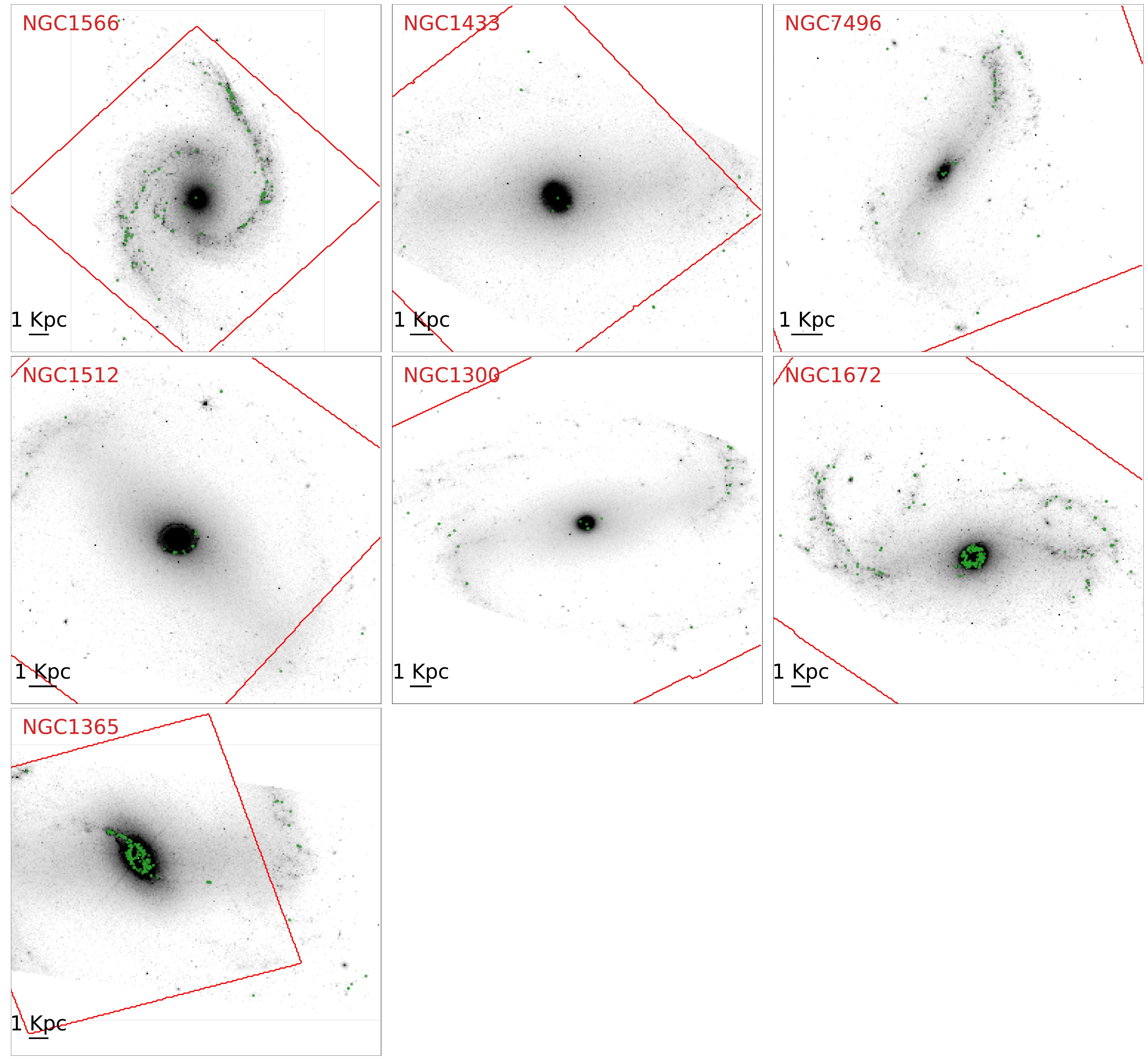}
    \caption{Figure \ref{fig:spatial_dist_1} continuation}
    \label{fig:spatial_dist_2}
\end{figure*}

\begin{figure*}[]
    \centering
    \includegraphics[width=2\columnwidth]{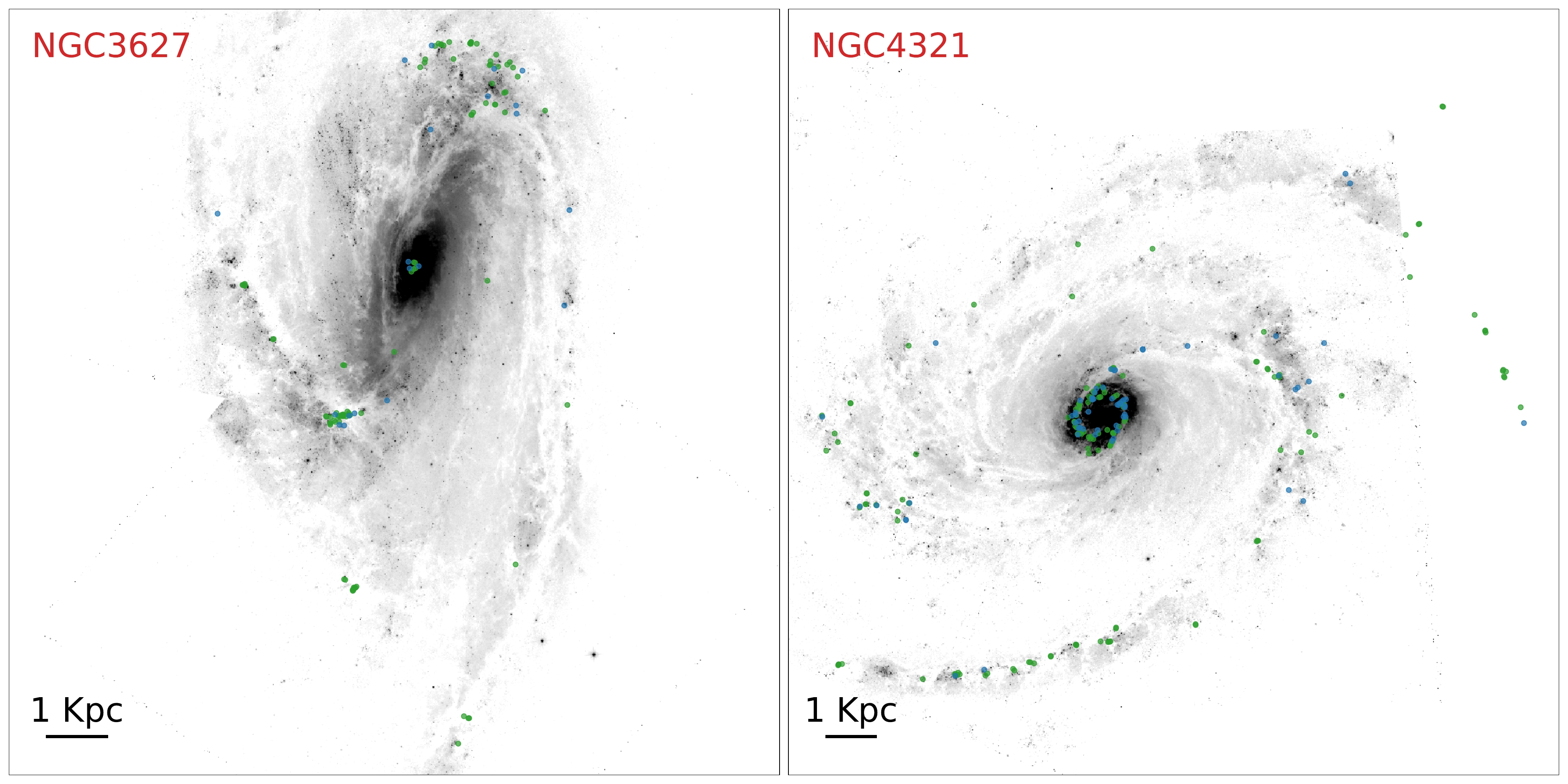}
    \caption{Spatial distribution of PAH emitters in NGC3627 and NGC4321 overlaid on the HST F555W image of the galaxies. We observe that the PAH emitters are mainly located in and around the dust lanes.  Here we show both 5$\sigma$ (green) and 3-5$\sigma$ PAH (blue) samples. }
    \label{fig:spatial_dist_f555w}
\end{figure*}

\subsection{Spatial Distribution}
\label{sec:spatial}

The PHANGS-JWST Cycle 1 galaxy sample provides an excellent opportunity to study the locations of compact PAH emitters within galaxies, given its inclusion of a diverse range of galaxy morphologies and masses. Fig.~\ref{fig:spatial_dist_1} overlays the PAH emitter population on the detection image (F335M) for each galaxy. 
The PAH emitters are generally found in the spiral arms of the galaxies, or concentrated on the bar ends. They also delineate outer and inner star-forming rings such as in NGC~3351, NGC~4321, and NGC~1672.  The figures provide assurance that the adopted selection criteria yield samples that are not merely capturing diffuse PAH emission but are instead tracing compact regions of star formation.  It is interesting to note that in almost all galaxies the bar is free of PAH emitters with three exceptions, in the bars of NGC~5068, NGC~1087, and NGC~1385. Interestingly, these galaxies also exhibit a more flocculent morphology, and are the lower mass galaxies in the sample \citep[Table~2 of][]{phangs-jwst}. The environment here may not be as turbulent as in galaxies with clearly defined spiral arms, potentially influencing the facilitation of star formation.

Fig.~\ref{fig:spatial_dist_f555w} shows the distribution of PAH emitters, but this time focusing on just two galaxies, NGC~3627 and NGC~4321, to illustrate the location of the PAH emitters relative to dust which is apparent as attenuation features 
in the HST F555W image. The PAH emitters are primarily located in the dust lanes within the spiral arms of these galaxies. While we only show these two cases, this is observed in most galaxies. These images present also the distribution of the 5$\sigma$ (green) and 3-5$\sigma$ PAH (blue) PAH emitter selections, showing that they exhibit similar spatial distributions.



\subsection{3.3~$\mu$\MakeLowercase{m} PAH Luminosities}
\label{emission_flux}

Another basic property of the emitters is their 3.3~$\mu$m PAH luminosity.
To compute the 3.3~$\mu$m PAH emission flux we use the flux density obtained from the $\rm F300M$ photometry to estimate the contribution from the continuum to the F335M band. We decided to use only $\rm F300M$ for the estimation of the continuum instead of a combination of the $\rm F300M$ and $\rm F360M$ bands, as $\rm F360M$ is contaminated by PAH emission \citep{2023ApJ...944L...7S}.  
The median 3.3~$\mu$m PAH luminosity of the aggregate sample across all 19 galaxies is 3.5$\times$10$^{35}$erg$\;$s$^{-1}$ with a range of 2.5$\times$10$^{34}$ (IC~5332)  to 1.5$\times$10$^{37}$ erg$\;$s$^{-1}$ (NGC~1365).

To put these values into context, we estimate the luminosity of 30~Doradus (a.k.a the Tarantula nebula) based on its integrated flux in the 6.2$\mu$m feature of 2.5$\times$10$^{-14}$W$\;$m$^{-2}$ \citep{galliano08}. This number is fairly uncertain since the feature to continuum ratio is very low in 30~Doradus, and the emission is not compact; we have chosen the “peak” number, but there is extended emission nearly an order of magnitude brighter around the peak.  We choose the 6.2$\mu$m feature because theory suggests that the 3.3/6.2 ratio (0.3-0.9) is less sensitive to PAH size than other ratios, with the majority of the energy of both of these short-wavelength features being emitted by smaller PAHs \citep{marag23}. This results in an estimated compact 3.3$\mu$m luminosity of 5$\times$10$^{36}$erg$\;$s$^{-1}$.  Although this value is higher than the median luminosity across the entire sample of galaxies presented here, it falls within the overall luminosity range. The central young star cluster associated with 30~Doradus, NGC~2070, is relatively uncommon in nearby galaxies; it has a large number of young massive ionizing stars and a large overall stellar mass \citep[$\sim$10$^5$M$_{\odot}$][]{Brandl96, Dominguez23}, which is significantly more massive than the cluster population in the PHANGS-HST galaxies (median of $\sim$10$^4$ M$_{\odot}$). We found higher luminosities than this value just in two galaxies or our sample NGC~1672 and NGC~1365. Consistently these galaxies also present the highest SFR and L$_{CO}$ (Fig.~\ref{N_pah_galaxies_properties}).

\begin{figure}[h]
    \centering
    \includegraphics[width=\columnwidth]{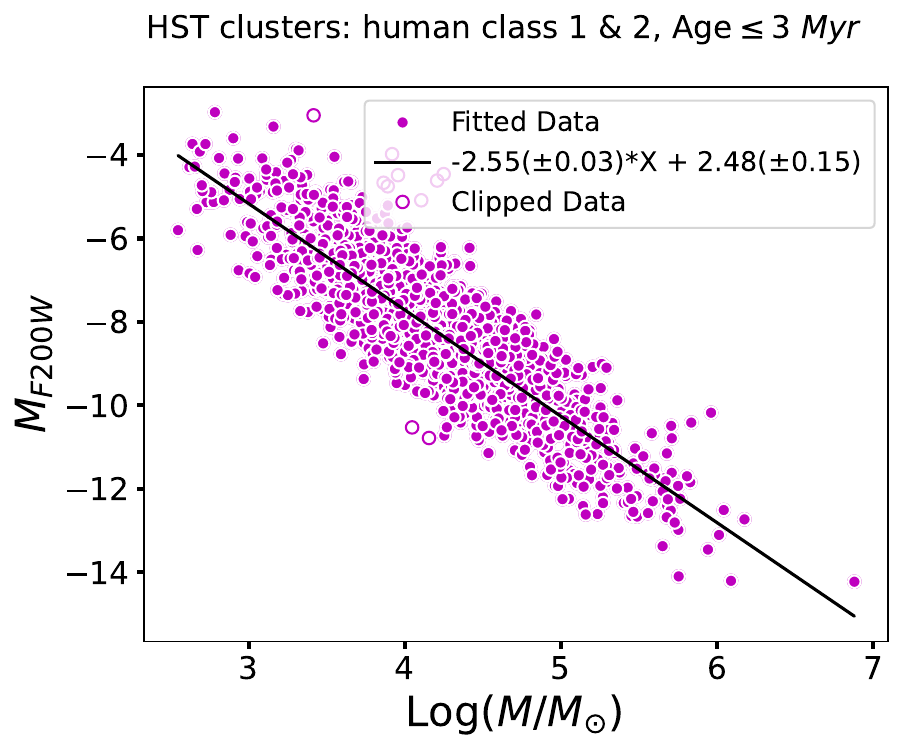}
    \caption{Mass-to-light ratio in the F200W band for the HST-cluster catalogs Class 1 and 2 with ages$\leq$3~Myr. The relation derived here was used to estimate the masses of the PAH emitters.}
    \label{fig:mass-light}
\end{figure} 

\subsection{Stellar Mass Estimates}
\label{sec:stellarmass}

To estimate the stellar masses of the PAH emitters, we use a mass-to-light ratio 
derived in F200W for the HST clusters from \citep{Maschmann2024,thilker24} with ages$\leq$3~Myr. If the PAH emitters are subject to larger amounts of dust attenuation than the optically detected clusters, then the computed masses will underestimate the true stellar mass; nevertheless they should be useful an estimate for a first examination of the properties of the sample. 

To determine the mass-to-light ratio, we fit a line to the F200W absolute magnitude M$_{F200W}$ as a function of the logarithm of cluster stellar mass, where the cluster stellar masses are computed via SED fitting with HST photometry as in \citet{thilker24} (a brief overview is provided in the next section). We derived the following relation: ${\rm M}_{\rm F200W} = -2.55\pm0.03~\log(M/M_\odot) +2.48\pm0.15$ (Fig.~\ref{fig:mass-light}). 
We applied this relationship to our list of PAH emitters and obtained masses in the range 
700 $\rm M_{\odot}$ (NGC~5068) to 6.5 $\rm \times 10^5 M_\odot$ (NGC~1365), with a median of 3.4$\rm \times 10^4 M_\odot$. The obtained mass distribution is presented in Fig.~\ref{fig:LF}.


\section{Comparison with PHANGS-HST cluster catalogs}
\label{sec:HST-clusters-comparison}

\subsection{HST Star Clusters on the F300M-F335M CMD}
Here, we examine the 3.3$\mu$m PAH properties of optically detected star clusters in the PHANGS-HST catalogs \citep[][]{Maschmann2024}.  The goal is to use the star cluster ages to gain insight into the duration of compact PAH emission associated with young stellar populations.  The HST star cluster ages adopted in this analysis are taken from \citet{thilker24}, and are computed through SED fitting of NUV-U-B-V-I HST photometry with priors on model age, reddening, and metallicity determined from ground-based H$\alpha$ imaging and other observable parameters providing categorical insight (morphology, color-color information).  The JWST photometry follows the same procedure as described in Sect.~\ref{Sect:Potometry} with apertures centered on the position of each optical cluster.  PHANGS-HST provides two types of catalogs: catalogs in which a subset of cluster candidates have been visually inspected and morphologically classified by a human, and catalogs where all candidates have been classified by neural networks.  The latter, which we refer to as the machine-learning catalogs, are on average $\sim$1 mag deeper than the human-classified catalogs.  In the following analysis, we examine clusters classified as Class 1 (single-peak and symmetric) and Class 2 (single-peak and elongated or asymmetric) in both human-classified and machine-learning catalogs.

\begin{figure*}
    \centering
    \includegraphics[width=1.8\columnwidth]{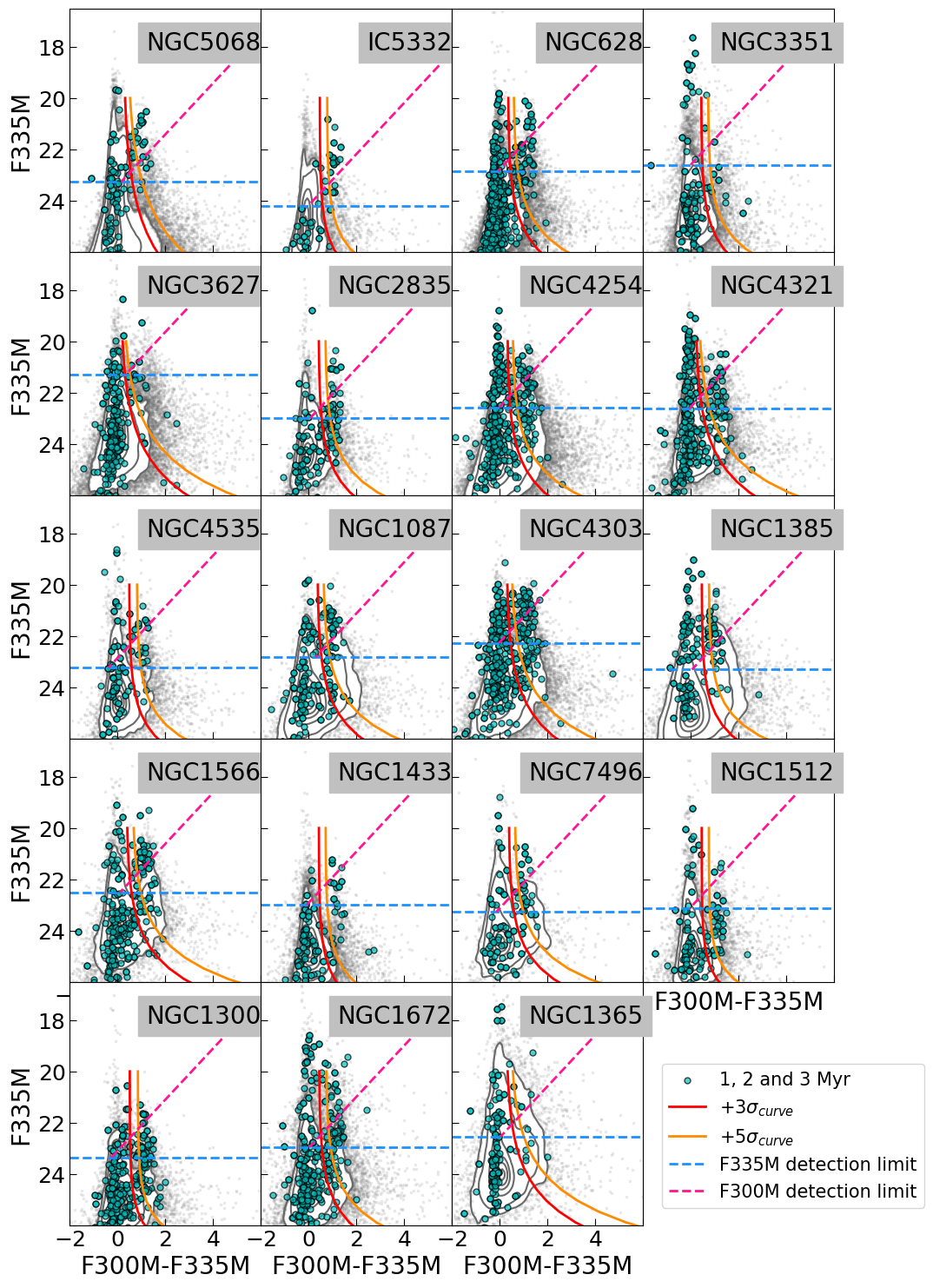}
    \caption{F335M vs. F300M-F335M CMDs. The contours represent the 5th, 10th, 30th, 50th, 70th, and 90th density percentiles of all the sources detected over the F335M images (see Sect~\ref{sec:detection}). The small grey dots indicate regions where the density is lower than the 5th percentile. Cyan circles are PHANGS-HST clusters with ages of 1, 2 and 3 Myr. The red and orange curves indicate +3$\sigma$ and +5$\sigma$ in color dispersion respectively. The dashed horizontal lines indicate the adopted F335M (light blue) and F300M (pink) detection limits for the PAH emitters sample, derived from 5$\sigma$ measurements relative to the mean obtained via random apertures (see Sec.~\ref{sec:CMD}).
    The plots are organized by the galaxies' increasing distance from the upper left corner to the bottom right}
    \label{fig:CMD_young_clusters}
\end{figure*}

\begin{figure*}
    \centering
    \includegraphics[width=1.8\columnwidth]{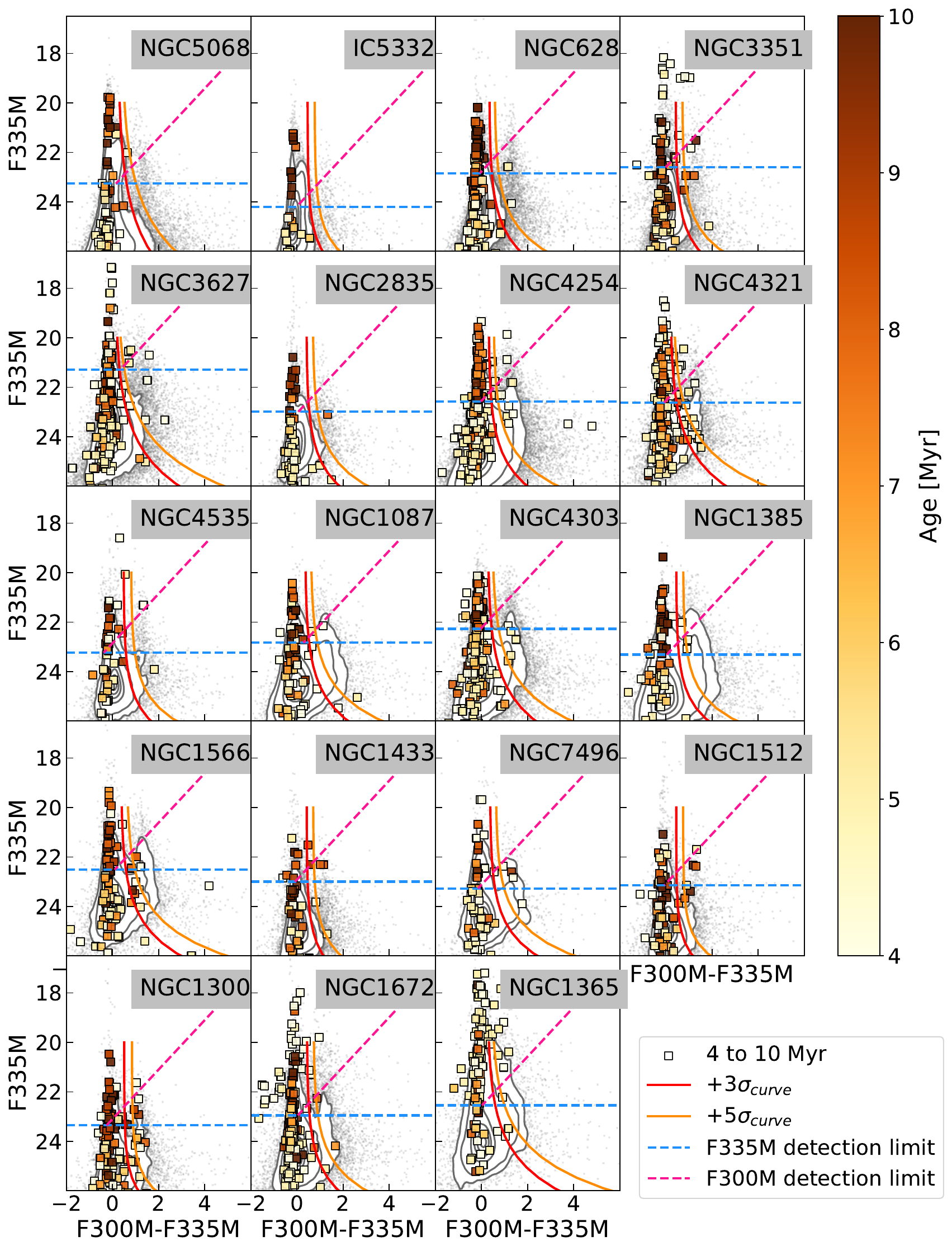}
    \caption{F335M vs. F300M-F335M CMDs. The contours represent the 5th, 10th, 30th, 50th, 70th, and 90th density percentiles of all the sources detected over the F335M images (see Sect~\ref{sec:detection}). The small grey dots indicate regions where the density is lower than the 5th percentile.
    The color squares represent PHANGS-HST clusters aged 3–10 Myr, color-coded by age. Notably, most clusters within this age range exhibit a $\mathrm{F300M-F335M}$ color index close to zero.
    The red and orange curves indicate +3$\sigma$ and +5$\sigma$ in color dispersion respectively. The dashed horizontal lines indicate the adopted F335M (light blue) and F300M (pink) detection limits for the PAH emitters sample, derived from 5$\sigma$ measurements relative to the mean obtained via random apertures (see Sec.~\ref{sec:CMD}). The plots are organized by the galaxies' increasing distance from the upper left corner to the bottom right.}
    \label{fig:CMD_4_10age_clusters}
\end{figure*}

Two sets of F335M vs. F300M-F335M CMDs are presented, which overlay the HST clusters in different age ranges on the F335M detections for each galaxy as initially presented in Fig.~\ref{fig:selection}.
Figs.~\ref{fig:CMD_young_clusters}-\ref{fig:CMD_4_10age_clusters} show young clusters with ages $\leq$3 Myr, and 4-10 Myr respectively. 
The specific number of HST clusters in different age bins in each galaxy whose colors fall in the F300M-F335M PAH-excess selection region ($>3\sigma$) over the total number of HST clusters above the F300M and F335M detection limit threshold are listed in Table~\ref{tab:PAH-HST-fractions}.

Depending on the galaxy, between 12\% and 93\% (with a median of $\sim$30\%) of clusters $\leq$3 Myr fall within the PAH emitters region of the F300M-F335M color-magnitude space, i.e., with a color excess larger than 3$\sigma$ (red curve in CMDs of Figs.~\ref{fig:CMD_young_clusters} and  \ref{fig:CMD_4_10age_clusters}). By contrast, only $\sim$  13\% of HST clusters with 4-5~Myr ages fall within the PAH emitter region, (Fig.~\ref{fig:CMD_4_10age_clusters} and Table~\ref{tab:PAH-HST-fractions}) and very few clusters older than this present  $\rm F300M-F335M >3\sigma$.
%
It is intriguing that so many clusters younger than 3~Myr do not show significant PAH feature emission i.e. F300M-F335M near zero.  This may be due to the selection effects inherent in the HST cluster catalog, namely that clusters detected in blue or even UV filters are not heavily extinguished, and thus may be associated with lower than average circum- and interstellar medium.  Other likely effects are inaccurate age dating (the HST clusters are in fact older), or more interesting if true, strong feedback clearing or destroying the PAHs in these clusters. 
Neverthelsss, the number of HST clusters that {\em do} show strong PAH emission visibly decreases with cluster age, and the timescale for clusters to lose that PAH excess emission is clearly less than 5~Myr.  In the last two columns of Table~\ref{tab:PAH-HST-fractions} we can see that the numbers of HST clusters older than 10~Myr with significant $\rm F300M-F335M$ color excess is almost cero, indicating an increasing lack of PAH emission with cluster age, and that very few if any clusters retain strong PAH emission at ages beyond a few Myr.

\begin{figure}
    \centering
    \includegraphics[width=\columnwidth]{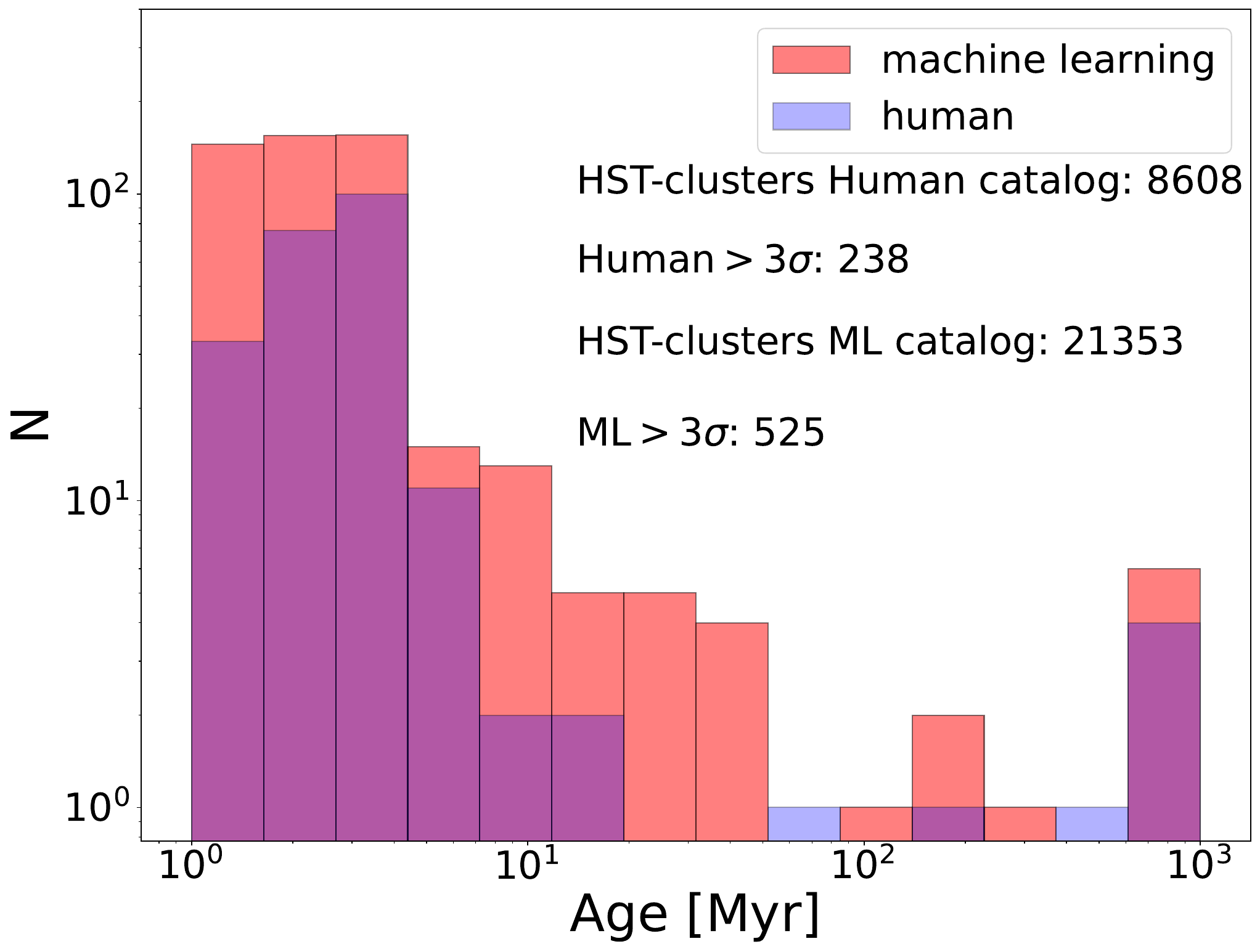}
    \caption{Age histograms for the PHANGS-HST clusters from \citep{Maschmann2024} with ages determined via SED fitting from \citep{thilker24} with $ \rm F300M -F335M > 3\sigma$ }
    \label{fig:cluster_age-hist_3sigmma}
\end{figure}

In Fig.~\ref{fig:cluster_age-hist_3sigmma} we present the age histograms of all HST clusters whose F300M and F335M photometry lies in the $>$3$\sigma$ PAH emitter region.   
This demonstrates again that most of PAH-emitting clusters are younger than 10~Myr with the exception of a very few clusters. We inspected the oldest objects across the 14 bands, and their brightness, and morphologies in the H$\alpha$, $\rm F335M$, and mid-infrared bands lead us to believe that they are likely very young clusters with inaccurately estimated ages via UV-optical SED fitting. 

We note that accurate age dating is difficult in the 5-10~Myr age range (optical stellar colors change rapidly as red supergiants emerge), and 3-5~Myr-old clusters with attenuation A$_V\gtrsim$ 1 have optical colors that become easily confused with the 5-10~Myr population. The UV-optical colors of clusters $\leq$ 3 Myr are also highly degenerate.  
Further investigation is required to determine if there really are some $>$5~Myr old clusters with extended PAH emission timescales and unique features, or if their ages were simply determined incorrectly.

\begin{table*}
\centering
\begin{tabular}{l|cc|cc|cc|cc|cc|cc|cc|}
Galaxy & \multicolumn{2}{c|}{PAH emitters}& \multicolumn{2}{c|}{HST Clusters $\le$ 3 Myr} & \multicolumn{2}{c|} {HST-PAH}  
& \multicolumn{2}{c|}{$\le$ 3 Myr} & \multicolumn{2}{c|}{ 4-5~Myr} & \multicolumn{2}{c|}{6-10~Myr} & \multicolumn{2}{c|}{ 11-1000~Myr} \\
    \cline{2-15}
     & 3$\sigma$-5$\sigma$ & $\rm >5\sigma$ & H & ML &  H & ML  &H & ML & H & ML & H & ML & H & ML \\
    \hline 
NGC5068 & 18 & 77 & 27 & 35 & 12 & 11 & 15/27 & 15/35 & 1/2 & 0/2 & 0/8 & 0/15 & 0/44 & 0/67 \\
IC5332 & 8 & 27 & 11 & 15 &  9 & 10 & 10/11 & 14/15 & - & - & 0/9 & 0/7 & 0/18 & 0/13 \\
NGC628 & 10 & 53 & 44 & 55 &  9 &  11 & 11/44 & 14/55 & 0/2 & 0/2 & 0/23 & 0/35 & 0/31 & 0/40 \\
NGC3351 & 31 & 41 & 17 & 15 &  0& 0 &2/17 & 1/15 & 4/17 & 3/16 & 0/10 & 0/9 & 1/5 & 1/6 \\
NGC3627 & 20 & 93 & 18 & 12 & 2& 2& 4/18 & 4/12 & 2/20 & 2/16 & 0/11 & 0/12 & 0/63 & 0/66 \\
NGC2835 & 4 & 38 & 14 & 13 &6 &7 &6/14 & 7/13 & - & - & 0/13 & 0/5 & 0/28 & 0/23 \\
NGC4254 & 60 & 181 & 64 & 72 &9  &12 & 13/64 & 19/72 & 3/17 & 4/21 & 0/45 & 0/57 & 0/76 & 0/110 \\
NGC4321 & 61 & 146 & 71 & 59 & 7 & 8 & 10/71 & 11/59 & 5/36 & 5/30 & 0/30 & 0/30 & 2/58 & 2/48 \\
NGC4535 & 11 & 45 & 23 & 18 &5 &3 & 8/23 & 7/18 & 2/8 & 1/7 & 0/10 & 0/6 & 1/26 & 1/23 \\
NGC1087 & 39 & 110 & 36 & 30 & 11& 10& 16/36 & 17/30 & 1/5 & 1/6 & 0/19 & 0/18 & 0/45 & 0/41 \\
NGC4303 & 47 & 131 & 68 & 71 & 19 & 25 & 28/68 & 31/71 & 0/11 & 0/11 & 0/39 & 0/38 & 0/100 & 0/113 \\
NGC1385 & 111 & 237 & 59 & 43 & 12 & 12 & 18/59 & 15/43 & 1/9 & 1/6 & 0/23 & 0/14 & 1/101 & 1/97 \\
NGC1566 & 35 & 131 & 38 & 31 &11 & 13 & 17/38 & 19/31 & 1/2 & 1/1 & 0/31 & 0/30 & 0/51 & 1/52 \\
NGC1433 & 6 & 12 & 6 & 6 & 2& 3& 2/6 & 3/6 & 0/2 & 0/2 & 2/7 & 2/6 & 0/4 & 1/4 \\
NGC7496 & 18 & 32 & 20 & 18 & 5& 8& 6/20 & 9/18 & 0/4 & 0/5 & 0/7 & 0/6 & 0/15 & 0/16 \\
NGC1512 & 15 & 19 & 16 & 10 &2 &2 & 2/16 & 2/10 & 1/8 & 1/4 & 0/5 & 0/11 & 1/14 & 0/10 \\
NGC1300 & 23 & 25 & 18 & 24 &4 & 5& 5/18 & 6/24 & 1/3 & 1/1 & 0/7 & 0/24 & 0/12 & 0/17 \\
NGC1672 & 119 & 230 & 43 & 48 & 9& 14 & 13/43 & 18/48 & 3/29 & 3/26 & 0/20 & 0/27 & 1/65 & 1/84 \\
NGC1365 & 87 & 188 & 31 & 28 &2 & 1 & 5/31 & 4/28 & 8/61 & 7/44 & 0/10 & 0/6 & 2/99 & 1/93 \\
\hline
Min: &  4 &  12 & 6 & 6 & 0& 0& 0.12 & 0.07 & 0 & 0 & 0& 0& 0& 0 \\
Max: & 119 & 237 & 71 & 72 & 19 & 25 & 0.91 & 0.93 & 0.5 & 1 & 0.29 & 0.33 & 0.2 & 0.25 \\ 
Median: & 23 & 77 & 27 & 28  & 7& 8& 0.3 & 0.38 &  0.12 & 0.14 & 0 & 0 & 0 & 0 \\ 
\end{tabular}
    \caption{Columns 2-3: Number of PAH emitters with 
    $3\sigma\leq$ F300M-F335M $<5\sigma$ (2) and F300M-F335 $\geq 5\sigma$ (3).
    Columns 4-5: Number of HST clusters from \citep{Maschmann2024,thilker24} younger than 3~Myr and brighter than our established detection threshold in F300M and F335M, in the HST Class 1 \& 2  human (4) machine learning (5). 
    Columns 6-7: Number of HST clusters younger than 3~Myr that are also found in the F300M-F335M $\geq5\sigma$ PAH emitters sample.
    Columns 8-9: Fraction of HST clusters $\leq$ 3~Myr with F300M-F335M $\geq3\sigma$.
    Columns 10-11: Fraction of HST clusters between 4 and 5~Myr with F300M-F335M $\geq3\sigma$. 
    Columns 12-13: Fraction of HST clusters between 6 and 10~Myr with F300M-F335M $\geq3\sigma$. 
    Columns 14-15: Fraction of HST clusters between 11 and 1000~Myr with F300M-F335M $>3\sigma$.}
\label{tab:PAH-HST-fractions}
\end{table*}

%


\subsection {Luminosity Functions}
\label{sec:LF}

\begin{figure}
    \centering
    \includegraphics[width=\columnwidth]{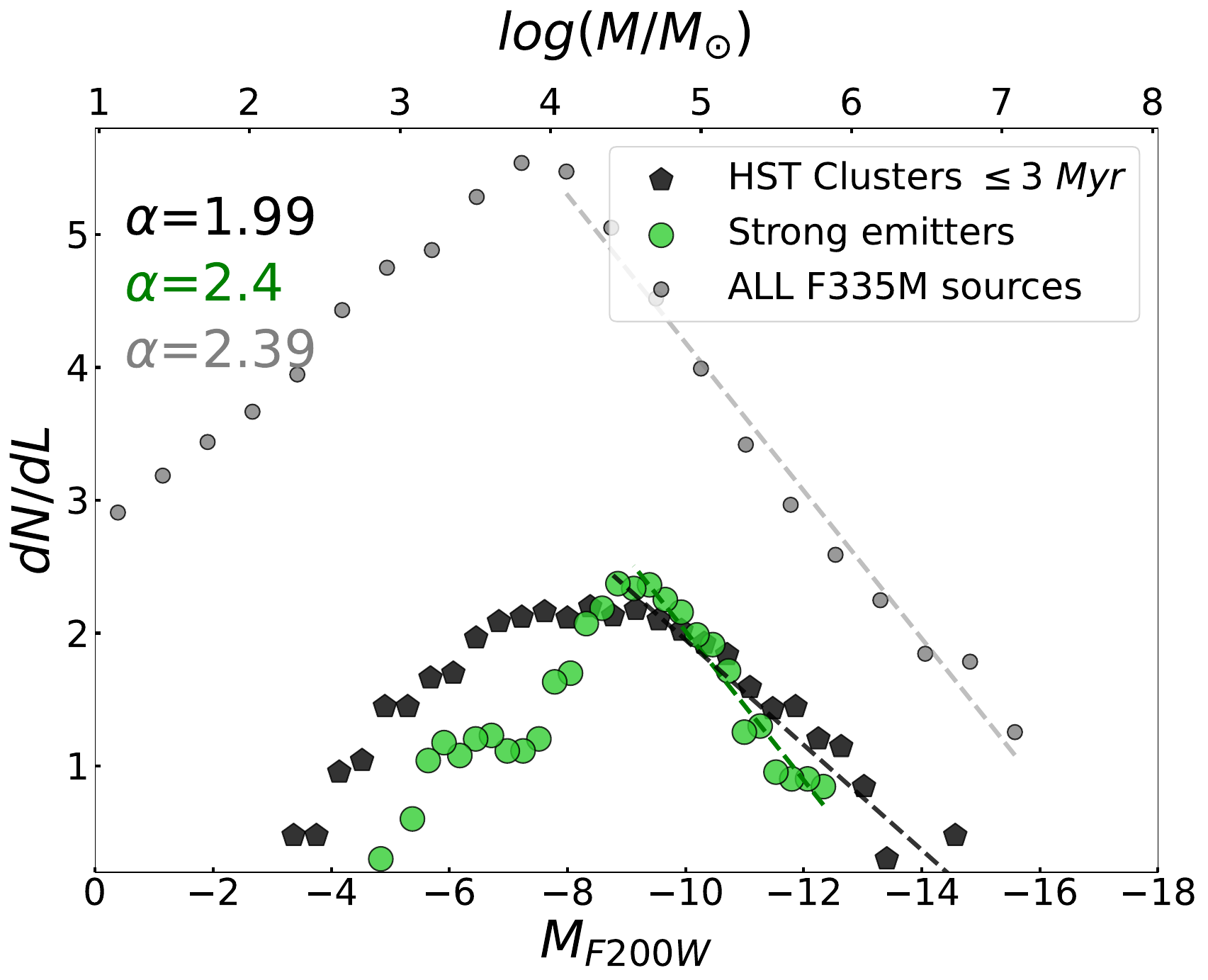}
    \caption{The LF for PAH emitters (green). We also present for comparison the LF for HST star clusters younger than 3~Myr (depicted in black) and the LF for all the sources detected over the F335M image in $\S$~\ref{sec:detection} (grey). The lines represent the best-fit linear models over the bright end of each population, with the respective fitted slope values displayed in the upper left corner. Masses derived from the method described in $\S$~\ref{sec:stellarmass} are depicted on the upper axis. The main distinction between PAH emitters and the HST clusters are the slope at the bright end.}
    \label{fig:LF}
\end{figure}

In Fig.~\ref{fig:LF} we present the F200W LF for the PAH emitters, together with the LF for HST-clusters younger than 3~Myr in black and the entire sample of objects detected in the F335M image in grey. 
The bright end of the HST cluster LF is 2 magnitudes more luminous, and extends to $\sim$-14 mag.
The LF of all F335M objects shows that objects of this magnitude and brighter are detected using our method, even though they are not selected as PAH emitters. This rules out the possibility that the difference is due to a source detection issue.
The LF can be described to first order by a power-law distribution, denoted as $dN/dL \propto L^{-\alpha}$ \citep{whitmore99,fall06}. The optically selected HST clusters present a LF slope of $\alpha=2$, consistent with the value for a scale-free distribution ($\alpha=2$) and with previous studies of star cluster LFs \citep[, e.g. ][]{larsen09,whitmore14a,adamo17,cook19,krumholz19}. However, the PAH emitters present a steeper distribution ($\alpha=2.4$) indicating a more rapid decrease in the number of bright clusters in this band. On the upper x-axis of this plot we also show the corresponding estimated masses as computed based on the F200W mass-to-light ratio for young clusters as described in Sec.~\ref{sec:stellarmass}. 

\subsection{Comparing the Spatial Distribution of PAH Emitters and HST Clusters}
\label{sec:spatial_dist_comparison}

In the right panel of Fig.~\ref{fig:example_region}, we compare the positions of the PAH emitters ($>5\sigma$ sample, depicted in green) with young clusters (age $<10^7$ Myr, shown in pink) over the northern area of the galaxy NGC~3627. Additionally, ALMA CO(2-1) intensity contours are overlaid in blue for comparison. We observe that the PAH emitters are located at the bar end. In this area gas are being compressed and heated, facilitating the star formation. We also observe the PAH emitters are within the ALMA contours, suggesting that they are still associated with molecular clouds. In contrast, the young HST clusters present a more disperse distribution. It is interesting to note that the few HST clusters within the ALMA contours are generally closer to the edges, suggesting a potential evolutionary sequence in the transition between PAH emitters and HST clusters. 

The 3.3~$\mu$m PAH emitters predominantly represent highly obscured, very young star-forming regions not detectable in optical wavelengths (as also illustrated by the comparison with the HST F555W imaging in Fig.~\ref{fig:spatial_dist_f555w}). The  HST clusters close to the edges of the ALMA contours might represent optically emerging star-forming regions. Here stellar feedback plays a crucial role in dispersing the surrounding ISM. These regions can be detected at optical wavelengths but may still be partially surrounded by gas and dust.
The more spread-out HST clusters are probably in a slightly more evolved state, having already swept away the material from the molecular cloud from which they formed, causing them to be very bright at bluer optical wavelengths and in the UV.
While we present only one galaxy here, similar patterns are observed across the 19 galaxies in our sample.

\section{Discussion}
\label{sec:Discussion}



\subsection{UV-IR SEDs of Compact 3.3$\mu$m PAH Emitters and Possible Evolutionary Sequence}
\label{sec:categories}

The 3.3~$\mu$m emission associated with young star clusters appears to persist for only a brief period, as evidenced by the near total absence of F335M color excess in HST clusters older than $\sim$3~Myr (Fig.~\ref{fig:CMD_young_clusters}, \ref{fig:CMD_4_10age_clusters}). Moreover, upon closer examination of the few oldest HST-clusters found in the F300M-F335M $>3\sigma$ region of the CMDs 
(clusters 11~Myr to 1~Gyr last two columns of Table.~\ref{tab:PAH-HST-fractions}), it was found that these clusters exhibited inaccurate SED-fitting derived ages. 

However, in this brief period when the cluster exhibits 3.3~$\mu$m emission, it will transition through various phases of its evolution, starting from heavily obscured, and therefore invisible at optical wavelengths to becoming detectable in the optical and UV. To begin to study the properties 
of clusters through these phases, we divided our samples of PAH emitters into four categories based on the detection of the objects in the UV-optical HST broad bands and the HST H$\alpha$ narrow band, and form their SEDs from the 14 bands of HST$+$JWST data available.  The categories are:


\begin{enumerate}
    \item Not detected in PHANGS-HST imaging: $S/N<5$ for aperture photometry in all five PHANGS-HST UV-optical bands (F275W, F336W, F438W, F555W, F814W) and narrowband (F657/8). Objects 3 and 4 are in Fig~\ref{fig:example_region} are in this category.
    \item Only nebular emission detected (i.e., H$\alpha$): $S/N<5$ for aperture photometry in all 5 HST broad bands band, but $S/N>5$ in narrowband. Object 5 in Fig.~\ref{fig:example_region} is in this category.
   \item Young sources with detection of stellar photospheric emission: $S/N>5$ for aperture photometry in at least one of the HST broadbands and H$\alpha$ flux with $S/N>5$. 
   Object 1 in Fig.~\ref{fig:example_region} is in this category.
   \item Older sources:  $S/N>5$ aperture photometry in at least 2 HST broadbands, but H$\alpha$ flux $S/N<5$.
\end{enumerate}

\begin{figure*}
    \centering
    \includegraphics[width=2\columnwidth]{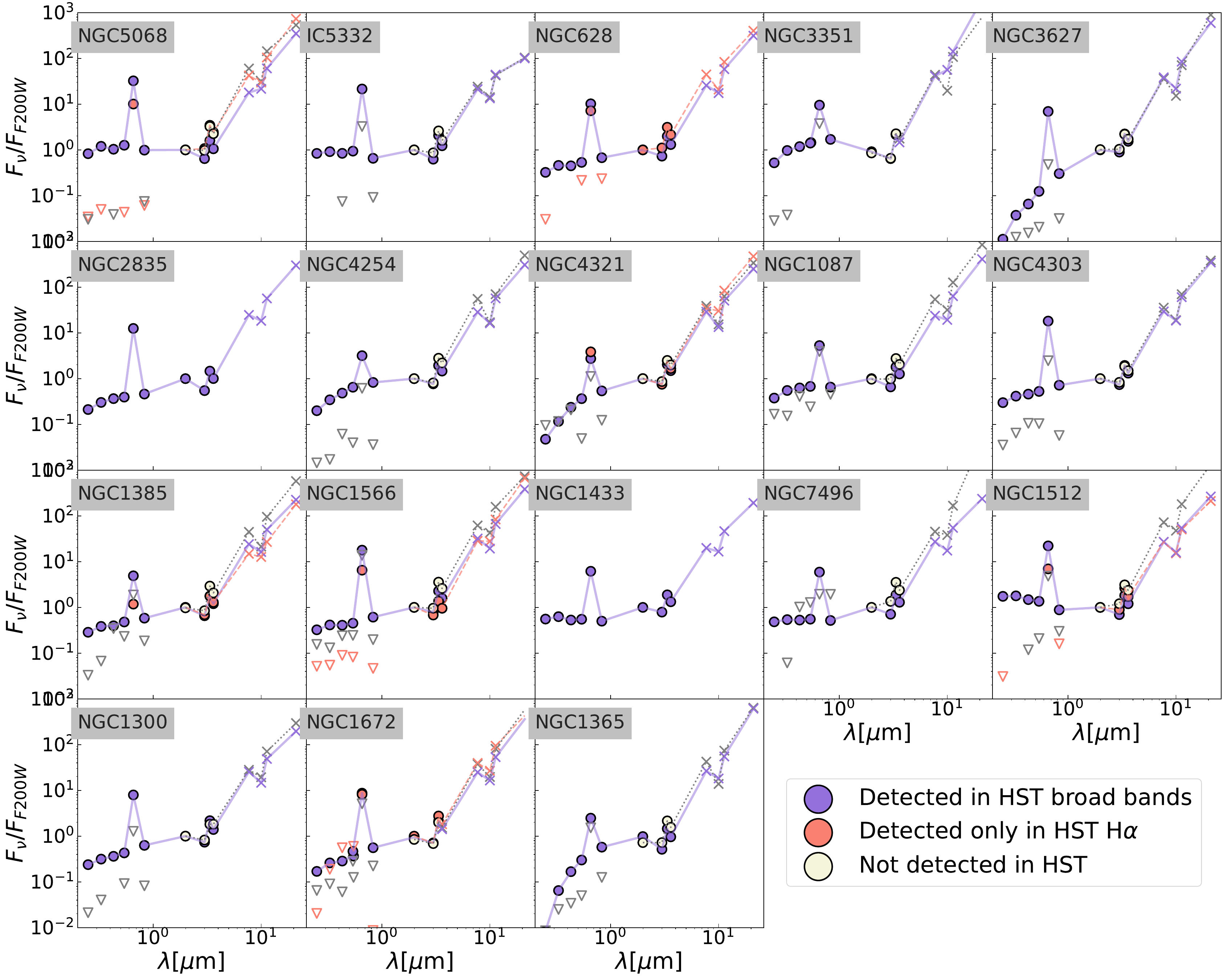}
    \caption{Median values of the SED normalized to the flux density in F200W for compact 3.3~$\mu$m PAH emitters in the `not detected in HST', `only detected in H$\alpha$' and `detected in HST' categories (see $S$\ref{sec:categories} for definition of these categories). These SEDs reveal three distinct characteristics indicative of very young objects. Emission features are observed in H$\alpha$ (for objects in the categories `only detected in H$\alpha$' and `detected in HST') and at 3.3~$\mu$m, and the apparent dip at 10~$\mu$m probably due to PAH emission at the side bands 7.7~$\mu$m and 11.3~$\mu$m. We also observe a general rising flux density with wavelength. Upper limits due to non-detections are shown by open downward pointing triangles, while the cases where there is a detection but it may be affected by neighboring sources (i.e. in the MIRI bands) are shown by Xs.}
    \label{fig:SED}
\end{figure*}

We built the observed spectral energy distribution (SED) from UV(2700~{\AA}) to 21~$\mu$m for objects in the `not detected in HST', `only detected in H$\alpha$' and `detected in HST' categories. In Fig.~\ref{fig:SED} we present median values normalized to the F200W flux density. As in Fig~\ref{fig:sed_example_region} we observe a rising IR distribution with the prominent H${\alpha}$ feature (in the `detected in HST' and `only H$\alpha$ categories), as well as the 3.3~$\mu$m feature and the 10~$\mu$m apparent dip. 
We can observe from these SEDs that objects in the first two categories (`Not detected in HST' and `Only detected H$\alpha$' ) are not found 
in all galaxies.


Another way to constrain the ages of the PAH emitters is to examine their H${\alpha}$ equivalent width (EW).
EWs were estimated as: $F_{H\alpha + N[II]}~\Delta_{NB}/F_{cont}$. 
Where $F_{H\alpha + N[II]}$ and $F_{cont}$ were derived employing a similar methodology as described in \citep{chandar24}. Briefly, we used the flux densities derived from the broadband F814W and F555W images to obtain the continuum flux contribution. To do this we convert flux density per unit of frequency into flux densities per unit of wavelength. Then we weighted the broadband flux densities based on the relative wavelength distance between the center of the broadband filter and the center of the narrow band filter:
\begin{equation}
F_{\rm cont}=F_\mathrm{F555W}W_\mathrm{F555W}+F_\mathrm{F814W}W_\mathrm{F814W}.
\end{equation}
Next, we subtracted the continuum contribution to the narrow band density flux:
\begin{equation}
 F_{H{\alpha}+[\rm N II]}=F_{\rm NB}-F_{\rm cont},
\end{equation}
where $F_{\rm NB}$ are the flux densities in the narrow band filter.
We converted from flux densities to flux multiplying by the narrow band filter width. The flux obtained following this procedure includes contributions from H${\alpha}$ $\lambda$6563 
as well as the [N II] doublet ($\lambda$6548, $\lambda$6583).



\begin{figure*}
    \centering
    \includegraphics[width=2\columnwidth]{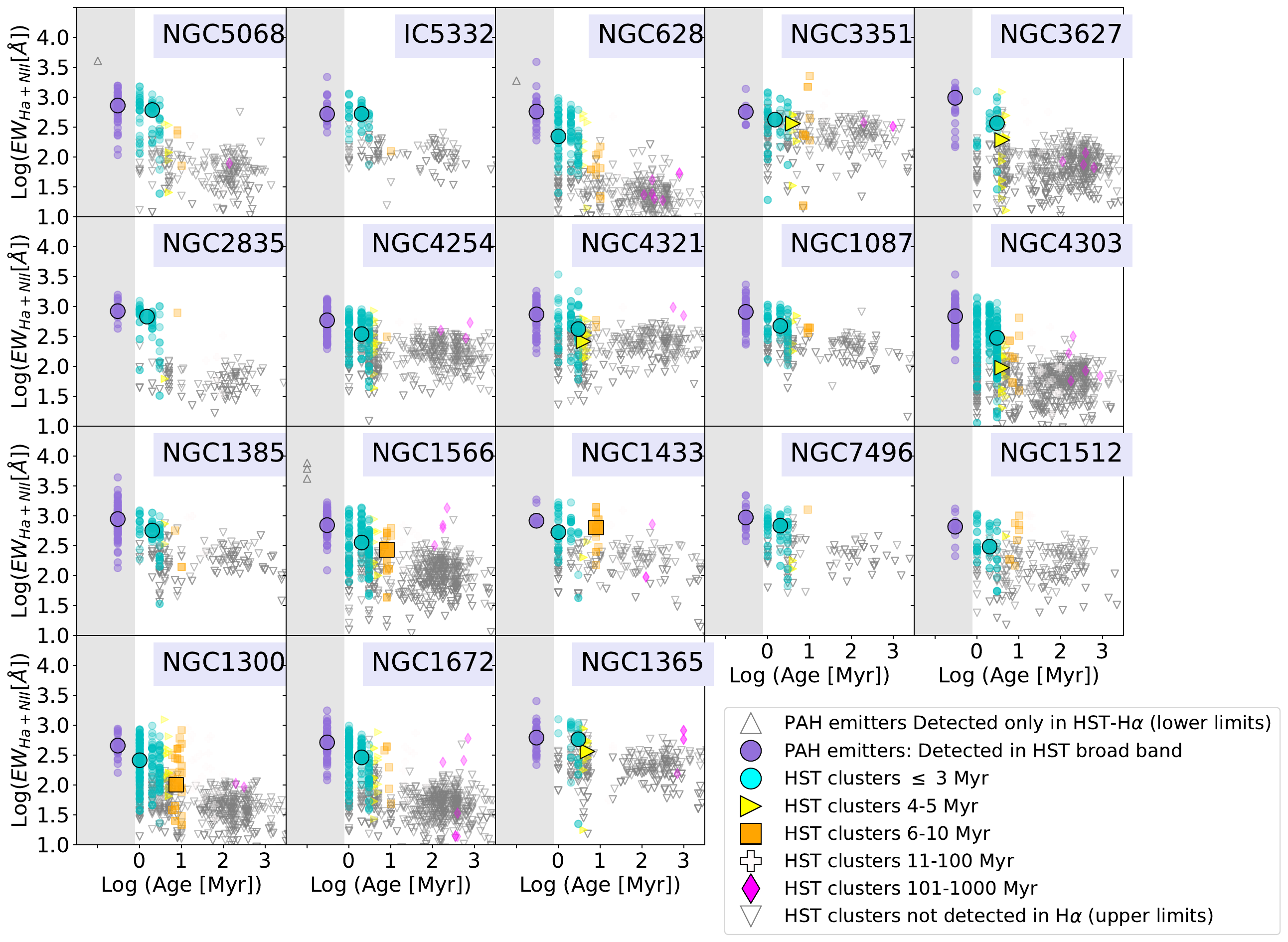}
    \caption{Evolution of H${\alpha}$ EW with Age. The plot presents the EW derived for HST Clusters and PAH Emitters. The grey region denotes the absence of estimated ages for the PAH emitters; however, they are positioned to the left of the HST clusters to suggest a potential evolutionary sequence (refer to text for details). Median values are shown with larger symbols when the sample includes at least 5 objects, while smaller symbols represent individual scatter. Galaxy plots are arranged by distance.}
    \label{fig:EW_Ha}
\end{figure*}

In Fig.~\ref{fig:EW_Ha} we show the H${\alpha} \rm +N[II]$ EW as a function of age, 
for objects in PAH emitters categories 2 and 3 and only lower limits (grey triangles pointing up) for the `only H$\alpha$' category as it was not possible to estimate the continuum flux since the objects were not detected in the F555W and F814W bands. 
Fig.~\ref{fig:EW_Ha} also shows the EW for HST clusters in different age bins: $\le$3~Myr, 4-5~Myr, and 6-10~Myr, 11-100~Myr and 101-1000~Myr.
Most of the HST clusters older than 3~Myr show only upper limits (grey triangles pointing down). The H$\alpha$ EW shows a clear decline with age.
In this figure, large symbols represent median values when the sample includes at least five objects, while small symbols indicate individual objects.
The gray zone in the plots shows a region in which we do not have estimated ages from UV-optical broadband SED fitting. However based on the CMDs presented in Figs.~\ref{fig:CMD_young_clusters} and \ref{fig:CMD_4_10age_clusters} where only  clusters younger than 3~Myr present an F335M color excess, we infer that the mean age of the 3.3~$\mu$m PAH emitter population should be younger than approximately 3~Myr. Assuming that PAH emitters detected in H$\alpha$ but not detected in the HST broad bands (the `only detected in H$\alpha$' category) are in an earlier phase than those PAH emitters detected in some of the HST broad bands the (`detected in HST' category), starting from the left of the plot we show the samples following this plausible evolutionary sequence: first the `only detected in H$\alpha$', then the `detected in HST' categories followed by the HST clusters in increasing order of age. The EW of the PAH emitters in the gray area (again, for which we cannot estimate the age through SED fitting) is on average higher (500-1000 \AA) than the optically detected PHANGS-HST clusters (200-900 \AA).  This provides direct evidence for them being younger than the youngest HST star clusters, and that they are in an earlier phase of evolution/dust clearing.

\begin{figure*}
    \centering
    \includegraphics[width=2\columnwidth]{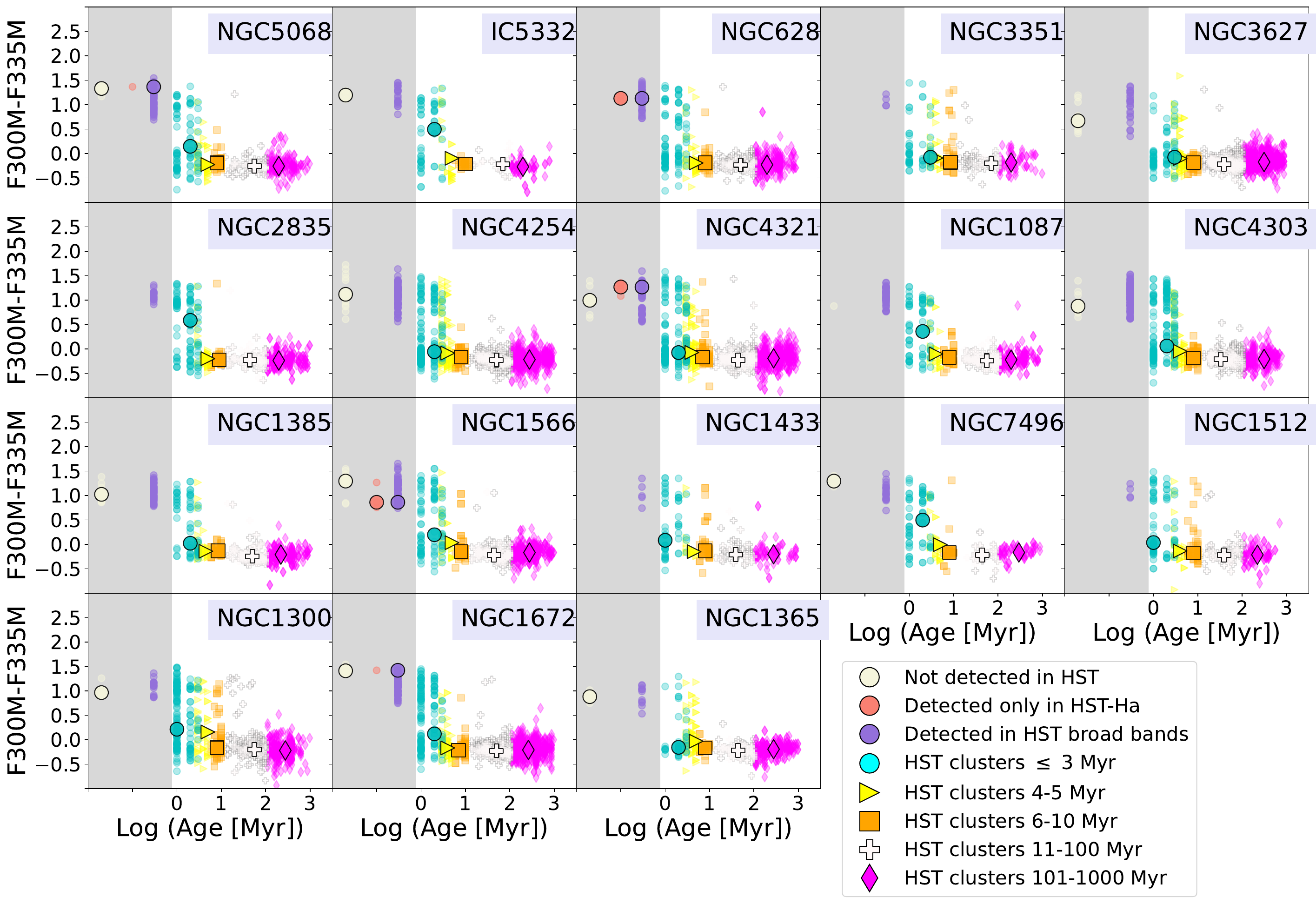}
    \caption{ $\rm F300M-F335M$ vs. Age.  The grey region denotes the absence of estimated ages for the PAH emitters; however, they are positioned to the left of the HST clusters to suggest a potential evolutionary sequence (refer to text for details). Median values are depicted with larger symbols, while smaller symbols represent individual scatter. PAH emitters in the different categories exhibit similar $\rm F300M-F335M$ colors, with a median around 1.5. In contrast, HST clusters older than 3~Myr show a median color around 0. HST clusters younger than 3Myr display a wide scatter in $\rm F300M-F335M$ values, falling between these two median values.  Galaxy plots are arranged by distance.}
    \label{fig:color-age}
\end{figure*}

Figure~\ref{fig:color-age} shows the color $\rm F300M-F335M$ as a function of age for PAH emitters and HST-clusters. As in Fig~\ref{fig:EW_Ha} the gray zone indicates the PAH emitters region for which we do not have estimated ages.  In contrast with the previous figure, this plot allows us to compare the `not detected in HST' objects with the other categories, and we place them to the left of the `only H$\alpha$' category, to suggest the earliest phase in a possible evolutionary sequence. 
Interestingly, this figure shows no significant variation in the $\rm F300M-F335M$ color for the PAH emitters in different categories.  However, it provides a complementary way of demonstrating the abrupt decay of the color for the HST clusters (as also illustrated in the CMDS in Figs.~\ref{fig:CMD_young_clusters}  and \ref{fig:CMD_4_10age_clusters}), with $F300M-F335M \sim 0$ for clusters older than 3~Myr. 

\subsection{How Many ``New'' Young Star Clusters are we Finding?}
\label{sec:newclusters}

To assess the potential increase in the catalogs of young star clusters resulting from the inclusion of newly detected 3.3~$\mu$m PAH emitters, we compared the number of PHANGS-HST clusters with ages $\leq 3$~Myr from the machine learning catalogs with the number of 3.3~$\mu$m PAH emitters $>5\sigma$ (columns 5 and 3 respectively of Table~\ref{tab:PAH-HST-fractions}).

We decided to use for the comparison the machine learning catalogs from \cite{Maschmann2024} (instead of the human catalogs) because they are more complete. 
From the HST sample we only considered clusters in the NIRCam FOV. We also accounted for overlaps between the two datasets, using a search radius of 0$\farcs$126 (2 pixels in the F335M image).  We found between 1 and 28 objects common in both samples in the different galaxies (column 7 of Table~\ref{tab:PAH-HST-fractions}). 
Subtracting these common objects 
we found that the number ``new'' clusters (number of PAH emitters not previously detected in the HST catalogs) shows a large variation from galaxy to galaxy. We found as few as 10 new young clusters in NGC~1433 and as many as 224 in NGC~1385. Accounting for these compact PAH sources could increase 
the number of young clusters significantly compared to optical catalogs. It is important to be precise when quoting an increase factor:  The number of PAH-emitting clusters not present in HST catalogs (Table~\ref{tab:PAH-HST-fractions}, column 3 minus column 6 or 7), plus the $<$3Myr-old HST clusters detected at 3$\mu$m (column 4 or 5), divided by that HST number, is a factor ranging from \factornewclustersmin to \factornewclustersmax (median 3.3x). 
If one does not require a 3$\mu$m detection to count ``known young HST clusters", then the increase factor represented by the 5$\sigma$ PAH emitters is between 1.1x and 3.3x (median 1.4x).

It is important to also note that these quoted numbers of ``new" (embedded) clusters only refer to the bright end of the LF, because of the magnitude limits for each galaxy imposed in $\S$\ref{sec:Identification}.  
At fainter magnitudes, both the PAH emitter and HST catalog samples suffer from incompleteness (evident in the LF turnover in Fig.~\ref{fig:LF}).  In addition, the PAH emitter sample may have increasing contamination from clumps of diffuse ISM. 
When we examined this fainter population, we identified a substantial number of contaminants: objects with a significant F300M–F335M color excess due to an unreliable detection in F300M. This issue led us to apply the current magnitude limits.
If one assumes that ~50\% of the fainter PAH-emitting sources are true clusters, and includes those fainter sources, the number of PAH emitters increases by $\sim$ 1.3 to 9
times (median 3.6x) the number reported in table~\ref{tab:PAH-HST-fractions}, but there is a similar increase in the number of HST clusters as one decreases the magnitude limit - for any population with a steep luminosity function, small changes in the completeness limits translate into horrifying changes in the number of sources in the sample. 
A critical science question that will need to be addressed by future work is whether the fraction of embedded clusters is different at lower masses.  
Establishing this clearly will require careful analysis of completeness, contamination, stochastic population of the cluster IMF, and how mass-to-light ratios vary between the embedded and HST-visible populations.

\subsection{Comparison with Other Studies}

Recent studies have leveraged JWST's unique capabilities to observe ISM emission from optically-thick regions at high resolution and advanced our understanding of the timescales over which young star clusters emerge from their dust-enshrouded molecular birth clouds.  
Prior to the current paper, this work has focused on selected individual galaxies, similar to our early paper on NGC~7496 \citep{2023ApJ...944L..26R}.  It is notable that all of these studies paint a consistent picture of these early phases of star cluster evolution, using complementary observational tracers and methods of analysis.  

All studies, including the present paper, have confirmed the short dust (or PAH in particular), clearing timescales ($\lesssim$3 Myr) reported by optical studies, which is physically significant because it indicates the importance of pre-supernovae (i.e., ``early") feedback in the star formation cycle.  

\citet{kim23} use a statistical 
analysis
to translate a observed spatial decorrelation between cold gas and star formation rate (SFR) tracers into timescale constraints for NGC~628 at $\sim1\arcsec$ resolution using PHANGS-JWST MIRI F2100W, ground-based H$\alpha$ narrowband, and PHANGS-ALMA CO(2-1) imaging.  They find that the heavily obscured phase of star formation (i.e., only detected in CO and 21 $\micron$, and invisible in H$\alpha$ emission) typically lasts for 2.3$^{+2.7}_{-1.4}$ Myr. In the same galaxy, \citet{pedrini24} leverage higher resolution NIRCam imaging (0\farcs15) to study the ages of clusters associated with different 3.3$\mu$m PAH morphologies (compact, extended, and open) of $\sim$1000 compact HII regions identified through JWST Pa$\alpha$ and Br$\alpha$ narrowband imaging. They find that compact regions, which presumably are the earliest stage in the sequence, are associated with clusters ages between 1 and 6 Myr, with a median of 4 Myr.  

Several other JWST studies focus on galaxies or regions of galaxies with high star formation intensities at z$\sim$0: the central starburst ring of NGC~3351 \citep{sun24}; the central kiloparsec of M82 \citep{levy24}; two cluster-rich luminous infrared galaxies in the Great Observatories All Sky LIRG Survey \citep{linden23,linden24}; and the starbursting barred (Seyfert2) spiral galaxy NGC~1365, the two most cluster-rich galaxies within $\sim$30 Mpc \citep{whitmore23}.  \citet{whitmore23, sun24, linden24} respectively report that clusters are completely obscured in the visible for 1.3 $\pm$ 0.7 Myr and either completely or partially obscured for 3.7 $\pm$ 1.1 Myr, that newly formed cluster become visible in the optical in $\sim2{-}3$ Myr, and that dust is cleared over a timescale of $<3{-}4$ Myr.

Recent studies consistently demonstrate that JWST's advanced capabilities enable the discovery of significantly more embedded young clusters than were previously cataloged using HST infrared-optical detections.
The key question of astrophysical significance is the fraction of young clusters that are sufficiently dust obscured so that they cannot be detected in the optical. 
While definitive answers and comparison across studies require careful analysis of detection thresholds and selection criteria to enable fair comparison of ``new" embedded clusters discovered in new JWST studies, we can attempt a first preliminary synthesis by taking the results reported in recent papers at face value.

\citet{levy24} report that the majority (87\%) of their $\sim$1400 massive ($>10^4$ M$_\odot$) star cluster candidates identified through NIRCam F250M imaging are new compared to previous optical catalogs, corresponding to a factor of 7--8 increase.  Similarly, \citet{linden24} find that their sample of dust-enshrouded YMCs ($\rm N = 116$) is larger by an order of magnitude relative to previous Hubble Space Telescope studies, and 16\% (an increasing factor of 1.16) of the sample is undetected at optical wavelengths.

In our previous work \citep{2023ApJ...944L..26R}, we found a total of 67 young clusters presenting 3.3$\mu$m PAH emission in NGC~7496, 59 of them were new detections in comparison with previous HST catalogs, producing a factor of 2 increase. We also found that $\sim$40\% of the detected sample (28 out of 67) were not detected in HST optical wavelengths.    
Similarly, in the present paper, we find an obscured (not optically detected) fraction of up to 40\% with large variations between galaxies and depending on magnitude cuts, similar to \citet{whitmore23} (16 out of 30 are new), and \citet{sun24} (8 out of 14 compact millimeter continuum sources not detected by HST). While in comparison with  previous HST catalogs the number of ``new" detections produce an increase factor ranging between \factornewclustersmin--\factornewclustersmax depending on the galaxy. This range is consistent with the factor reported in other studies, except \cite{linden24} which found a factor of 10 (with small number statistics).

The variations between 
different studies arise partly from the distinct identification methods and detection thresholds employed, as mentioned above. Additionally, they reflect intrinsic differences in the systems analyzed. For example, the high inclination of M82 \citep{levy24} and the dusty nature of VV~114 and NGC~3256 \citep{linden23,linden24} would both lead to more extinction and thus higher fraction of embedded clusters than the relatively face-on galaxy centers studied by \citet{sun24}, \citet{whitmore23}, and in this study.





\section{Summary and conclusions}
\label{sec:Conclusions}

We have expanded our initial study of dusty clusters in NGC~7496, as traced by compact 3.3~$\mu$m PAH emission \citep{2023ApJ...944L..26R} to include the full set of 19 PHANGS-JWST spiral galaxies observed during the first year of JWST science operations \citep{phangs-jwst, phangs-jwst-pipeline}. The successful deployment of JWST has enabled rapid advances in our understanding of the properties, formation, and evolution of star clusters in their earliest dust-embedded stages, with early work focusing on individual nearby galaxies. This paper is the first, to our knowledge, to conduct a comprehensive census of dusty clusters across a representative sample of the nearby galaxy population.

\begin{itemize}

    \item Based on the $\rm F300M-F335M >5\sigma$ color excess and detection limits in both filters, we identified 1816 compact 3.3~$\mu$m PAH emitters across the 19 galaxies (green points Fig.~\ref{fig:selection}), with per-galaxy counts ranging from 12 (NGC~1433) to 237 (NGC~1385) and a median of 77 (Table~\ref{tab:PAH-HST-fractions}).
    
    \item Objects with 3-5$\sigma$ color excess were also examined (blue points Fig.~\ref{fig:selection}), and these shared similar characteristics with the $>5\sigma$ emitters, though potentially more susceptible to contamination from continuum sources, as expected.

    \item A general selection criterion for compact 3.3~$\mu$m PAH emitters was established, with a median color threshold of $\rm F300M–F335M = 0.67$ at F335M $= 20$, though highly dependent on the background within each galaxy.

    \item The spectral energy distributions (SEDs, Fig.~\ref{fig:sed_example_region} and \ref{fig:SED}) of the PAH emitters display features of very young dusty objects: red mid-IR colors, strong $H\alpha$ and 3.3~$\mu$m emissions, and an apparent dip at 10~$\mu$m due to PAH emission in the flanking F770W and F1130W filters.

    \item Concentration index (CI) analysis indicates that 87\% of PAH emitters resemble extended objects like star clusters (Fig.~\ref{fig:hist_ci}).  Examination of their spatial distribution within the host galaxy show that they are primarily located in dust lanes, spiral arms, bar ends, inner star-forming rings, and galaxy centers (Figs.~\ref{fig:spatial_dist_1} and \ref{fig:spatial_dist_f555w}).

    \item The 3.3~$\mu$m PAH luminosities range from $2.5\times10^{34}$ to $1.5\times10^{37}$ erg$\;$s$^{-1}$, with a median of $3.5\times10^{35}$ erg$\;$s$^{-1}$. Masses, estimated based on F200W mass-to-light ratios for clusters younger than 3~Myr, span from $700$ to $6.5\times10^{5}  \rm M_{\odot}$, with a median of $3.4\times10^{4} \rm M_{\odot}$ (Fig.~\ref{fig:LF}).

    \item The F200W luminosity function of PAH emitters (Fig.~\ref{fig:LF}) shows a steeper bright end compared to PHANGS-HST clusters younger than 3~Myr, indicating a rapid decrease in the number of bright clusters with 2$\mu$m luminosity.

    \item In general, only optically selected clusters from the PHANGS-HST catalog which are younger than 3~Myr display 3.3~$\mu$m emission (Fig.~\ref{fig:CMD_young_clusters}), suggesting a short lifespan ($\lesssim$3~Myr) for PAH emission at the cluster scale.  This provides one constraint on the duration of the dust-embedded phase of star clusters.

    \item The compact PAH emitters exhibit high observed (i.e., no dust correction) H$\alpha+ \rm [NII]$ equivalent widths (median 700~\AA), 1–2.8 times greater than the youngest optically detected PHANGS-HST clusters (Fig.~\ref{fig:EW_Ha}).  This supports the conclusion that the 3.3~$\mu$m PAH emitters are on average younger than optically selected clusters and represent an earlier phase in cluster evolution.

    \item The overlap between PAH emitters and optically selected clusters from PHANGS-HST catalogs is limited ($\lesssim$10\%), yielding 1645 new objects across the 19 galaxies.  For bright clusters in both samples, the number of embedded cluster candidates identified by PAH emission relative to HST-selected clusters younger than 3~Myr varies between a factor of \factornewclustersmin and \factornewclustersmax per galaxy (see \ref{sec:newclusters} for details and caveats).

    \item We compare our results with a number of recent JWST papers on dust embedded clusters in individual nearby galaxies, which also identify significant new populations of dusty clusters and generally paint a consistent picture of a short-lived compact PAH emission and dust embedded phase.

   \end{itemize}

 Future analyses of completeness and the stochastic initial mass function (IMF) in small clusters will clarify the contribution of embedded clusters to galaxy-wide star formation. JWST photometry and the 3.3~$\mu$m PAH emission feature offer a powerful new tool for this assessment.

\section{Acknowledgments}
The authors would like to thank the anonymous referee for constructive comments that helped improve the quality of this work.
This work is based on observations made with the NASA/ESA/CSA James Webb Space Telescope (program \#2107) and the NASA/ESA Hubble Space Telescope (program \#15654 \& \#17126).   
The data were obtained from the Mikulski Archive for Space Telescopes at the Space Telescope Science Institute, which is operated by the Association of Universities for Research in Astronomy, Inc., under NASA contract NAS 5-03127 for JWST and 5-26555 for HST.
The specific observations analyzed can be accessed via \dataset[DOI: 10.17909/t9-r08f-dq31]{https://doi.org/10.17909/t9-r08f-dq31};
\dataset[DOI: 10.17909/jray-9798]{https://dx.doi.org/10.17909/jray-9798};
\dataset[DOI: DOI/ew88-jt15]{https://archive.stsci.edu/doi/resolve/resolve.html?doi=10.17909/ew88-jt15}.

This paper makes use of the following ALMA data: ADS/JAO.ALMA\#2015.00956.S. ALMA is a partnership of ESO (representing its member states), NSF (USA) and NINS (Japan), together with NRC (Canada), NSTC and ASIAA (Taiwan), and KASI (Republic of Korea), in cooperation with the Republic of Chile. The Joint ALMA Observatory is operated by ESO, AUI/NRAO and NAOJ.

JS acknowledges support by the National Aeronautics and Space Administration (NASA) through the NASA Hubble Fellowship grant HST-HF2-51544 awarded by the Space Telescope Science Institute (STScI), which is operated by the Association of Universities for Research in Astronomy, Inc., under contract NAS~5-26555.

MB gratefully acknowledges support from the ANID BASAL project FB210003 and from the FONDECYT regular grant 1211000.

This work was supported by the French government through the France 2030 investment plan managed by the National Research Agency (ANR), as part of the Initiative of Excellence of Université Côte d’Azur under reference number ANR-15-IDEX-01.

AW acknowledges UNAM and the PASPA of DGAPA.

RCL acknowledges partial support for this work provided by a National Science Foundation (NSF) Astronomy and Astrophysics Postdoctoral Fellowship under award AST-2102625.

KG is supported by the Australian Research Council through the Discovery Early Career Researcher Award (DECRA) Fellowship (project number DE220100766) funded by the Australian Government. 
KG is supported by the Australian Research Council Centre of Excellence for All Sky Astrophysics in 3 Dimensions (ASTRO~3D), through project number CE170100013. 

RSK acknowledges financial support from the European Research Council via the ERC Synergy Grant ``ECOGAL'' (project ID 855130),  from the German Excellence Strategy via the Heidelberg Cluster of Excellence (EXC 2181 - 390900948) ``STRUCTURES'', and from the German Ministry for Economic Affairs and Climate Action in project ``MAINN'' (funding ID 50OO2206). RSK is grateful for computing resources provided by the Ministry of Science, Research and the Arts (MWK) of the State of Baden-W\"{u}rttemberg through bwHPC and the German Science Foundation (DFG) through grants INST 35/1134-1 FUGG and 35/1597-1 FUGG, and also for data storage at SDS@hd funded through grants INST 35/1314-1 FUGG and INST 35/1503-1 FUGG. RSK also thanks the Harvard-Smithsonian Center for Astrophysics and the Radcliffe Institute for Advanced Studies for their hospitality during his sabbatical, and the 2024/25 Class of Radcliffe Fellows for a great community and highly interesting and stimulating discussions.


%

\vspace{5mm}
\facilities{JWST, HST}


 \software{Astropy \citep{astropy:2013, astropy:2018, astropy:2022}.
This research made use of Photutils, an Astropy package for
detection and photometry of astronomical sources \citep{larry_bradley_2022_6825092}
AstroPy (astropy.org)
          }



\appendix

\section{IR colors of PHANGS-HST red evolved star candidates}
\label{append:AGBs}

A population that serves as a useful point of reference for this study is red evolved stars.   This population includes asymptotic giant branch stars (AGBs) and red supergiants (RSGs), and candidates can be readily selected from existing PHANGS-HST V and I band (F555W and F814W) DOLPHOT catalogs \citep{dolphin_dolphot_2016, 2022Thilker}.  Such samples of luminous red point sources will be bright in our JWST images, and they allow us to (1) determine MIRI color criteria to help remove old stellar populations from our 3.3~$\mu$m PAH sample (Sec.~\ref{sec:filtering}), and (2) characterize the NIRCam concentration indices of point sources so that we can ascertain the compactness of the 3.3~$\mu$m PAH emitters (Sec.~\ref{sec:concentration_index}).  

\begin{figure}
    \centering
    \includegraphics[width=2in]{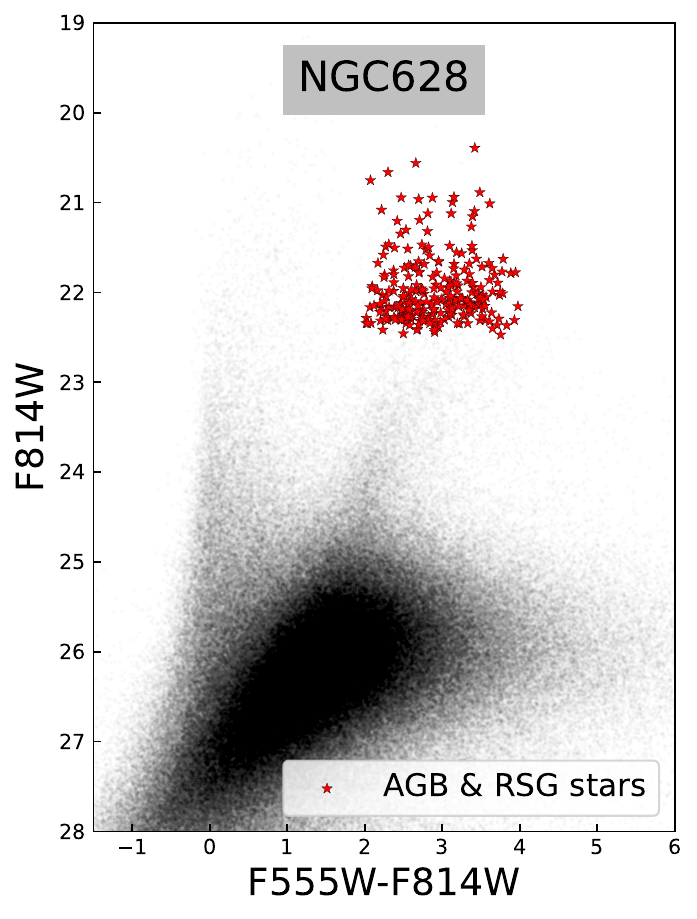}
    \caption{An example F555W-F814W vs. F814W color-magnitude diagram for the galaxy NGC~628. The black points are sources from the PHANGS-HST DOLPHOT catalog \citep{2022Thilker}, while the red points show a candidate sample of red evolved stars, used in this paper to help determine MIRI criteria to remove these sources from our 3.3~$\mu$m PAH sample, and to calculate the NIRCam concentration indices of point sources as a reference for evaluating the compactness of the 3.3~$\mu$m PAH emitters.}
    \label{fig:CMD_agb}
\end{figure}

Selection of candidate samples of red evolved stars in each galaxy is based on DOLPHOT output parameters, and location in the V-I color-magnitude diagram (Figure~\ref{fig:CMD_agb}). Specifically, we use the following DOLPHOT criteria: ${\rm S/N_{F814W}} \ge 10$, ${\rm crowd_{F814W}} \le 0.01$, ${\rm PhotQualFlag_{F814W}} = 0$, ${\rm OBJTYPE}=1$, $-0.025 < {\rm sharp_{F814W}} \le 0.01$, and $-0.15 < {\rm round_{F814W}} \le 0.4$, which ensures the identification of bright, uncrowded point sources which are well-fit by the HST PSF.  A description of these parameters can be found in the documentation for {\tt DOLPHOT} available at \url{http://americano.dolphinsim.com/dolphot/}.  The sharpness and roundness limits were adjusted as needed for each galaxy. In terms of the color-magnitude criteria, we selected sources with F814W magnitudes $3-5$ mag brighter than the foreground extinction-corrected tip of the red giant branch \citep{anand20} and considered only those sources with colors within the range $ \rm 2 \le F555W-F814W < 4$ mag (Figure~\ref{fig:CMD_agb}).

\begin{figure*}
    \centering
    \includegraphics[width=5in]{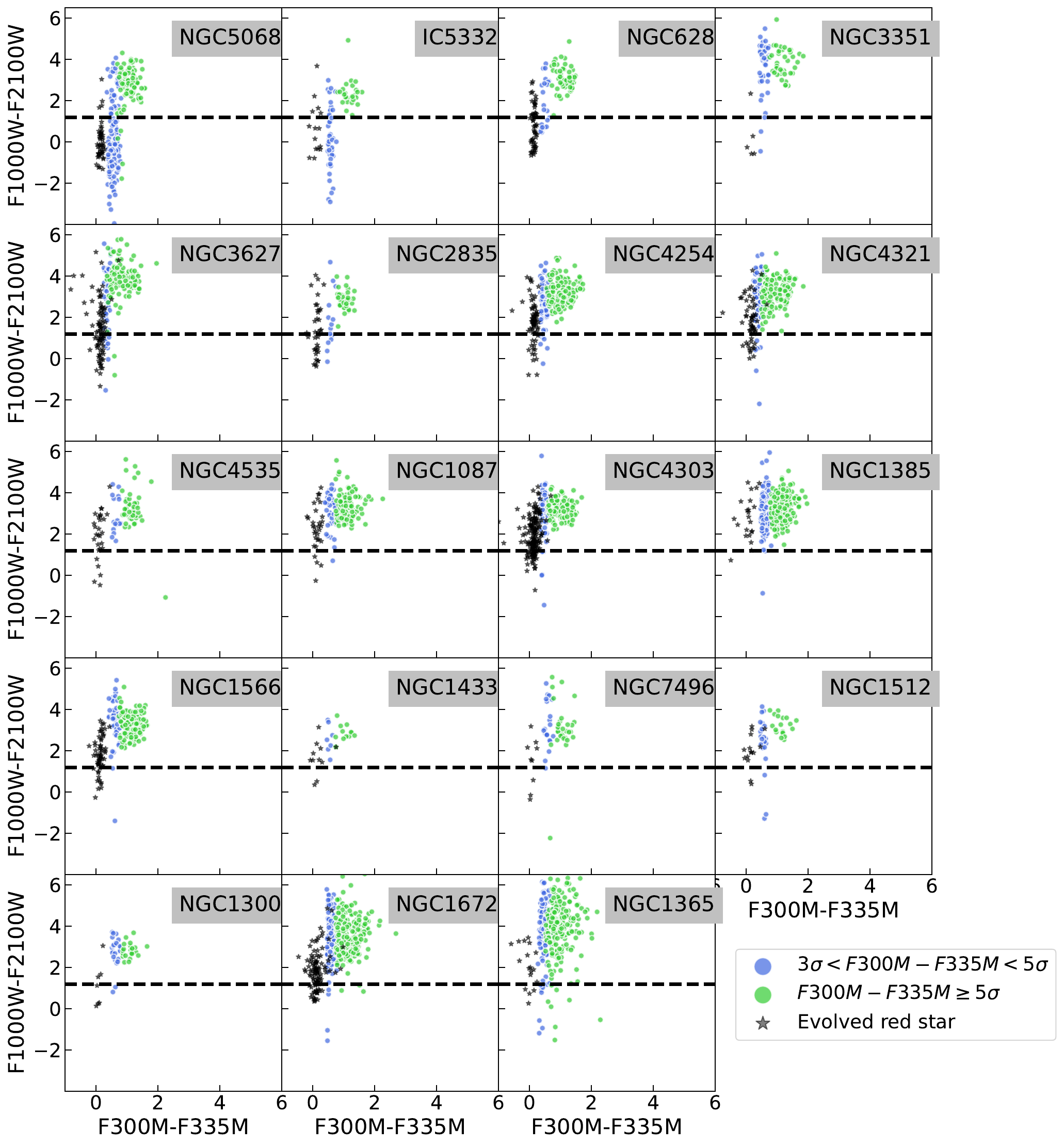}
    \caption{F1000W-F2100W vs. F300M-F335M color-color diagram for $>$5$\sigma$ (green) and 3-5$\sigma$ (blue) PAH emitters. The HST candidate samples of red evolved stars are plotted if their F1000W and F2100W photometry are above their respective 5$\sigma$ point source limits \citep[][Table 4]{phangs-jwst}.  The horizontal line corresponds to a value of F1000W-F2100W=1.2 (or equivalently $F_\mathrm{F2100W}/F_\mathrm{F1000W}=3$. )  To remove red IR-bright stars, we exclude objects with $F_\mathrm{F2100W}/F_\mathrm{F1000W}<3$ from the PAH emitter sample analyzed here. }
    \label{fig:TCD_21_10_300_335}
\end{figure*}

In Fig.~\ref{fig:TCD_21_10_300_335} we show the 
F1000W-F2100W vs.
F300M-F335M
color-color diagram for $>$5$\sigma$ (green) and 3-5$\sigma$ (blue) 3.3$\mu$m PAH emitters. The HST  red evolved star candidates are plotted if their F1000W and F2100W photometry is above the 5$\sigma$ point source limit \citep[23.2 and 21.6, repsectively][Table 4]{phangs-jwst}.  

We observe that the vast majority of the $>5\sigma$ PAH sample has F1000W-F2100W$>$ 1.2 ($F_\mathrm{F2100W}/F_\mathrm{F1000W}\geq3$, shown with a horizontal line). In the nearest galaxies (first row of panels where all galaxies have distances closer than 10 Mpc), we observe that a significant fraction of the 3-5$\sigma$ PAH emitters falls below this value. 
We anticipate a higher rate of contamination from individual stars in these nearby galaxies.  This is corroborated by the location of the candidate sample of evolved red stars in the nearest galaxy in the sample (NGC~5068, 5.2 Mpc), which is mostly below $F_\mathrm{F2100W}/F_\mathrm{F1000W}=3$.  As the galaxy distance increases, the F1000W and F2100W photometry (0\farcs328 and 0\farcs674 PSF FWHM, respectively) will not reflect individual luminous old stars, or even clusters dominated by such stars, and the F1000W-F2100W color of the sources gradually becomes larger.  The location of SAGE LMC stars \citep{Jones2017} in Figure~\ref{fig:CMD_21_10_jones_strong} clearly shows that individual stars cannot be detected in the more distant galaxies.

Nearly all sources in the 3.3$\mu$m PAH emitter sample ($\sim$99\%) are detected in F1000W and F2100W, although it should be noted that precise photometry for an individual star cluster can be challenging due to complex diffuse background emission and the coarser angular resolution.  Fully optimized MIRI photometry for star clusters is beyond the scope of this work, but the point source sensitivity limit in the least populated regions of the images \cite[][Table 4]{phangs-jwst} motivates a fairly conservative flux ratio cut. 
However, the F1000W and F2100W fluxes are {\em only} being used to exclude individual evolved stars from our sample.  Confusion and inclusion of emission from neighboring sources and/or diffuse ISM will increase $F_{F2100W}/F_{F1000W}$, since the resolution is twice as poor at F2100W as F1000W.  Thus, cutting only those sources with low $F_{F2100W}/F_{F1000W}$ is valid to remove contaminants, even with the relatively poor resolution at those wavelengths.

Based on these observations, and to be conservative in the criteria to exclude individual evolved stars from the sample, we restrict our analysis to objects with a ratio of $F_\mathrm{F2100W}/F_\mathrm{F1000W} > 3$. 

\section{Photometric Errors}
\label{append:Errors}

The photometric errors in the HST H$\alpha$, JWST F300M, and longer wavelength bands were estimated using the difference between the flux density values obtained by subtracting to the 0.1 and 0.9 quantiles of the annulus, as described in $\S$~\ref{Sect:Potometry}. This method enable to capture small variations in the local structure of the background.

\begin{figure*}
    \centering
    \includegraphics[width=5in]{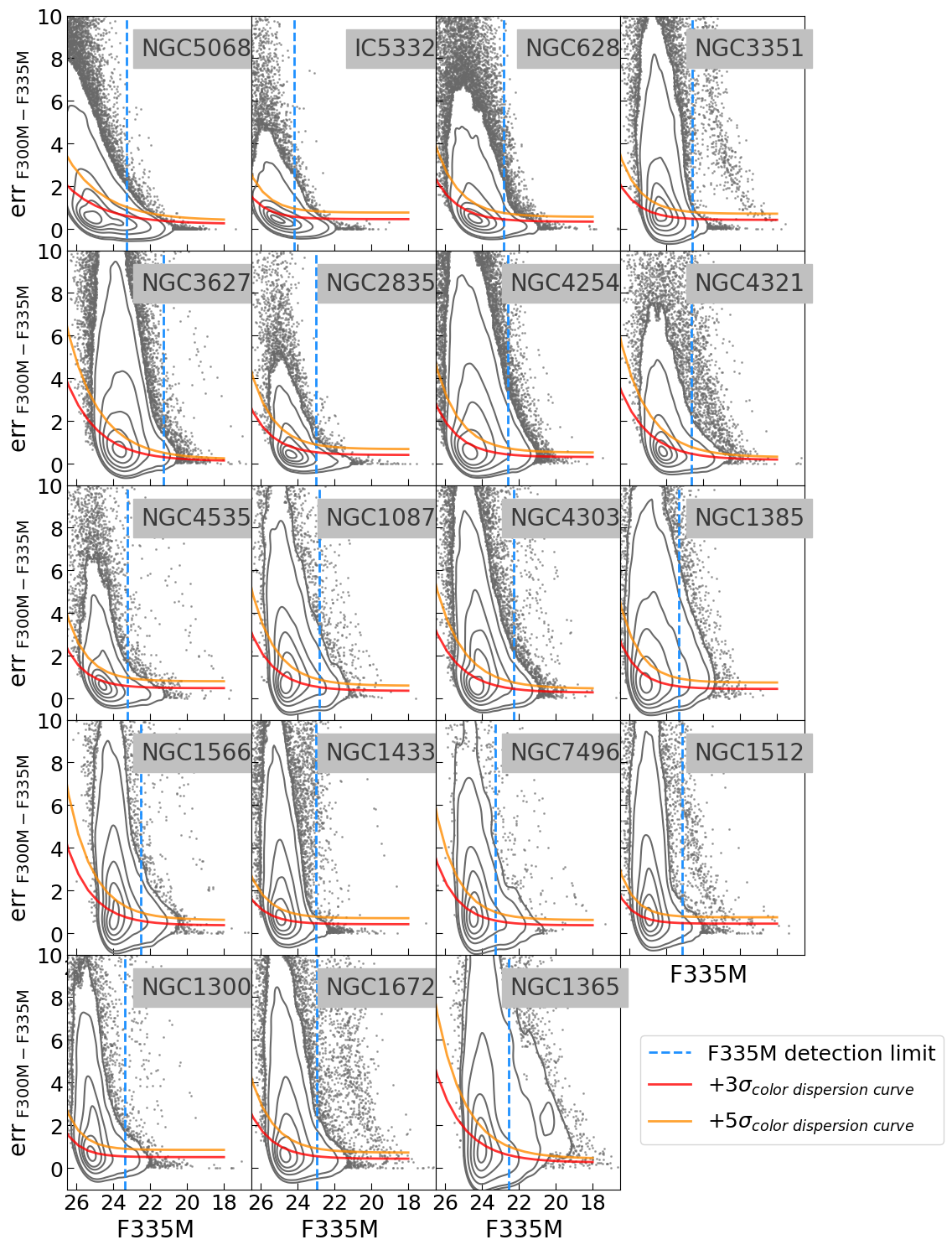}
    \caption{$\rm F300M-F335M$ photometric errors. The contours represent the 5th, 10th, 30th, 50th, 70th, and 90th density percentiles of all sources detected in the F335M images. Small gray dots mark regions with densities below the 5th percentile. The blue dashed line indicates the F335M detection limit, derived from random apertures ($\S$~\ref{sec:Identification}). The red and orange curves indicate the 3$\sigma$ and 5$\sigma$ color ($\rm F300M - F335M$) dispersion respectively (see fig.~\ref{fig:selection}).}
    \label{F300M-F335M_error}
\end{figure*}

In Fig.~\ref{F300M-F335M_error}, we present the photometric errors for the color $\rm F300M-F335M$. The blue dashed line represents the 5$\sigma$ detection limit derived from random aperture measurements over the F335M image. We note that this limit does not vary from galaxy to galaxy in the same way as the individual uncertainties, i.e., the variation in the detection limit across galaxies does not follow the same pattern as the average of the individual errors. This discrepancy arises because the detection limit is estimated using random apertures placed across the entire image, while the F335M detections ($\S$~\ref{sec:detection}) are concentrated primarily in the main body of galaxies, where the background is brighter, resulting in higher uncertainties.

\bibliography{all}
\bibliographystyle{aasjournal}



\end{document}

%% file: authors.tex
\author[0000-0002-0579-6613]{M. Jimena Rodríguez}
\email{jrodriguez@stsci.edu}
\affiliation{Space Telescope Science Institute, 3700 San Martin Drive, Baltimore, MD 21218, USA}
\affiliation{Instituto de Astrofísica de La Plata, CONICET--UNLP, Paseo del Bosque S/N, B1900FWA La Plata, Argentina }

\author[0000-0003-0946-6176]{Janice C. Lee}
\affiliation{Space Telescope Science Institute, 3700 San Martin Drive, Baltimore, MD 21218, USA}
\affiliation{Steward Observatory, University of Arizona, 933 N Cherry Ave, Tucson, AZ 85721, USA}

\author[0000-0002-4663-6827]{Remy Indebetouw}
\affiliation{University of Virginia Astronomy Department, P.O. Box 400325, Charlottesville, VA, 22904, USA}
\affiliation{National Radio Astronomy Observatory, 520 Edgemont Rd, Charlottesville, VA 22903, USA}

\author[0000-0002-3784-7032]{B. C. Whitmore}
\affiliation{Space Telescope Science Institute, 3700 San Martin Drive, Baltimore, MD 21218, USA}

\author[0000-0001-6038-9511]{Daniel Maschmann}
\affiliation{Steward Observatory, University of Arizona, 933 N Cherry Ave, Tucson, AZ 85721, USA}

\author[0000-0002-0012-2142]{Thomas G. Williams}
\affiliation{Sub-department of Astrophysics, Department of Physics, University of Oxford, Keble Road, Oxford OX1 3RH, UK}
\affiliation{Max-Planck-Institut f\"ur Astronomie, K\"onigstuhl 17, D-69117 Heidelberg, Germany}

\author[0000-0003-0085-4623]{Rupali~Chandar}
\affiliation{Ritter Astrophysical Research Center, University of Toledo, Toledo, OH 43606, USA}

\author[0000-0003-0410-4504]{A.~T.~Barnes}
\affiliation{European Southern Observatory (ESO), Karl-Schwarzschild-Stra{\ss}e 2, 85748 Garching, Germany}

\author[0000-0001-9852-9954]{Oleg~Y.~Gnedin}
\affiliation{Department of Astronomy, University of Michigan, Ann Arbor, MI 48109, USA}

\author[0000-0002-4378-8534]{Karin M. Sandstrom}
\affiliation{Department of Astronomy \& Astrophysics, University of California, San Diego, 9500 Gilman Drive, San Diego, CA 92093, USA}

\author[0000-0002-5204-2259]{Erik~Rosolowsky}
\affiliation{Department of Physics, University of Alberta, Edmonton, AB T6G 2E1, Canada}

\author[0000-0002-2545-1700]{Adam K. Leroy}
\affiliation{Department of Astronomy, The Ohio State University, Columbus, Ohio 43210, USA}

\author[0000-0002-8528-7340]{David A. Thilker}
\affiliation{Department of Physics and Astronomy, The Johns Hopkins University, Baltimore, MD 21218 USA}

\author[0000-0003-4770-688X]{Hwihyun~Kim}
\affiliation{Gemini Observatory/NSF’s NOIRLab, 950 N. Cherry Avenue, Tucson, AZ, USA}

\author[0000-0003-0378-4667]{Jiayi Sun}
\altaffiliation{NASA Hubble Fellow}
\affiliation{Department of Astrophysical Sciences, Princeton University, 4 Ivy Lane, Princeton, NJ 08544, USA}

\author[0000-0002-0560-3172]{Ralf S.\ Klessen}
\affiliation{Universit\"{a}t Heidelberg, Zentrum f\"{u}r Astronomie, Institut f\"{u}r Theoretische Astrophysik, Albert-Ueberle-Stra{\ss}e 2, D-69120 Heidelberg, Germany}
\affiliation{Universit\"{a}t Heidelberg, Interdisziplin\"{a}res Zentrum f\"{u}r Wissenschaftliches Rechnen, Im Neuenheimer Feld 205, D-69120 Heidelberg, Germany}

\author[0000-0002-9768-0246]{Brent Groves}
\affiliation{International Centre for Radio Astronomy Research, University of Western Australia, 7 Fairway, Crawley, 6009 WA, Australia}

\author[0000-0001-8289-3428]{Aida Wofford}\affiliation{Instituto de Astronom\'{i}a, Universidad Nacional Aut\'{o}noma de M\'{e}xico, Unidad Acad\'{e}mica en Ensenada, Km 103 Carr. Tijuana-Ensenada, Ensenada 22860, M\'{e}xico}
\affiliation{Department of Astronomy \& Astrophysics, University of California San Diego, 9500 Gilman Drive, La Jolla, CA 92093, USA}

\author[0000-0003-0946-6176]{Médéric Boquien}
\affiliation{Université Côte d'Azur, Observatoire de la Côte d'Azur, CNRS, Laboratoire Lagrange, 06000, Nice, France}

\author[0000-0002-5782-9093]{Daniel~A.~Dale}
\affiliation{Department of Physics and Astronomy, University of Wyoming, Laramie, WY 82071, USA}

\author[0000-0003-2508-2586]{Rebecca C. Levy}
\altaffiliation{NSF Astronomy and Astrophysics Postdoctoral Fellow}
\affiliation{Steward Observatory, University of Arizona, 933 N Cherry Ave, Tucson, AZ 85721, USA}
\affiliation{Space Telescope Science Institute, 3700 San Martin Drive, Baltimore, MD 21218, USA}

\author[0000-0002-4781-7291]{Sumit K. Sarbadhicary}
\affiliation{Department of Physics, The Ohio State University, Columbus, Ohio 43210, USA}
\affiliation{Center for Cosmology \& Astro-Particle Physics, The Ohio State University, Columbus, Ohio 43210, USA}
\affiliation{Department of Astronomy, The Ohio State University, Columbus, Ohio 43210, USA}

\author[0000-0001-7130-2880]{Leonardo \'Ubeda}
\affiliation{Space Telescope Science Institute, 3700 San Martin Drive, Baltimore, MD 21218, USA}

\author[0000-0003-3917-6460]{Kirsten~L.~Larson}
\affiliation{AURA for the European Space Agency (ESA), Space Telescope Science Institute, 3700 San Martin Drive, Baltimore, MD 21218, USA}

\author[0000-0001-8348-2671]{Kelsey~E.~Johnson}
\affiliation{Department of Astronomy, University of Virginia, Charlottesville, VA 22904, USA}

\author[0000-0003-0166-9745]{Frank Bigiel}
\affiliation{Argelander-Institut f\"ur Astronomie, Universit\"at Bonn, Auf dem H\"ugel 71, 53121 Bonn, Germany}

\author[0000-0002-8806-6308]{Hamid Hassani}
\affiliation{Department of Physics, University of Alberta, Edmonton, AB T6G 2E1, Canada}

\author[0000-0002-3247-5321]{Kathryn~Grasha}
\altaffiliation{ARC DECRA Fellow}
\affiliation{Research School of Astronomy and Astrophysics, Australian National University, Canberra, ACT 2611, Australia}   
\affiliation{ARC Centre of Excellence for All Sky Astrophysics in 3 Dimensions (ASTRO 3D), Australia}

%% file: main.bbl
\begin{thebibliography}{}
\expandafter\ifx\csname natexlab\endcsname\relax\def\natexlab#1{#1}\fi
\providecommand{\url}[1]{\href{#1}{#1}}
\providecommand{\dodoi}[1]{doi:~\href{http://doi.org/#1}{\nolinkurl{#1}}}
\providecommand{\doeprint}[1]{\href{http://ascl.net/#1}{\nolinkurl{http://ascl.net/#1}}}
\providecommand{\doarXiv}[1]{\href{https://arxiv.org/abs/#1}{\nolinkurl{https://arxiv.org/abs/#1}}}

\bibitem[{{Adamo} {et~al.}(2017){Adamo}, {Ryon}, {Messa}, {Kim}, {Grasha}, {Cook}, {Calzetti}, {Lee}, {Whitmore}, {Elmegreen}, {Ubeda}, {Smith}, {Bright}, {Runnholm}, {Andrews}, {Fumagalli}, {Gouliermis}, {Kahre}, {Nair}, {Thilker}, {Walterbos}, {Wofford}, {Aloisi}, {Ashworth}, {Brown}, {Chandar}, {Christian}, {Cignoni}, {Clayton}, {Dale}, {de Mink}, {Dobbs}, {Elmegreen}, {Evans}, {Gallagher}, {Grebel}, {Herrero}, {Hunter}, {Johnson}, {Kennicutt}, {Krumholz}, {Lennon}, {Levay}, {Martin}, {Nota}, {{\"O}stlin}, {Pellerin}, {Prieto}, {Regan}, {Sabbi}, {Sacchi}, {Schaerer}, {Schiminovich}, {Shabani}, {Tosi}, {Van Dyk}, \& {Zackrisson}}]{adamo17}
{Adamo}, A., {Ryon}, J.~E., {Messa}, M., {et~al.} 2017, \apj, 841, 131, \dodoi{10.3847/1538-4357/aa7132}

\bibitem[{{Anand} {et~al.}(2020){Anand}, {Lee}, {Van Dyk}, {Leroy}, {Rosolowsky}, {Schinnerer}, {Larson}, {Kourkchi}, {Kreckel}, {Scheuermann}, {Rizzi}, {Thilker}, {Tully}, {Bigiel}, {Blanc}, {Boquien}, {Chandar}, {Dale}, {Emsellem}, {Deger}, {Glover}, {Grasha}, {Groves}, {K{\ensuremath{\sim}}lessen}, {Kruijssen}, {Querejeta}, {S{\'a}nchez-Bl{\'a}zquez}, {Schruba}, {Turner}, {Ubeda}, {Williams}, \& {Whitmore}}]{anand20}
{Anand}, G.~S., {Lee}, J.~C., {Van Dyk}, S.~D., {et~al.} 2020, \mnras, \dodoi{10.1093/mnras/staa3668}

\bibitem[{{Astropy Collaboration} {et~al.}(2013){Astropy Collaboration}, {Robitaille}, {Tollerud}, {Greenfield}, {Droettboom}, {Bray}, {Aldcroft}, {Davis}, {Ginsburg}, {Price-Whelan}, {Kerzendorf}, {Conley}, {Crighton}, {Barbary}, {Muna}, {Ferguson}, {Grollier}, {Parikh}, {Nair}, {Unther}, {Deil}, {Woillez}, {Conseil}, {Kramer}, {Turner}, {Singer}, {Fox}, {Weaver}, {Zabalza}, {Edwards}, {Azalee Bostroem}, {Burke}, {Casey}, {Crawford}, {Dencheva}, {Ely}, {Jenness}, {Labrie}, {Lim}, {Pierfederici}, {Pontzen}, {Ptak}, {Refsdal}, {Servillat}, \& {Streicher}}]{astropy:2013}
{Astropy Collaboration}, {Robitaille}, T.~P., {Tollerud}, E.~J., {et~al.} 2013, \aap, 558, A33, \dodoi{10.1051/0004-6361/201322068}

\bibitem[{{Astropy Collaboration} {et~al.}(2018){Astropy Collaboration}, {Price-Whelan}, {Sip{\H{o}}cz}, {G{\"u}nther}, {Lim}, {Crawford}, {Conseil}, {Shupe}, {Craig}, {Dencheva}, {Ginsburg}, {Vand erPlas}, {Bradley}, {P{\'e}rez-Su{\'a}rez}, {de Val-Borro}, {Aldcroft}, {Cruz}, {Robitaille}, {Tollerud}, {Ardelean}, {Babej}, {Bach}, {Bachetti}, {Bakanov}, {Bamford}, {Barentsen}, {Barmby}, {Baumbach}, {Berry}, {Biscani}, {Boquien}, {Bostroem}, {Bouma}, {Brammer}, {Bray}, {Breytenbach}, {Buddelmeijer}, {Burke}, {Calderone}, {Cano Rodr{\'\i}guez}, {Cara}, {Cardoso}, {Cheedella}, {Copin}, {Corrales}, {Crichton}, {D'Avella}, {Deil}, {Depagne}, {Dietrich}, {Donath}, {Droettboom}, {Earl}, {Erben}, {Fabbro}, {Ferreira}, {Finethy}, {Fox}, {Garrison}, {Gibbons}, {Goldstein}, {Gommers}, {Greco}, {Greenfield}, {Groener}, {Grollier}, {Hagen}, {Hirst}, {Homeier}, {Horton}, {Hosseinzadeh}, {Hu}, {Hunkeler}, {Ivezi{\'c}}, {Jain}, {Jenness}, {Kanarek}, {Kendrew}, {Kern}, {Kerzendorf}, {Khvalko}, {King}, {Kirkby}, {Kulkarni},
  {Kumar}, {Lee}, {Lenz}, {Littlefair}, {Ma}, {Macleod}, {Mastropietro}, {McCully}, {Montagnac}, {Morris}, {Mueller}, {Mumford}, {Muna}, {Murphy}, {Nelson}, {Nguyen}, {Ninan}, {N{\"o}the}, {Ogaz}, {Oh}, {Parejko}, {Parley}, {Pascual}, {Patil}, {Patil}, {Plunkett}, {Prochaska}, {Rastogi}, {Reddy Janga}, {Sabater}, {Sakurikar}, {Seifert}, {Sherbert}, {Sherwood-Taylor}, {Shih}, {Sick}, {Silbiger}, {Singanamalla}, {Singer}, {Sladen}, {Sooley}, {Sornarajah}, {Streicher}, {Teuben}, {Thomas}, {Tremblay}, {Turner}, {Terr{\'o}n}, {van Kerkwijk}, {de la Vega}, {Watkins}, {Weaver}, {Whitmore}, {Woillez}, {Zabalza}, \& {Astropy Contributors}}]{astropy:2018}
{Astropy Collaboration}, {Price-Whelan}, A.~M., {Sip{\H{o}}cz}, B.~M., {et~al.} 2018, \aj, 156, 123, \dodoi{10.3847/1538-3881/aabc4f}

\bibitem[{{Astropy Collaboration} {et~al.}(2022){Astropy Collaboration}, {Price-Whelan}, {Lim}, {Earl}, {Starkman}, {Bradley}, {Shupe}, {Patil}, {Corrales}, {Brasseur}, {N{"o}the}, {Donath}, {Tollerud}, {Morris}, {Ginsburg}, {Vaher}, {Weaver}, {Tocknell}, {Jamieson}, {van Kerkwijk}, {Robitaille}, {Merry}, {Bachetti}, {G{"u}nther}, {Aldcroft}, {Alvarado-Montes}, {Archibald}, {B{'o}di}, {Bapat}, {Barentsen}, {Baz{'a}n}, {Biswas}, {Boquien}, {Burke}, {Cara}, {Cara}, {Conroy}, {Conseil}, {Craig}, {Cross}, {Cruz}, {D'Eugenio}, {Dencheva}, {Devillepoix}, {Dietrich}, {Eigenbrot}, {Erben}, {Ferreira}, {Foreman-Mackey}, {Fox}, {Freij}, {Garg}, {Geda}, {Glattly}, {Gondhalekar}, {Gordon}, {Grant}, {Greenfield}, {Groener}, {Guest}, {Gurovich}, {Handberg}, {Hart}, {Hatfield-Dodds}, {Homeier}, {Hosseinzadeh}, {Jenness}, {Jones}, {Joseph}, {Kalmbach}, {Karamehmetoglu}, {Ka{l}uszy{'n}ski}, {Kelley}, {Kern}, {Kerzendorf}, {Koch}, {Kulumani}, {Lee}, {Ly}, {Ma}, {MacBride}, {Maljaars}, {Muna}, {Murphy}, {Norman}, {O'Steen},
  {Oman}, {Pacifici}, {Pascual}, {Pascual-Granado}, {Patil}, {Perren}, {Pickering}, {Rastogi}, {Roulston}, {Ryan}, {Rykoff}, {Sabater}, {Sakurikar}, {Salgado}, {Sanghi}, {Saunders}, {Savchenko}, {Schwardt}, {Seifert-Eckert}, {Shih}, {Jain}, {Shukla}, {Sick}, {Simpson}, {Singanamalla}, {Singer}, {Singhal}, {Sinha}, {Sip{H{o}}cz}, {Spitler}, {Stansby}, {Streicher}, {{{S}}umak}, {Swinbank}, {Taranu}, {Tewary}, {Tremblay}, {Val-Borro}, {Van Kooten}, {Vasovi{'c}}, {Verma}, {de Miranda Cardoso}, {Williams}, {Wilson}, {Winkel}, {Wood-Vasey}, {Xue}, {Yoachim}, {Zhang}, {Zonca}, \& {Astropy Project Contributors}}]{astropy:2022}
{Astropy Collaboration}, {Price-Whelan}, A.~M., {Lim}, P.~L., {et~al.} 2022, apj, 935, 167, \dodoi{10.3847/1538-4357/ac7c74}

\bibitem[{{Belfiore} {et~al.}(2023){Belfiore}, {Leroy}, {Williams}, {Barnes}, {Bigiel}, {Boquien}, {Cao}, {Chastenet}, {Congiu}, {Dale}, {Egorov}, {Eibensteiner}, {Emsellem}, {Glover}, {Groves}, {Hassani}, {Klessen}, {Kreckel}, {Neumann}, {Neumann}, {Querejeta}, {Rosolowsky}, {Sanchez-Blazquez}, {Sandstrom}, {Schinnerer}, {Sun}, {Sutter}, \& {Watkins}}]{belfiore23}
{Belfiore}, F., {Leroy}, A.~K., {Williams}, T.~G., {et~al.} 2023, \aap, 678, A129, \dodoi{10.1051/0004-6361/202347175}

\bibitem[{{Bertin} \& {Arnouts}(1996)}]{1996A&AS..117..393B}
{Bertin}, E., \& {Arnouts}, S. 1996, \aaps, 117, 393, \dodoi{10.1051/aas:1996164}

\bibitem[{Bradley {et~al.}(2022)Bradley, Sipőcz, Robitaille, Tollerud, Vinícius, Deil, Barbary, Wilson, Busko, Donath, Günther, Cara, Lim, Meßlinger, Conseil, Bostroem, Droettboom, Bray, Bratholm, Barentsen, Craig, Rathi, Pascual, Perren, Georgiev, de~Val-Borro, Kerzendorf, Bach, Quint, \& Souchereau}]{larry_bradley_2022_6825092}
Bradley, L., Sipőcz, B., Robitaille, T., {et~al.} 2022, astropy/photutils: 1.5.0, 1.5.0,  Zenodo, \dodoi{10.5281/zenodo.6825092}

\bibitem[{{Brandl} {et~al.}(1996){Brandl}, {Sams}, {Bertoldi}, {Eckart}, {Genzel}, {Drapatz}, {Hofmann}, {Loewe}, \& {Quirrenbach}}]{Brandl96}
{Brandl}, B., {Sams}, B.~J., {Bertoldi}, F., {et~al.} 1996, \apj, 466, 254, \dodoi{10.1086/177507}

\bibitem[{{Brown} \& {Gnedin}(2021)}]{brown2021}
{Brown}, G., \& {Gnedin}, O.~Y. 2021, arXiv e-prints, arXiv:2106.12420.
\newblock \doarXiv{2106.12420}

\bibitem[{{Chandar} {et~al.}(2024){Chandar}, {Barnes}, \& {Thilker}}]{chandar24}
{Chandar}, R., {Barnes}, A., \& {Thilker}, D. 2024

\bibitem[{{Chandar} {et~al.}(2010){Chandar}, {Whitmore}, {Kim}, {Kaleida}, {Mutchler}, {Calzetti}, {Saha}, {O'Connell}, {Balick}, {Bond}, {Carollo}, {Disney}, {Dopita}, {Frogel}, {Hall}, {Holtzman}, {Kimble}, {McCarthy}, {Paresce}, {Silk}, {Trauger}, {Walker}, {Windhorst}, \& {Young}}]{chandar10b}
{Chandar}, R., {Whitmore}, B.~C., {Kim}, H., {et~al.} 2010, \apj, 719, 966, \dodoi{10.1088/0004-637X/719/1/966}

\bibitem[{{Chevance} {et~al.}(2020){Chevance}, {Kruijssen}, {Hygate}, {Schruba}, {Longmore}, {Groves}, {Henshaw}, {Herrera}, {Hughes}, {Jeffreson}, {Lang}, {Leroy}, {Meidt}, {Pety}, {Razza}, {Rosolowsky}, {Schinnerer}, {Bigiel}, {Blanc}, {Emsellem}, {Faesi}, {Glover}, {Haydon}, {Ho}, {Kreckel}, {Lee}, {Liu}, {Querejeta}, {Saito}, {Sun}, {Usero}, \& {Utomo}}]{Chevance20}
{Chevance}, M., {Kruijssen}, J.~M.~D., {Hygate}, A. P.~S., {et~al.} 2020, \mnras, 493, 2872, \dodoi{10.1093/mnras/stz3525}

\bibitem[{{Cook} {et~al.}(2019){Cook}, {Lee}, {Adamo}, {Kim}, {Chandar}, {Whitmore}, {Mok}, {Ryon}, {Dale}, {Calzetti}, {Andrews}, {Aloisi}, {Ashworth}, {Bright}, {Brown}, {Christian}, {Cignoni}, {Clayton}, {da Silva}, {de Mink}, {Dobbs}, {Elmegreen}, {Elmegreen}, {Evans}, {Fumagalli}, {Gallagher}, {Gouliermis}, {Grasha}, {Grebel}, {Herrero}, {Hunter}, {Jensen}, {Johnson}, {Kahre}, {Kennicutt}, {Krumholz}, {Lee}, {Lennon}, {Linden}, {Martin}, {Messa}, {Nair}, {Nota}, {{\"O}stlin}, {Parziale}, {Pellerin}, {Regan}, {Sabbi}, {Sacchi}, {Schaerer}, {Schiminovich}, {Shabani}, {Slane}, {Small}, {Smith}, {Smith}, {Taibi}, {Thilker}, {de la Torre}, {Tosi}, {Turner}, {Ubeda}, {Van Dyk}, {Walterbos}, \& {Wofford}}]{cook19}
{Cook}, D.~O., {Lee}, J.~C., {Adamo}, A., {et~al.} 2019, \mnras, 484, 4897, \dodoi{10.1093/mnras/stz331}

\bibitem[{{Deger} {et~al.}(2022){Deger}, {Lee}, {Whitmore}, {Thilker}, {Boquien}, {Chandar}, {Dale}, {Ubeda}, {White}, {Grasha}, {Glover}, {Schruba}, {Barnes}, {Klessen}, {Kruijssen}, {Rosolowsky}, \& {Williams}}]{deger22}
{Deger}, S., {Lee}, J.~C., {Whitmore}, B.~C., {et~al.} 2022, \mnras, 510, 32, \dodoi{10.1093/mnras/stab3213}

\bibitem[{Dolphin(2016)}]{dolphin_dolphot_2016}
Dolphin, A. 2016, Astrophysics Source Code Library, ascl:1608.013.
\newblock \url{https://ui.adsabs.harvard.edu/abs/2016ascl.soft08013D}

\bibitem[{{Dom{\'\i}nguez} {et~al.}(2023){Dom{\'\i}nguez}, {Pellegrini}, {Klessen}, \& {Rahner}}]{Dominguez23}
{Dom{\'\i}nguez}, R., {Pellegrini}, E.~W., {Klessen}, R.~S., \& {Rahner}, D. 2023, \mnras, 520, 5600, \dodoi{10.1093/mnras/stad482}

\bibitem[{{Draine} \& {Hensley}(2021)}]{dh21_dielectric}
{Draine}, B.~T., \& {Hensley}, B.~S. 2021, \apj, 909, 94, \dodoi{10.3847/1538-4357/abd6c6}

\bibitem[{{Emsellem} {et~al.}(2022){Emsellem}, {Schinnerer}, {Santoro}, {Belfiore}, {Pessa}, {McElroy}, {Blanc}, {Congiu}, {Groves}, {Ho}, {Kreckel}, {Razza}, {Sanchez-Blazquez}, {Egorov}, {Faesi}, {Klessen}, {Leroy}, {Meidt}, {Querejeta}, {Rosolowsky}, {Scheuermann}, {Anand}, {Barnes}, {Be{\v{s}}li{\'c}}, {Bigiel}, {Boquien}, {Cao}, {Chevance}, {Dale}, {Eibensteiner}, {Glover}, {Grasha}, {Henshaw}, {Hughes}, {Koch}, {Kruijssen}, {Lee}, {Liu}, {Pan}, {Pety}, {Saito}, {Sandstrom}, {Schruba}, {Sun}, {Thilker}, {Usero}, {Watkins}, \& {Williams}}]{phangs-muse}
{Emsellem}, E., {Schinnerer}, E., {Santoro}, F., {et~al.} 2022, \aap, 659, A191, \dodoi{10.1051/0004-6361/202141727}

\bibitem[{{Fall}(2006)}]{fall06}
{Fall}, S.~M. 2006, \apj, 652, 1129, \dodoi{10.1086/508404}

\bibitem[{{Galliano} {et~al.}(2008){Galliano}, {Dwek}, \& {Chanial}}]{galliano08}
{Galliano}, F., {Dwek}, E., \& {Chanial}, P. 2008, \apj, 672, 214, \dodoi{10.1086/523621}

\bibitem[{{Groenewegen}(2022)}]{2022Groenewegen}
{Groenewegen}, M.~A.~T. 2022, \aap, 659, A145, \dodoi{10.1051/0004-6361/202142648}

\bibitem[{{Grudi{\'c}} {et~al.}(2021){Grudi{\'c}}, {Guszejnov}, {Hopkins}, {Offner}, \& {Faucher-Gigu{\`e}re}}]{grudic21}
{Grudi{\'c}}, M.~Y., {Guszejnov}, D., {Hopkins}, P.~F., {Offner}, S. S.~R., \& {Faucher-Gigu{\`e}re}, C.-A. 2021, \mnras, 506, 2199, \dodoi{10.1093/mnras/stab1347}

\bibitem[{{Hannon} {et~al.}(2019){Hannon}, {Lee}, {Whitmore}, {Chandar}, {Adamo}, {Mobasher}, {Aloisi}, {Calzetti}, {Cignoni}, {Cook}, {Dale}, {Deger}, {Della Bruna}, {Elmegreen}, {Gouliermis}, {Grasha}, {Grebel}, {Herrero}, {Hunter}, {Johnson}, {Kennicutt}, {Kim}, {Sacchi}, {Smith}, {Thilker}, {Turner}, {Walterbos}, \& {Wofford}}]{hannon19}
{Hannon}, S., {Lee}, J.~C., {Whitmore}, B.~C., {et~al.} 2019, \mnras, 490, 4648, \dodoi{10.1093/mnras/stz2820}

\bibitem[{{Hannon} {et~al.}(2022){Hannon}, {Lee}, {Whitmore}, {Mobasher}, {Thilker}, {Chandar}, {Adamo}, {Wofford}, {Orozco-Duarte}, {Calzetti}, {Della Bruna}, {Kreckel}, {Groves}, {Barnes}, {Boquien}, {Belfiore}, \& {Linden}}]{hannon22}
---. 2022, \mnras, 512, 1294, \dodoi{10.1093/mnras/stac550}

\bibitem[{{Hannon} {et~al.}(2023){Hannon}, {Whitmore}, {Lee}, {Thilker}, {Deger}, {Huerta}, {Wei}, {Mobasher}, {Klessen}, {Boquien}, {Dale}, {Chevance}, {Grasha}, {Sanchez-Blazquez}, {Williams}, {Scheuermann}, {Groves}, {Kim}, {Kruijssen}, \& {The Phangs-HST Team}}]{hannon23}
{Hannon}, S., {Whitmore}, B.~C., {Lee}, J.~C., {et~al.} 2023, \mnras, 526, 2991, \dodoi{10.1093/mnras/stad2238}

\bibitem[{Hollyhead {et~al.}(2015)Hollyhead, Bastian, Adamo, Silva-Villa, Dale, Ryon, \& Gazak}]{hollyhead15}
Hollyhead, K., Bastian, N., Adamo, A., {et~al.} 2015, Monthly Notices of the Royal Astronomical Society, 449, 1106, \dodoi{10.1093/mnras/stv331}

\bibitem[{{Hopkins} {et~al.}(2014){Hopkins}, {Kere{\v s}}, {O{\~n}orbe}, {Faucher-Gigu{\`e}re}, {Quataert}, {Murray}, \& {Bullock}}]{hopkins14}
{Hopkins}, P.~F., {Kere{\v s}}, D., {O{\~n}orbe}, J., {et~al.} 2014, \mnras, 445, 581, \dodoi{10.1093/mnras/stu1738}

\bibitem[{Imanishi {et~al.}(2010)Imanishi, Nakagawa, Shirahata, Ohyama, \& Onaka}]{Imanishi_2010}
Imanishi, M., Nakagawa, T., Shirahata, M., Ohyama, Y., \& Onaka, T. 2010, The Astrophysical Journal, 721, 1233, \dodoi{10.1088/0004-637X/721/2/1233}

\bibitem[{{Jones} {et~al.}(2017){Jones}, {Meixner}, {Justtanont}, \& {Glasse}}]{Jones2017}
{Jones}, O.~C., {Meixner}, M., {Justtanont}, K., \& {Glasse}, A. 2017, \apj, 841, 15, \dodoi{10.3847/1538-4357/aa6bf6}

\bibitem[{{Kemper} {et~al.}(2010){Kemper}, {Woods}, {Antoniou}, {Bernard}, {Blum}, {Boyer}, {Chan}, {Chen}, {Cohen}, {Dijkstra}, {Engelbracht}, {Galametz}, {Galliano}, {Gielen}, {Gordon}, {Gorjian}, {Harris}, {Hony}, {Hora}, {Indebetouw}, {Jones}, {Kawamura}, {Lagadec}, {Lawton}, {Leisenring}, {Madden}, {Marengo}, {Matsuura}, {McDonald}, {McGuire}, {Meixner}, {Mulia}, {O'Halloran}, {Oliveira}, {Paladini}, {Paradis}, {Reach}, {Rubin}, {Sandstrom}, {Sargent}, {Sewilo}, {Shiao}, {Sloan}, {Speck}, {Srinivasan}, {Szczerba}, {Tielens}, {van Aarle}, {Van Dyk}, {van Loon}, {Van Winckel}, {Vijh}, {Volk}, {Whitney}, {Wilkins}, \& {Zijlstra}}]{Kemper2010}
{Kemper}, F., {Woods}, P.~M., {Antoniou}, V., {et~al.} 2010, \pasp, 122, 683, \dodoi{10.1086/653438}

\bibitem[{{Kennicutt} \& {Evans}(2012)}]{kenicutt12}
{Kennicutt}, R.~C., \& {Evans}, N.~J. 2012, \araa, 50, 531, \dodoi{10.1146/annurev-astro-081811-125610}

\bibitem[{{Kim} {et~al.}(2023){Kim}, {Chevance}, {Kruijssen}, {Barnes}, {Bigiel}, {Blanc}, {Boquien}, {Cao}, {Congiu}, {Dale}, {Egorov}, {Faesi}, {Glover}, {Grasha}, {Groves}, {Hassani}, {Hughes}, {Klessen}, {Kreckel}, {Larson}, {Lee}, {Leroy}, {Liu}, {Longmore}, {Meidt}, {Pan}, {Pety}, {Querejeta}, {Rosolowsky}, {Saito}, {Sandstrom}, {Schinnerer}, {Smith}, {Usero}, {Watkins}, \& {Williams}}]{kim23}
{Kim}, J., {Chevance}, M., {Kruijssen}, J.~M.~D., {et~al.} 2023, \apjl, 944, L20, \dodoi{10.3847/2041-8213/aca90a}

\bibitem[{{Kim} {et~al.}(2012){Kim}, {Im}, {Lee}, {Lee}, {Jun}, {Nakagawa}, {Matsuhara}, {Wada}, {Oyabu}, {Takagi}, {Inami}, {Ohyama}, {Yamada}, {Helou}, {Armus}, \& {Shi}}]{kim12}
{Kim}, J.~H., {Im}, M., {Lee}, H.~M., {et~al.} 2012, \apj, 760, 120, \dodoi{10.1088/0004-637X/760/2/120}

\bibitem[{{Klessen} \& {Glover}(2016)}]{2016SAAS...43...85K}
{Klessen}, R.~S., \& {Glover}, S. C.~O. 2016, Saas-Fee Advanced Course, 43, 85, \dodoi{10.1007/978-3-662-47890-5_2}

\bibitem[{{Krumholz} {et~al.}(2019){Krumholz}, {McKee}, \& {Bland -Hawthorn}}]{krumholz19}
{Krumholz}, M.~R., {McKee}, C.~F., \& {Bland -Hawthorn}, J. 2019, \araa, 57, 227, \dodoi{10.1146/annurev-astro-091918-104430}

\bibitem[{{Larsen}(2009)}]{larsen09}
{Larsen}, S.~S. 2009, \aap, 494, 539, \dodoi{10.1051/0004-6361:200811212}

\bibitem[{{Lee} {et~al.}(2012){Lee}, {Ly}, {Spitler}, {Labb{\'e}}, {Salim}, {Persson}, {Ouchi}, {Dale}, {Monson}, \& {Murphy}}]{jlee12}
{Lee}, J.~C., {Ly}, C., {Spitler}, L., {et~al.} 2012, \pasp, 124, 782, \dodoi{10.1086/666528}

\bibitem[{{Lee} {et~al.}(2022){Lee}, {Whitmore}, {Thilker}, {Deger}, {Larson}, {Ubeda}, {Anand}, {Boquien}, {Chandar}, {Dale}, {Emsellem}, {Leroy}, {Rosolowsky}, {Schinnerer}, {Schmidt}, {Lilly}, {Turner}, {Van Dyk}, {White}, {Barnes}, {Belfiore}, {Bigiel}, {Blanc}, {Cao}, {Chevance}, {Congiu}, {Egorov}, {Glover}, {Grasha}, {Groves}, {Henshaw}, {Hughes}, {Klessen}, {Koch}, {Kreckel}, {Kruijssen}, {Liu}, {Lopez}, {Mayker}, {Meidt}, {Murphy}, {Pan}, {Pety}, {Querejeta}, {Razza}, {Saito}, {S{\'a}nchez-Bl{\'a}zquez}, {Santoro}, {Sardone}, {Scheuermann}, {Schruba}, {Sun}, {Usero}, {Watkins}, \& {Williams}}]{phangs-hst}
{Lee}, J.~C., {Whitmore}, B.~C., {Thilker}, D.~A., {et~al.} 2022, \apjs, 258, 10, \dodoi{10.3847/1538-4365/ac1fe5}

\bibitem[{{Lee} {et~al.}(2023){Lee}, {Sandstrom}, {Leroy}, {Thilker}, {Schinnerer}, {Rosolowsky}, {Larson}, {Egorov}, {Williams}, {Schmidt}, {Emsellem}, {Anand}, {Barnes}, {Belfiore}, {Be{\v{s}}li{\'c}}, {Bigiel}, {Blanc}, {Bolatto}, {Boquien}, {den Brok}, {Cao}, {Chandar}, {Chastenet}, {Chevance}, {Chiang}, {Congiu}, {Dale}, {Deger}, {Eibensteiner}, {Faesi}, {Glover}, {Grasha}, {Groves}, {Hassani}, {Henny}, {Henshaw}, {Hoyer}, {Hughes}, {Jeffreson}, {Jim{\'e}nez-Donaire}, {Kim}, {Kim}, {Klessen}, {Koch}, {Kreckel}, {Kruijssen}, {Li}, {Liu}, {Lopez}, {Maschmann}, {Chen}, {Meidt}, {Murphy}, {Neumann}, {Neumayer}, {Pan}, {Pessa}, {Pety}, {Querejeta}, {Pinna}, {Rodr{\'\i}guez}, {Saito}, {S{\'a}nchez-Bl{\'a}zquez}, {Santoro}, {Sardone}, {Smith}, {Sormani}, {Scheuermann}, {Stuber}, {Sutter}, {Sun}, {Teng}, {Tre{\ss}}, {Usero}, {Watkins}, {Whitmore}, \& {Razza}}]{phangs-jwst}
{Lee}, J.~C., {Sandstrom}, K.~M., {Leroy}, A.~K., {et~al.} 2023, \apjl, 944, L17, \dodoi{10.3847/2041-8213/acaaae}

\bibitem[{{Leger} {et~al.}(1989){Leger}, {D'Hendecourt}, \& {Defourneau}}]{1989A&A...216..148L}
{Leger}, A., {D'Hendecourt}, L., \& {Defourneau}, D. 1989, \aap, 216, 148

\bibitem[{{Leroy} {et~al.}(2021){Leroy}, {Schinnerer}, {Hughes}, {Rosolowsky}, {Pety}, {Schruba}, {Usero}, {Blanc}, {Chevance}, {Emsellem}, {Faesi}, {Herrera}, {Liu}, {Meidt}, {Querejeta}, {Saito}, {Sandstrom}, {Sun}, {Williams}, {Anand}, {Barnes}, {Behrens}, {Belfiore}, {Benincasa}, {Be{\v{s}}li{\'c}}, {Bigiel}, {Bolatto}, {den Brok}, {Cao}, {Chandar}, {Chastenet}, {Chiang}, {Congiu}, {Dale}, {Deger}, {Eibensteiner}, {Egorov}, {Garc{\'\i}a-Rodr{\'\i}guez}, {Glover}, {Grasha}, {Henshaw}, {Ho}, {Kepley}, {Kim}, {Klessen}, {Kreckel}, {Koch}, {Kruijssen}, {Larson}, {Lee}, {Lopez}, {Machado}, {Mayker}, {McElroy}, {Murphy}, {Ostriker}, {Pan}, {Pessa}, {Puschnig}, {Razza}, {S{\'a}nchez-Bl{\'a}zquez}, {Santoro}, {Sardone}, {Scheuermann}, {Sliwa}, {Sormani}, {Stuber}, {Thilker}, {Turner}, {Utomo}, {Watkins}, \& {Whitmore}}]{phangs-alma}
{Leroy}, A.~K., {Schinnerer}, E., {Hughes}, A., {et~al.} 2021, \apjs, 257, 43, \dodoi{10.3847/1538-4365/ac17f3}

\bibitem[{{Levy} {et~al.}(2024){Levy}, {Bolatto}, {Mayya}, {Cuevas-Otahola}, {Tarantino}, {Boyer}, {Boogaard}, {B{\"o}ker}, {Cronin}, {Dale}, {Donaghue}, {Emig}, {Fisher}, {Glover}, {Herrera-Camus}, {Jim{\'e}nez-Donaire}, {Klessen}, {Lenki{\'c}}, {Leroy}, {De Looze}, {Meier}, {Mills}, {Ott}, {Rela{\~n}o}, {Veilleux}, {Villanueva}, {Walter}, \& {van der Werf}}]{levy24}
{Levy}, R.~C., {Bolatto}, A.~D., {Mayya}, D., {et~al.} 2024, \apjl, 973, L55, \dodoi{10.3847/2041-8213/ad7af3}

\bibitem[{{Li}(2020)}]{li20}
{Li}, A. 2020, Nature Astronomy, 4, 339, \dodoi{10.1038/s41550-020-1051-1}

\bibitem[{{Linden} {et~al.}(2023){Linden}, {Evans}, {Armus}, {Rich}, {Larson}, {Lai}, {Privon}, {U}, {Inami}, {Bohn}, {Song}, {Barcos-Mu{\~n}oz}, {Charmandaris}, {Medling}, {Stierwalt}, {Diaz-Santos}, {B{\"o}ker}, {van der Werf}, {Aalto}, {Appleton}, {Brown}, {Hayward}, {Howell}, {Iwasawa}, {Kemper}, {Frayer}, {Law}, {Malkan}, {Marshall}, {Mazzarella}, {Murphy}, {Sanders}, \& {Surace}}]{linden23}
{Linden}, S.~T., {Evans}, A.~S., {Armus}, L., {et~al.} 2023, \apjl, 944, L55, \dodoi{10.3847/2041-8213/acb335}

\bibitem[{{Linden} {et~al.}(2024){Linden}, {Lai}, {Evans}, {Armus}, {Larson}, {Rich}, {U}, {Privon}, {Inami}, {Song}, {Bianchin}, {Bohn}, {Buiten}, {Sanchez-Garc{\'\i}a}, {Kader}, {Lenki{\'c}}, {Medling}, {B{\"o}ker}, {D{\'\i}az-Santos}, {Charmandaris}, {Barcos-Mu{\~n}oz}, {van der Werf}, {Stierwalt}, {Aalto}, {Appleton}, {Hayward}, {Howell}, {Malkan}, {Mazzarella}, {Murphy}, \& {Surace}}]{linden24}
{Linden}, S.~T., {Lai}, T., {Evans}, A.~S., {et~al.} 2024, \apjl, 974, L27, \dodoi{10.3847/2041-8213/ad7eae}

\bibitem[{{Ly} {et~al.}(2011){Ly}, {Lee}, {Dale}, {Momcheva}, {Salim}, {Staudaher}, {Moore}, \& {Finn}}]{ly11}
{Ly}, C., {Lee}, J.~C., {Dale}, D.~A., {et~al.} 2011, \apj, 726, 109, \dodoi{10.1088/0004-637X/726/2/109}

\bibitem[{{Maragkoudakis} {et~al.}(2023){Maragkoudakis}, {Peeters}, \& {Ricca}}]{marag23}
{Maragkoudakis}, A., {Peeters}, E., \& {Ricca}, A. 2023, \mnras, 520, 5354, \dodoi{10.1093/mnras/stad465}

\bibitem[{{Martins} {et~al.}(2005){Martins}, {Schaerer}, \& {Hillier}}]{martins05}
{Martins}, F., {Schaerer}, D., \& {Hillier}, D.~J. 2005, \aap, 436, 1049, \dodoi{10.1051/0004-6361:20042386}

\bibitem[{{Maschmann} {et~al.}(2024){Maschmann}, {Lee}, {Thilker}, {Whitmore}, {Deger}, {Boquien}, {Chandar}, {Dale}, {Wofford}, {Hannon}, {Larson}, {Leroy}, {Schinnerer}, {Rosolowsky}, {Ubeda}, {Barnes}, {Emsellem}, {Grasha}, {Groves}, {Kim}, {Klessen}, {Kreckel}, {Levy}, {Pinna}, {Rodriguez}, {Tian}, \& {Williams}}]{Maschmann2024}
{Maschmann}, D., {Lee}, J.~C., {Thilker}, D.~A., {et~al.} 2024, arXiv e-prints, arXiv:2403.04901, \dodoi{10.48550/arXiv.2403.04901}

\bibitem[{{Mori} {et~al.}(2014){Mori}, {Onaka}, {Sakon}, {Ishihara}, {Shimonishi}, {Ohsawa}, \& {Bell}}]{Mori2014}
{Mori}, T.~I., {Onaka}, T., {Sakon}, I., {et~al.} 2014, \apj, 784, 53, \dodoi{10.1088/0004-637X/784/1/53}

\bibitem[{{Ohsawa} {et~al.}(2013){Ohsawa}, {Onaka}, {Sakon}, {Mori}, {Yamamura}, {Matsuura}, {Kaneda}, {Bernard-Salas}, {Bern{\'e}}, \& {Joblin}}]{2013Ohsawa}
{Ohsawa}, R., {Onaka}, T., {Sakon}, I., {et~al.} 2013, in Proceedings of The Life Cycle of Dust in the Universe: Observations, 126, \dodoi{10.22323/1.207.0126}

\bibitem[{{Ossenkopf} \& {Henning}(1994)}]{1994A&A...291..943O}
{Ossenkopf}, V., \& {Henning}, T. 1994, \aap, 291, 943

\bibitem[{{Pedrini} {et~al.}(2024){Pedrini}, {Adamo}, {Calzetti}, {Bik}, {Gregg}, {Linden}, {Bajaj}, {Ryon}, {Ali}, {Bortolini}, {Correnti}, {Elmegreen}, {Elmegreen}, {Gallagher}, {Grasha}, {Gutermuth}, {Johnson}, {Melinder}, {Messa}, {{\"O}stlin}, {Sabbi}, {Smith}, {Tosi}, \& {Vieira}}]{pedrini24}
{Pedrini}, A., {Adamo}, A., {Calzetti}, D., {et~al.} 2024, \apj, 971, 32, \dodoi{10.3847/1538-4357/ad534d}

\bibitem[{{Peeters} {et~al.}(2002){Peeters}, {Mart{\'\i}n-Hern{\'a}ndez}, {Damour}, {Cox}, {Roelfsema}, {Baluteau}, {Tielens}, {Churchwell}, {Kessler}, {Mathis}, {Morisset}, \& {Schaerer}}]{peeters-iso}
{Peeters}, E., {Mart{\'\i}n-Hern{\'a}ndez}, N.~L., {Damour}, F., {et~al.} 2002, \aap, 381, 571, \dodoi{10.1051/0004-6361:20011516}

\bibitem[{{Portegies Zwart} {et~al.}(2010){Portegies Zwart}, {McMillan}, \& {Gieles}}]{pz10}
{Portegies Zwart}, S.~F., {McMillan}, S. L.~W., \& {Gieles}, M. 2010, \araa, 48, 431, \dodoi{10.1146/annurev-astro-081309-130834}

\bibitem[{{Rieke} \& {Lebofsky}(1985)}]{RiekeLebofsky}
{Rieke}, G.~H., \& {Lebofsky}, M.~J. 1985, \apj, 288, 618, \dodoi{10.1086/162827}

\bibitem[{{Rodr{\'\i}guez} {et~al.}(2023){Rodr{\'\i}guez}, {Lee}, {Whitmore}, {Thilker}, {Maschmann}, {Chandar}, {Deger}, {Boquien}, {Dale}, {Larson}, {Williams}, {Kim}, {Schinnerer}, {Rosolowsky}, {Leroy}, {Emsellem}, {Sandstrom}, {Kruijssen}, {Grasha}, {Watkins}, {Barnes}, {Sormani}, {Kim}, {Anand}, {Chevance}, {Bigiel}, {Klessen}, {Hassani}, {Liu}, {Faesi}, {Cao}, {Belfiore}, {Pessa}, {Kreckel}, {Groves}, {Pety}, {Indebetouw}, {Egorov}, {Blanc}, {Saito}, \& {Hughes}}]{2023ApJ...944L..26R}
{Rodr{\'\i}guez}, M.~J., {Lee}, J.~C., {Whitmore}, B.~C., {et~al.} 2023, \apjl, 944, L26, \dodoi{10.3847/2041-8213/aca653}

\bibitem[{{Ryon} {et~al.}(2017){Ryon}, {Gallagher}, {Smith}, {Adamo}, {Calzetti}, {Bright}, {Cignoni}, {Cook}, {Dale}, {Elmegreen}, {Fumagalli}, {Gouliermis}, {Grasha}, {Grebel}, {Kim}, {Messa}, {Thilker}, \& {Ubeda}}]{ryon17}
{Ryon}, J.~E., {Gallagher}, J.~S., {Smith}, L.~J., {et~al.} 2017, \apj, 841, 92, \dodoi{10.3847/1538-4357/aa719e}

\bibitem[{{Sandstrom} {et~al.}(2023){Sandstrom}, {Chastenet}, {Sutter}, {Leroy}, {Egorov}, {Williams}, {Bolatto}, {Boquien}, {Cao}, {Dale}, {Lee}, {Rosolowsky}, {Schinnerer}, {Barnes}, {Belfiore}, {Bigiel}, {Chevance}, {Grasha}, {Groves}, {Hassani}, {Hughes}, {Klessen}, {Kruijssen}, {Larson}, {Liu}, {Lopez}, {Meidt}, {Murphy}, {Sormani}, {Thilker}, \& {Watkins}}]{2023ApJ...944L...7S}
{Sandstrom}, K.~M., {Chastenet}, J., {Sutter}, J., {et~al.} 2023, \apjl, 944, L7, \dodoi{10.3847/2041-8213/acb0cf}

\bibitem[{{Schinnerer} \& {Leroy}(2024)}]{evaadam24}
{Schinnerer}, E., \& {Leroy}, A.~K. 2024, arXiv e-prints, arXiv:2403.19843, \dodoi{10.48550/arXiv.2403.19843}

\bibitem[{{Schutte} {et~al.}(1993){Schutte}, {Tielens}, \& {Allamandola}}]{1993ApJ...415..397S}
{Schutte}, W.~A., {Tielens}, A.~G.~G.~M., \& {Allamandola}, L.~J. 1993, \apj, 415, 397, \dodoi{10.1086/173173}

\bibitem[{{Smith} {et~al.}(2007){Smith}, {Draine}, {Dale}, {Moustakas}, {Kennicutt}, {Helou}, {Armus}, {Roussel}, {Sheth}, {Bendo}, {Buckalew}, {Calzetti}, {Engelbracht}, {Gordon}, {Hollenbach}, {Li}, {Malhotra}, {Murphy}, \& {Walter}}]{Smith2007}
{Smith}, J.~D.~T., {Draine}, B.~T., {Dale}, D.~A., {et~al.} 2007, \apj, 656, 770, \dodoi{10.1086/510549}

\bibitem[{{Spoon} {et~al.}(2000){Spoon}, {Koornneef}, {Moorwood}, {Lutz}, \& {Tielens}}]{spoon2000}
{Spoon}, H.~W.~W., {Koornneef}, J., {Moorwood}, A.~F.~M., {Lutz}, D., \& {Tielens}, A.~G.~G.~M. 2000, \aap, 357, 898, \dodoi{10.48550/arXiv.astro-ph/0003457}

\bibitem[{{Sturm} {et~al.}(2000){Sturm}, {Lutz}, {Tran}, {Feuchtgruber}, {Genzel}, {Kunze}, {Moorwood}, \& {Thornley}}]{sturm2000}
{Sturm}, E., {Lutz}, D., {Tran}, D., {et~al.} 2000, \aap, 358, 481, \dodoi{10.48550/arXiv.astro-ph/0002195}

\bibitem[{{Sun} {et~al.}(2024){Sun}, {He}, {Batschkun}, {Levy}, {Emig}, {Rodr{\'\i}guez}, {Hassani}, {Leroy}, {Schinnerer}, {Ostriker}, {Wilson}, {Bolatto}, {Mills}, {Rosolowsky}, {Lee}, {Dale}, {Larson}, {Thilker}, {Ubeda}, {Whitmore}, {Williams}, {Barnes}, {Bigiel}, {Chevance}, {Glover}, {Grasha}, {Groves}, {Henshaw}, {Indebetouw}, {Jim{\'e}nez-Donaire}, {Klessen}, {Koch}, {Liu}, {Mathur}, {Meidt}, {Menon}, {Neumann}, {Pinna}, {Querejeta}, {Sormani}, \& {Tress}}]{sun24}
{Sun}, J., {He}, H., {Batschkun}, K., {et~al.} 2024, \apj, 967, 133, \dodoi{10.3847/1538-4357/ad3de6}

\bibitem[{{Thilker} {et~al.}(2024){Thilker}, {Lee}, {Whitmore}, \& {Maschmann}}]{thilker24}
{Thilker}, D., {Lee}, J., {Whitmore}, B., \& {Maschmann}, D. 2024

\bibitem[{{Thilker} {et~al.}(2022){Thilker}, {Whitmore}, {Lee}, {Deger}, {Chandar}, {Larson}, {Hannon}, {Ubeda}, {Dale}, {Glover}, {Grasha}, {Klessen}, {Kruijssen}, {Rosolowsky}, {Schruba}, {White}, \& {Williams}}]{2022Thilker}
{Thilker}, D.~A., {Whitmore}, B.~C., {Lee}, J.~C., {et~al.} 2022, \mnras, 509, 4094, \dodoi{10.1093/mnras/stab3183}

\bibitem[{{Tielens}(2008)}]{2008ARA&A..46..289T}
{Tielens}, A.~G.~G.~M. 2008, \araa, 46, 289, \dodoi{10.1146/annurev.astro.46.060407.145211}

\bibitem[{{Turner} {et~al.}(2021){Turner}, {Dale}, {Lee}, {Boquien}, {Chandar}, {Deger}, {Larson}, {Mok}, {Thilker}, {Ubeda}, {Whitmore}, {Belfiore}, {Bigiel}, {Blanc}, {Emsellem}, {Grasha}, {Groves}, {Klessen}, {Kreckel}, {Kruijssen}, {Leroy}, {Rosolowsky}, {Sanchez-Blazquez}, {Schinnerer}, {Schruba}, {Van Dyk}, \& {Williams}}]{turner21}
{Turner}, J.~A., {Dale}, D.~A., {Lee}, J.~C., {et~al.} 2021, arXiv e-prints, arXiv:2101.02134.
\newblock \doarXiv{2101.02134}

\bibitem[{{Wei} {et~al.}(2020){Wei}, {Huerta}, {Whitmore}, {Lee}, {Hannon}, {Chandar}, {Dale}, {Larson}, {Thilker}, {Ubeda}, {Boquien}, {Chevance}, {Kruijssen}, {Schruba}, {Blanc}, \& {Congiu}}]{wei20}
{Wei}, W., {Huerta}, E.~A., {Whitmore}, B.~C., {et~al.} 2020, \mnras, 493, 3178, \dodoi{10.1093/mnras/staa325}

\bibitem[{{Whitmore} {et~al.}(2014{\natexlab{a}}){Whitmore}, {Chandar}, {Bowers}, {Larsen}, {Lindsay}, {Ansari}, \& {Evans}}]{whitmore14a}
{Whitmore}, B.~C., {Chandar}, R., {Bowers}, A.~S., {et~al.} 2014{\natexlab{a}}, \aj, 147, 78, \dodoi{10.1088/0004-6256/147/4/78}

\bibitem[{{Whitmore} {et~al.}(1999){Whitmore}, {Zhang}, {Leitherer}, {Fall}, {Schweizer}, \& {Miller}}]{whitmore99}
{Whitmore}, B.~C., {Zhang}, Q., {Leitherer}, C., {et~al.} 1999, \aj, 118, 1551, \dodoi{10.1086/301041}

\bibitem[{{Whitmore} {et~al.}(2014{\natexlab{b}}){Whitmore}, {Brogan}, {Chandar}, {Evans}, {Hibbard}, {Johnson}, {Leroy}, {Privon}, {Remijan}, \& {Sheth}}]{whitmore14}
{Whitmore}, B.~C., {Brogan}, C., {Chandar}, R., {et~al.} 2014{\natexlab{b}}, \apj, 795, 156, \dodoi{10.1088/0004-637X/795/2/156}

\bibitem[{{Whitmore} {et~al.}(2021){Whitmore}, {Lee}, {Chandar}, {Thilker}, {Hannon}, {Wei}, {Huerta}, {Bigiel}, {Boquien}, {Chevance}, {Dale}, {Deger}, {Grasha}, {Klessen}, {Kruijssen}, {Larson}, {Mok}, {Rosolowsky}, {Schinnerer}, {Schruba}, {Ubeda}, {Van Dyk}, {Watkins}, \& {Williams}}]{whitmore21}
{Whitmore}, B.~C., {Lee}, J.~C., {Chandar}, R., {et~al.} 2021, \mnras, 506, 5294, \dodoi{10.1093/mnras/stab2087}

\bibitem[{{Whitmore} {et~al.}(2023{\natexlab{a}}){Whitmore}, {Chandar}, {Rodr{\'\i}guez}, {Lee}, {Emsellem}, {Floyd}, {Kim}, {Kruijssen}, {Mok}, {Sormani}, {Boquien}, {Dale}, {Faesi}, {Henny}, {Hannon}, {Thilker}, {White}, {Barnes}, {Bigiel}, {Chevance}, {Henshaw}, {Klessen}, {Leroy}, {Liu}, {Maschmann}, {Meidt}, {Rosolowsky}, {Schinnerer}, {Sun}, {Watkins}, \& {Williams}}]{2023ApJ...944L..14W}
{Whitmore}, B.~C., {Chandar}, R., {Rodr{\'\i}guez}, M.~J., {et~al.} 2023{\natexlab{a}}, \apjl, 944, L14, \dodoi{10.3847/2041-8213/acae94}

\bibitem[{{Whitmore} {et~al.}(2023{\natexlab{b}}){Whitmore}, {Chandar}, {Rodr{\'\i}guez}, {Lee}, {Emsellem}, {Floyd}, {Kim}, {Kruijssen}, {Mok}, {Sormani}, {Boquien}, {Dale}, {Faesi}, {Henny}, {Hannon}, {Thilker}, {White}, {Barnes}, {Bigiel}, {Chevance}, {Henshaw}, {Klessen}, {Leroy}, {Liu}, {Maschmann}, {Meidt}, {Rosolowsky}, {Schinnerer}, {Sun}, {Watkins}, \& {Williams}}]{whitmore23}
---. 2023{\natexlab{b}}, \apjl, 944, L14, \dodoi{10.3847/2041-8213/acae94}

\bibitem[{{Williams} {et~al.}(2024){Williams}, {Lee}, {Larson}, {Leroy}, {Sandstrom}, {Schinnerer}, {Thilker}, {Belfiore}, {Egorov}, {Rosolowsky}, {Sutter}, {DePasquale}, {Pagan}, {Anand}, {Barnes}, {Bigiel}, {Boquien}, {Cao}, {Chastenet}, {Chevance}, {Chown}, {Dale}, {Eibensteiner}, {Emsellem}, {Faesi}, {Glover}, {Grasha}, {Hannon}, {Hassani}, {Henshaw}, {Jim{\'e}nez-Donaire}, {Kim}, {Klessen}, {Koch}, {Li}, {Liu}, {Meidt}, {M{\'e}ndez-Delgado}, {Murphy}, {Neumann}, {Neumann}, {Neumayer}, {Oakes}, {Pathak}, {Pety}, {Pinna}, {Querejeta}, {Ramambason}, {Romanelli}, {Sormani}, {Stuber}, {Sun}, {Teng}, {Usero}, {Watkins}, \& {Weinbeck}}]{phangs-jwst-pipeline}
{Williams}, T.~G., {Lee}, J.~C., {Larson}, K.~L., {et~al.} 2024, arXiv e-prints, arXiv:2401.15142, \dodoi{10.48550/arXiv.2401.15142}

\bibitem[{{Yamada} {et~al.}(2013){Yamada}, {Oyabu}, {Kaneda}, {Yamagishi}, {Ishihara}, {Kim}, \& {Im}}]{2013PASJ...65..103Y}
{Yamada}, R., {Oyabu}, S., {Kaneda}, H., {et~al.} 2013, \pasj, 65, 103, \dodoi{10.1093/pasj/65.5.103}

\bibitem[{{Zhang} \& {Ho}(2023)}]{zhang23}
{Zhang}, L., \& {Ho}, L.~C. 2023, \apj, 943, 60, \dodoi{10.3847/1538-4357/acab60}

\end{thebibliography}
